\documentclass[11pt]{report}
\usepackage{graphicx}
\usepackage[utf8]{inputenc} 
\pdfoutput=1

\usepackage{fancyhdr}

\usepackage{amsmath,amsfonts,amssymb,epsfig,empheq}
\usepackage{slashed,xcolor}
\usepackage[symbol]{footmisc}

\numberwithin{equation}{section}

\usepackage{graphicx}
\usepackage{cancel}
\usepackage{cite}
\usepackage{color}
\usepackage{dsfont}
\usepackage{amssymb}
\usepackage{booktabs,multirow,tabularx}
\usepackage{lscape}
\usepackage{float} % improved interface for floating objects
\usepackage{subfig} % supersedes the subfigure package
\usepackage{array}
\newcolumntype{?}{!{\vrule width 1pt}}
\usepackage{lineno}
\makeatletter
\def\hlinewd#1{%
\noalign{\ifnum0=`}\fi\hrule \@height #1 %
\futurelet\reserved@a\@xhline}
\makeatother

\usepackage{xspace}

\newcommand{\mgamclong}{{\sc\small MadGraph5\_aMC@NLO}}
\newcommand{\mgamcshort}{{\sc\small MG5\_aMC}}
\newcommand{\amcatnlo}{{\sc\small aMC@NLO}}
\newcommand{\mg}{{\sc\small MadGraph5}}
\newcommand{\cuttools}{{\sc CutTools}}
\newcommand{\samurai}{{\sc Samurai}}
\newcommand{\ninja}{{\sc Ninja}}
\newcommand{\collier}{{\sc Collier}}
\newcommand{\iregi}{{\sc Iregi}}
\newcommand{\pjfry}{{\sc PJFry++}}
\newcommand{\golem}{{\sc Golem95}}
\newcommand{\ml}{{\sc MadLoop}}
\newcommand{\aloha}{{\sc Aloha}}
\newcommand{\helas}{{\sc Helas}}

\newcommand{\mfks}{{\sc MadFKS}}
\newcommand{\recola}{{\sc Recola}}
\newcommand{\mcfm}{{\sc MCFM}}
\newcommand{\sherpa}{{\sc Sherpa}}
\newcommand{\alpgen}{{\sc Alpgen}}
\newcommand{\openloops}{{\sc OpenLoops}}
\newcommand{\gosam}{{\sc GoSam}}
\newcommand{\feynrules}{{\sc FeynRules}}
\newcommand{\nloct}{{\sc NloCt}}
\newcommand{\luxqed}{{\sc LUXqed}}
\newcommand\prompt{{\tt MG5\_aMC>}}
\newcommand{\nmax}{{\tt n_{\tt max}}}
\newcommand{\mmax}{{\tt m_{\tt max}}}
\newcommand{\lhapdf}{{\sc LHAPDF6}}
\newcommand{\fastjet}{{\sc FastJet}}
\newcommand{\maddipole}{{\sc MadDipole}}

\def\gev{\ \mathrm{GeV}}

\newcommand{\exclude}[1]{}
\long\def\symbolfootnote[#1]#2{\begingroup%
\def\thefootnote{\fnsymbol{footnote}}\footnote[#1]{#2}\endgroup}

\def\ov{\overline}

\def\msbar{{\ov {\rm MS}}}

\newcommand\sss{}
\newcommand\mydot{\!\cdot\!}

\newcommand\ep{\epsilon}
\def\beq{\begin{equation}}
\def\eeq{\end{equation}}
\def\beqn{\begin{eqnarray}}
\def\eeqn{\end{eqnarray}}
\def\abs#1{\left|#1\right|}
\newcommand\allproc{{\cal R}}

\newcommand\allprocnpo{\allproc_{n+1}}
\newcommand\allprocn{\allproc_{n}}

\newcommand\nini{n_{\sss I}}
\newcommand\nlight{n_{\sss L}}
\newcommand\nlightB{\nlight^{\sss (B)}}
\newcommand\nlightR{\nlight^{\sss (R)}}

\newcommand\nheavy{n_{\sss H}}

\newcommand\avg{{\cal N}}
\newcommand\ident{{\cal I}}
\newcommand\Ione{\ident_1}
\newcommand\Itwo{\ident_2}
\newcommand\amp{{\cal A}}

\newcommand\ampnt{\amp^{(n,0)}}
\newcommand\ampnUV{\amp^{(n,\mathrm{UV})}}
\newcommand\ampnRT{\amp^{(n,R_2)}}
\newcommand\ampnpot{\amp^{(n+1,0)}}
\newcommand\ampnl{\amp^{(n,1)}}
\newcommand\ampnll{\amp^{(n,\ell)}}
\newcommand\JetsB{J^{\nlightB}}

\def\remove#1#2{#1\hspace{-#2truecm}\backslash}

\newcommand\isubrmv{\remove{i}{0.125}}

\newcommand\FKSpairs{{\cal P}_{\sss\rm FKS}}

\newcommand\xicut{\xi_{cut}}
\newcommand\ximax{\xi_{\rm max}}
\newcommand\deltaO{\delta_{\sss O}}
\newcommand\deltaI{\delta_{\sss I}}

\newcommand\Qop{\vec{Q}}

\newcommand\ampsq{{\cal M}}

\newcommand\ampsqnt{\ampsq^{(n,0)}}
\newcommand\ampsqnttilde{\tilde{\ampsq}^{(n,0)}}

\newcommand\ampsqnpot{\ampsq^{(n+1,0)}}
\newcommand\ampsqnl{\ampsq^{(n,1)}}

\newcommand\xii{\xi_i}
\newcommand\yij{y_{ij}}
\newcommand\phii{\varphi_i}

\newcommand\xic{\left(\frac{1}{\xii}\right)_c}

\newcommand\omyijd{\left(\frac{1}{1-\yij}\right)_\delta}

\newcommand\Sfun{{\cal S}}
\newcommand\Sfunij{\Sfun_{ij}}

\newcommand\stepf{\Theta}
\newcommand\NC{N_{\sss c}}

\newcommand\phsp{\mathrm{d}\phi}

\newcommand\phspn{\phsp_{n}}
\newcommand\phspnpo{\phsp_{n+1}}
\newcommand\tphsp{\mathrm{d}\widetilde{\phi}}

\newcommand\tphspnij{\tphsp_{n}^{ij}}

\newcommand{\boldirrep}{\mathbf}
\newcommand{\irrepbase}[1]{\ensuremath{\boldirrep{#1}}}

\newlength{\irrepwidth}
\newlength{\irrepbarthickness}
\newlength{\irrepbarheight}
\newcommand{\irrepbarbase}[1]{%
    \settoheight{\irrepbarheight}{\irrepbase{#1}}%
    \settowidth{\irrepwidth}{\irrepbase{#1}}%
    \makebox[0pt][l]{\irrepbase{#1}}%
    \rule[1.2\irrepbarheight]{\irrepwidth}{\irrepbarthickness}%
}

\def\primes#1#2{\count0=#1 \loop \ifnum\count0>0 \advance\count0 by -1 #2\repeat}
\newcommand{\irrep}[2][0]{\ensuremath{\irrepbase{#2}^{\primes{#1}{\prime}}}}
\newcommand{\irrepbar}[2][0]{\ensuremath{\irrepbarbase{#2}^{\primes{#1}{\prime}}}}

\def\eg{{\it e.g.}}
\def\ie{{\it i.e.}}

\def\cf{{\it cf.}}

\def\twomat[#1,#2][#3,#4]{\left( \begin{array}{cc} #1 & #2 \\ #3 & #4 \end{array} \right)}
\def\threemat[#1,#2,#3][#4,#5,#6][#7,#8,#9]{\left( \begin{array}{ccc} #1 & #2 & #3\\ #4 & #5 & #6 \\ #7 & #8 & #9 \end{array} \right)}
\def\twovec[#1,#2]{\left( \begin{array}{c} #1  \\ #2 \end{array} \right)}
\def\thv[#1,#2,#3]{\left( \begin{array}{c} #1 \\ #2 \\ #3 \end{array} \right)}
\def\twv[#1,#2]{\left( \begin{array}{c} #1 \\ #2 \end{array} \right)}

\begin{document}

\begin{titlepage}

% Upper part of the page
\begin{center}
\includegraphics[height=0.08\textheight]{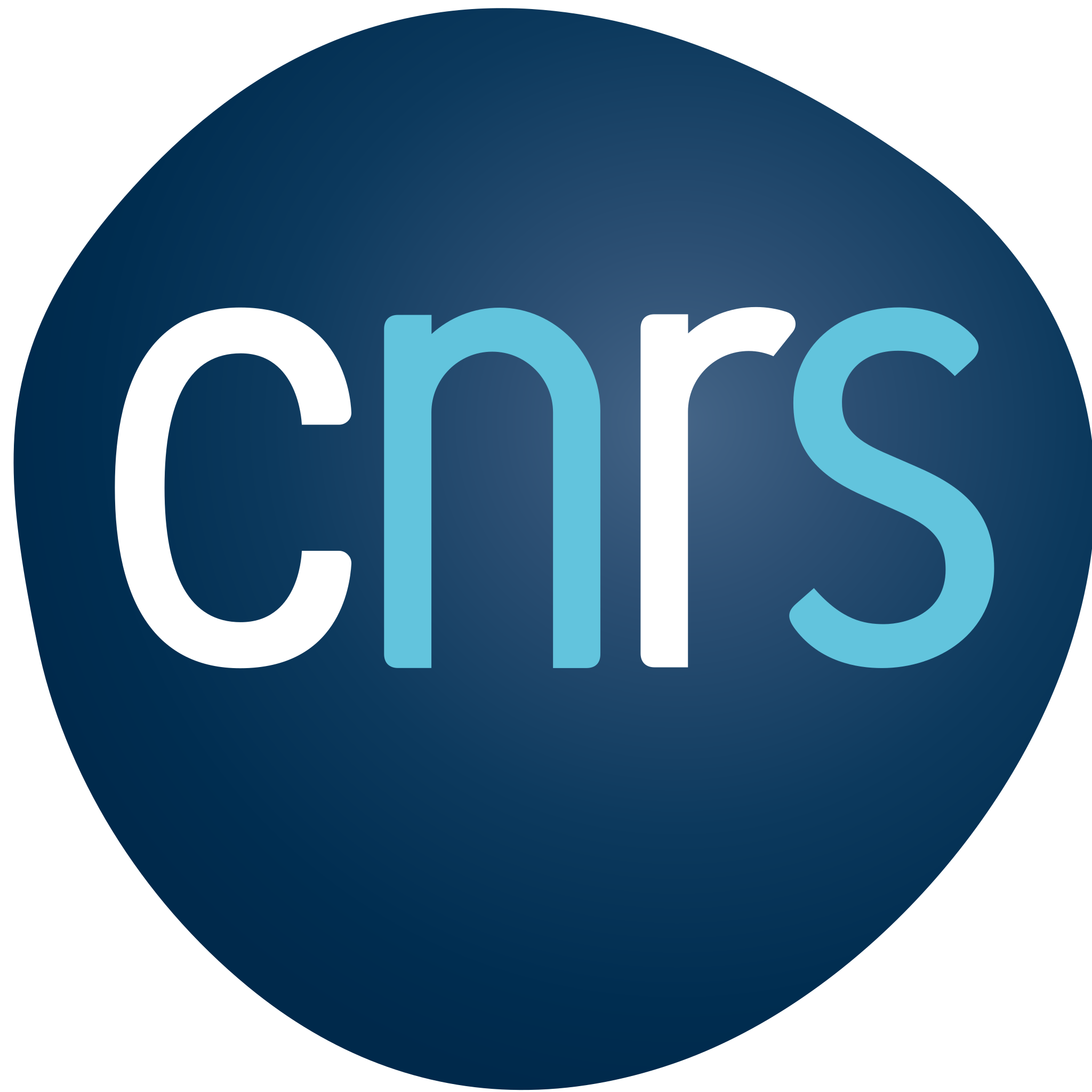}\hfill\includegraphics[height=0.08\textheight]{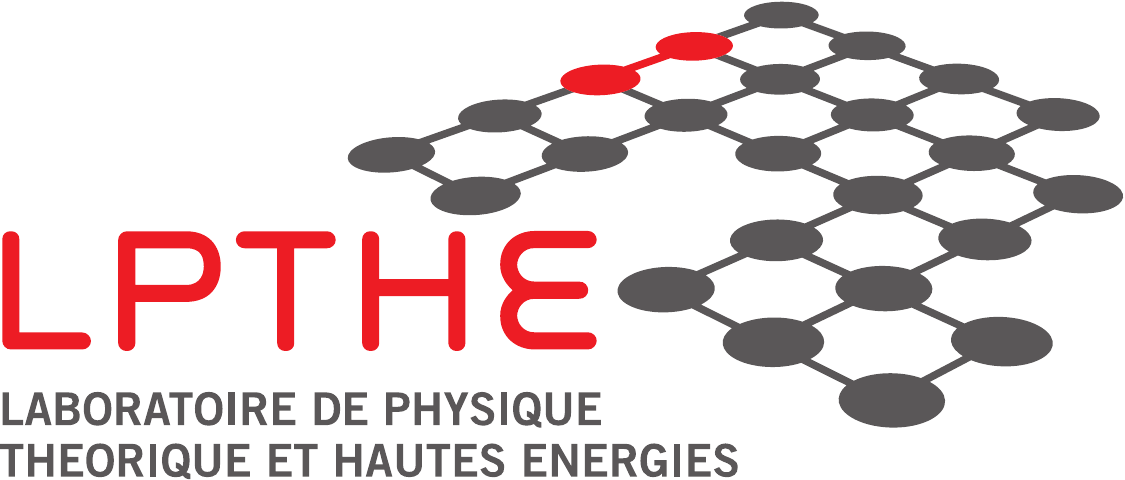}\hfill\includegraphics[height=0.1\textheight]{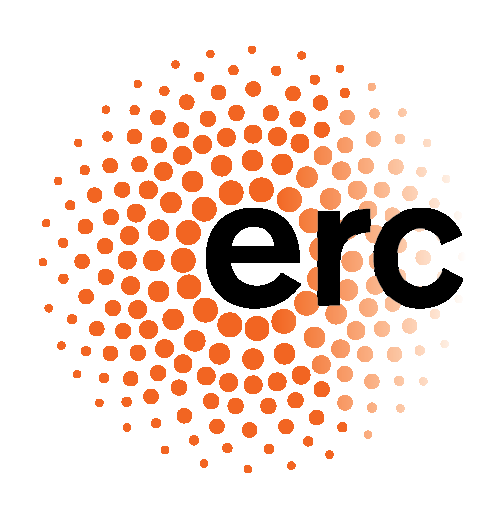}\hfill\includegraphics[height=0.08\textheight]{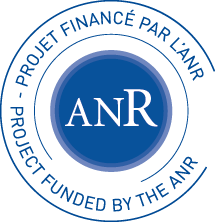}\hfill\includegraphics[height=0.08\textheight]{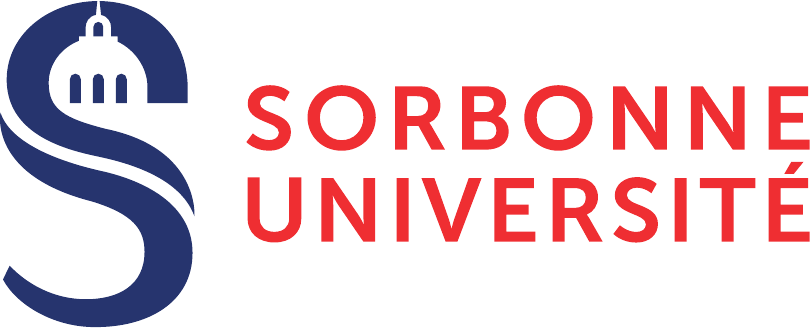}
\end{center}

\begin{center}
\vspace{0.8cm}
  
\centerline{\LARGE \scshape Th\`ese d'habilitation de Sorbonnne Universit\'e}
\vspace{0.5cm}
\centerline{\Large Habilitation \`a Diriger des Recherches}
\centerline{\Large Sp\'ecialit\'e : Physique}
%\centerline{\large \'Ecole doctorale n 564: Physique en \^Ile-de-France}
\vspace{0.8cm}
%\centerline{Sur le th\`eme : }
%\vskip .5cm
{\LARGE \bf Automation of Electroweak Corrections}

\vspace{1cm}

%\centerline{présentée par}

\centerline{\Large \bf Hua-Sheng Shao}

\vspace*{5mm}

{\sl Laboratoire de Physique Th\'eorique et Hautes Energies (LPTHE),\\ UMR 7589,
Sorbonne Universit\'e et CNRS, 4 place Jussieu, 75252 Paris Cedex 05, France.  }
\end{center}

\vspace{0.3cm}

\centerline{\large Soutenue le 03 d\'ecembre 2024} 
\centerline{\large devant le jury compos\'e de}

\vspace{0.3cm}

\hspace{2cm}\begin{minipage}{2.5in}
\textbf{Prof. Fawzi BOUDJEMA} \\
{\small LAPTh, Annecy}\\
\textbf{Prof. Elena FERREIRO}\\
{\small University of Santiago de Compostela}\\
\textbf{Prof. Anna KULESZA}\\
{\small University of M\"unster}\\
\textbf{Prof. Bertrand LAFORGE}\\
{\small Sorbonne University}\\
\textbf{Prof. Michela PETRINI}\\
{\small Sorbonne University}\\
\textbf{Prof. Gavin SALAM}\\
{\small University of Oxford}
\end{minipage}
\hfill
\begin{minipage}{1.3in}
Rapporteur\\
\phantom{{\small LAPTh, Annecy}}\\
Examinatrice\\
\phantom{{\small University of Santiago de Compostela}}\\
Rapporteuse\\
\phantom{{\small University of M\"unster}}\\
Pr\'esident du jury\\
\phantom{{\small Sorbonne University}}\\
Examinatrice\\
\phantom{{\small Sorbonne University}}\\
Rapporteur\\
\phantom{{\small University of Oxford}}
\end{minipage}\hspace{2.6cm}

%\begin{flushleft}

%{\bf Fawzi BOUDJEMA}\hspace{5cm} Rapporteur\\
%{\bf Elena G. FERREIRO}\hspace{5cm} Examinatrice

%\end{flushleft}

\vspace{0.2cm}

\begin{center}

\today

%\vspace{1cm}
%\includegraphics[width=0.3\textwidth]{sorbonne-logo.pdf}
  
\end{center}

\symbolfootnote[0]{{\tt e-mail:}}
\symbolfootnote[0]{{\tt huasheng.shao@lpthe.jussieu.fr}}

\vspace{0.7cm}

%\abstract{I do some stuff}

\vfill

\end{titlepage}

\newpage
\thispagestyle{empty}
\mbox{}
\newpage

\setcounter{page}{0}
\pagenumbering{roman}
%\frontmatter
\tableofcontents

\setcounter{footnote}{0}

\chapter*{Abstract}
\addcontentsline{toc}{chapter}{Abstract}

This dissertation addresses a topic that I have worked on over the past decade: the automation of next-to-leading order electroweak corrections in the Standard Model of particle physics. After introducing the basic concepts and techniques of next-to-leading order QCD calculations that underpin the \mgamclong\ framework,  I present a few key features relevant to the automated  next-to-leading order electroweak contributions to short-distance cross sections, with an emphasis on the mixed QCD and electroweak coupling expansions. These include the FKS subtraction, the renormalization and electroweak input parameter schemes, and the complex mass scheme for dealing with unstable particles. Issues related to the initial or final photons and leptons are also discussed. Two remaining challenges are highlighted if one wishes to go beyond next-to-leading order computations. Some phenomenological applications at the LHC are given to demonstrate the relevance of electroweak corrections at colliders. Finally, an outlook on future studies concludes the dissertation.

\chapter*{R\'esum\'e}
 \addcontentsline{toc}{chapter}{R\'esum\'e}

Cette th\'ese traite d'un sujet sur lequel j'ai travaill\'e au cours de la dernière d\'ecennie : l'automatisation des corrections \'electrofaibles d'ordre sup\'erieur (NLO) dans le Mod\`ele Standard de la physique des particules. Apr\`es avoir introduit les concepts et techniques de base des calculs d'ordre sup\'erieur (NLO) en QCD qui sous-tendent le cadre \mgamclong, je pr\'esente quelques caract\'eristiques cl\'es pertinentes aux contributions \'electrofaibles automatis\'ees d'ordre sup\'erieur aux sections efficaces \`a courte distance, en mettant l'accent sur les expansions en couplage mixte QCD et \'electrofaible. Celles-ci incluent la soustraction FKS, les sch\'emas de renormalisation et des param\`etres d'entr\'ee \'electrofaibles, ainsi que le schéma de masse complexe pour traiter les particules instables. Les probl\`emes li\'es aux photons et leptons initiaux ou finaux sont \'egalement discut\'es. Deux d\'efis majeurs sont soulign\'es pour ceux qui souhaitent aller au-del\`a des calculs d'ordre sup\'erieur. Quelques applications ph\'enom\'enologiques au LHC sont pr\'esent\'ees pour d\'emontrer la pertinence des corrections \'electrofaibles dans les collisions de particules. Enfin, une perspective sur les \'etudes futures conclut cette th\`ese.

\chapter*{Acknowledgements}
 \addcontentsline{toc}{chapter}{Acknowledgements}

For humans, whose lifespans are relatively short, a decade is often a milestone worth commemorating. For many of us, ten years can signify a new beginning. Personally, ten years ago, I earned my PhD and started my postdoctoral research career at CERN. Looking back over the past decade, I can clearly see my growth and transformation.

First and foremost, I would like to express my heartfelt thanks to the jury members who reviewed my HDR dissertation and participated in my HDR defense: Fawzi Boudjema, Elena Ferreiro, Anna Kulesza, Bertrand Laforge, Michela Petrini, and Gavin Salam. They are all highly esteemed physicists with numerous responsibilities and commitments. I deeply appreciate them taking the time out of their busy schedules to carefully read my dissertation, attend my defense, and provide many valuable comments. My interactions with them have been immensely beneficial.

This dissertation is based on my research over the past decade, particularly the collaborative work with Rikkert Frederix, Stefano Frixione, Valentin Hirschi, Davide Pagani, and Marco Zaro. Without their outstanding contributions, most of the content in this dissertation would not exist! I am especially grateful to them—not only have I learned a great deal from our collaborations, but I have also greatly enjoyed the process. I also want to thank other members of the \mgamclong\ collaboration, particularly Fabio Maltoni, Olivier Mattelaer, and Paolo Torrielli. Interacting with them, whether about physics or other topics, has always been a pleasure. I would also like to express my gratitude to my collaborators David d'Enterria and Lukas Simon, who kindly answered my questions during the writing of this document.

I am grateful to Matteo Cacciari, Benjamin Fuks, Michelangelo Mangano, and Davide Pagani for reading this thesis and providing many invaluable suggestions for improvement. I also want to thank the members of the HDR Committee in Physics at Sorbonne University, especially Eli Ben-Haim, for their guidance during my HDR application process. I am indebted to my colleagues Matteo Cacciari, Marco Cirelli, Benoit Estienne, Benjamin Fuks, Mark Goodsell, Marco Picco, and Michela Petrini for their generous help and for answering my many questions during my HDR application. My thanks also extend to all my colleagues at the LPTHE and to those in other laboratories in the Paris area with whom I have interacted during this period.

Special thanks go to my two long-term collaborators in Paris, Benjamin Fuks and Jean-Philippe Lansberg. My collaborations with them over the past decade, both in scientific research and beyond, have transformed me from a fledgling postdoc into an independent PI.

Last but not least, I want to thank my family. I am deeply grateful to my wife, Ping, for her unwavering support, and to my daughters, Ana\"is, Alice, and Angela, for allowing their dad to work undisturbed in his room.

I acknowledge the support from the ERC (grant 101041109 ``BOSON") and the French ANR (grant ANR-20-CE31-0015, ``PrecisOnium''). Views and opinions expressed are however those of the authors only and do not necessarily reflect those of the European Union or the European Research Council Executive Agency. Neither the European Union nor the granting authority can be held responsible for them.

%\mainmatter

 \setcounter{footnote}{2}
 \cleardoublepage
 \pagenumbering{arabic}
 %\clearpage
\chapter{Introduction}
\label{SEC:INTRODUCTION}
\setcounter{page}{1}

An important and elegant guideline for advancing physics is the pursuit of unification. The idea behind unification is that, from an aesthetic and philosophical perspective, the vast diversity of phenomena observed in Nature is governed by a few simple, fundamental physics laws. This approach offers deep insights and connections between seemingly unrelated theories or phenomena,  expanding our knowledge and leading to entirely new predictions. Historically, Newton's theory of gravity unifies terrestrial and celestial motion. Maxwell's equations reveal that electricity and magnetism are two sides of the same coin, now known as the electromagnetic force. Einstein's special relativity unifies space and time into a single framework, spacetime. A more recent example is the unification of the electromagnetic and weak interactions into the electroweak (EW) interaction by Glashow~\cite{Glashow:1961tr}, Weinberg~\cite{Weinberg:1967tq}, and Salam~\cite{Salam:1968rm}. This is now known as the Standard Model (SM) of particle physics, which also incorporates the theory of the strong interaction, Quantum Chromodynamics (QCD)~\cite{Fritzsch:1973pi,Gross:1973ju,Gross:1973id,Politzer:1973fx,Gross:1974cs}. Each unification has revolutionized our understanding of Nature. There is no reason for the pursuit of unification to end here. 

In the past decades, especially in the pre-Large Hadron Collider (LHC) era, many believed that the path toward energies beyond the EW scale, around $100$ GeV, was clear. At a few TeV, supersymmetry was expected to show up to unify bosons and fermions. At 10$^{16}$ GeV, the strong and EW forces could merge into a single force in a grand unified theory. Finally, at the Planck scale of 10$^{19}$ GeV, a quantum gravity theory--often called a Theory of Everything--might unify the gravitational force with the other three fundamental forces.

Although the SM of particle physics is a highly successful theory for describing the strong and EW interactions among subatomic particles, withstanding almost all experimental tests, there are many plausible clues--beyond aesthetics arguments like unification--that suggest the SM is merely a low-energy effective theory of a yet-unknown, more fundamental framework. However, despite the completion of over half of LHC Run III and an integrated luminosity exceeding 180 fb$^{-1}$ at 13.6 TeV in proton-proton ($pp$) collisions, the discovery of physics beyond the Standard Model (BSM) remains elusive. This could be due either to a higher characteristic BSM energy scale than previously anticipated, placing it beyond the reach of the LHC, or to BSM signals that are particularly difficult to detect in direct searches. While it is currently unclear which of these two scenarios reflects reality, each suggests a different long-term strategy. In the first case, probing larger particle mass scales would require higher collider energies, whereas in the second, accumulating more data and reducing systematic uncertainties would be crucial. In both cases, indirect searches may play a significant role.

To understand and interpret experiments, theorists must continually improve both the scope and precision of their calculations. Scope has two main aspects. The first is the ability of computational tools to efficiently handle the matrix elements for processes in both the SM and new physics models. The second involves enabling fully realistic final-state simulations by combining these matrix elements with general-purpose parton shower Monte Carlo (PSMC) programs. Precision, in most applications, refers to calculating higher order quantum radiative corrections in perturbation theory. These quantum corrections generally encompass three types of computations (or their combinations).~\footnote{For proton-initiated processes, such as those at the LHC, another crucial fact in determining the precision of perturbative calculations is the parton distribution functions of proton. Their energy-scale evolution and extraction from experimental data require inputs from higher-order quantum corrections. However, this will not be discussed in this context.}
\begin{itemize}
\item Fixed order: A truncated series expansion of small coupling constants, such as the strong coupling constant $\alpha_s$ and the EW or QED coupling constant $\alpha$, within the matrix elements;
\item Resummation: An all-order calculation of certain universal terms (\eg, logarithms or $\pi^2$ terms) using renormalization group equations;
\item Parton shower: A numerical simulation that models the successive emission of radiation, such as quarks, gluons, or photons, from an energetic parton using Markov Chain Monte Carlo methods.
\end{itemize}
In the following, I will focus solely on fixed-order calculations, although the other two topics have been the subject of extensive research over the past decades and warrant separate discussions. 

Given the significant numerical value of $\alpha_s$ and the critical role of hadron collision physics in the LHC era, QCD corrections have been particularly prominent over the last decades. Fully differential next-to-leading order (NLO) results in $\alpha_s$, along with their matching to parton showers, have become standard and can now be automatically calculated using several public programs (see,\eg, the review~\cite{Proceedings:2018jsb}), even for processes with complex final states. Additionally, with customized approaches, increasingly more next-to-NLO (NNLO)~\cite{Zanderighi:2017,Heinrich:2020ybq} and even next-to-NNLO (N$^3$LO)~\cite{Anastasiou:2015vya,Anastasiou:2016cez,Dreyer:2016oyx,Mistlberger:2018etf,Dreyer:2018qbw,Dulat:2018bfe,Cieri:2018oms,Chen:2019lzz,Chen:2019fhs,Duhr:2020seh,Duhr:2020sdp,Chen:2021isd,Chen:2021vtu,Baglio:2022wzu} predictions are becoming available for key low-multiplicity processes, with varying levels of inclusiveness (also see the reviews~\cite{Heinrich:2020ybq,Huston:2023ofk,Andersen:2024czj}).

EW corrections, which account for quantum radiative effects involving weak and electromagnetic interactions, have become increasingly important to consider in full generality. First, based on the values of $\alpha_s\sim 0.1$ and $\alpha\sim 0.01$, one can expect that the sizes of NNLO QCD and NLO EW corrections are numerically comparable. Second, LHC measurements have reached deeper into the multi-TeV region, which has never been explored before, and would be particularly sensitive to BSM effects if BSM physics is realized as described in the first scenario. In this region, the naive scaling behavior of $\alpha$ could actually be violated, leading EW corrections to grow faster than their QCD counterparts due to the presence of the so-called EW Sudakov logarithms~\cite{Kuroda:1990wn,Degrassi:1992ue,Ciafaloni:1998xg,Ciafaloni:2000df,Denner:2000jv,Denner:2001gw}, a point I will discuss later in section \ref{sec:EWSudakovLog}.~\footnote{Due to the presence of EW Sudakov logarithms, EW corrections are also considered necessary in dark matter indirect detection~\cite{Ciafaloni:2010ti}.} Unlike in QCD, EW Sudakov logarithms cannot be fully eliminated even in sufficiently inclusive observables because the initial state (\eg, protons at the LHC) is not a singlet under the weak SU(2)$_L$ gauge group. 

Furthermore, an electron-positron ($e^-e^+$) collider has been prioritized for studying EW, Higgs, and top quark physics in the future of particle physics following the LHC~\cite{CERNNextCollider:2024}. In this context, the incorporation of EW corrections in physics simulations is crucial. A particularly significant example is the large electromagnetic logarithm from initial photon radiation due to the tiny electron mass, which highlights the prominent role of EW corrections at $e^-e^+$ colliders. While it is clear that the decision to conduct a fully-fledged higher-order computation for a given observable must be made after careful consideration of the benefits versus costs, increasing precision in theoretical predictions is always advantageous. 

The combination of the necessity for NLO EW corrections and the demand for flexibility and applicability to arbitrary processes makes automation a natural solution, significantly reducing resource costs. This is particularly compelling given the remarkable success of automation in QCD, where NLO results are now produced on a massive scale and form the backbone of the ATLAS and CMS proton-proton event simulations. While automation of NLO EW corrections is not yet as comprehensive as in QCD, steady progress has been made in automating both one-loop and real-emission contributions. Collaborative efforts in this direction include tools such as \gosam~\cite{Cullen:2011ac,GoSam:2014iqq} in conjunction with \maddipole~\cite{Frederix:2008hu,Gehrmann:2010ry} or \sherpa~\cite{Gleisberg:2008ta,Schonherr:2017qcj,Sherpa:2019gpd}, as well as \openloops~\cite{Cascioli:2011va,Kallweit:2014xda,Buccioni:2019sur} and \recola~\cite{Actis:2012qn,Actis:2016mpe,Biedermann:2017yoi} working alongside \sherpa\ or private phase-space integrators, and \mgamclong\ (\mgamcshort\  for short henceforth)~\cite{Alwall:2014hca,Frederix:2018nkq}. Recent results produced by these tools clearly demonstrate how automation enables the tackling of complex problems that would be impractical to solve using traditional methods.

This dissertation explores the development and implementation of computational tools that automate the calculation of EW corrections, primarily within the \mgamcshort\ framework. I will discuss both the theoretical foundations and phenomenological applications relevant to the LHC and future colliders. The dissertation synthesizes lectures and talks I have given, along with several publications I co-authored in recent years.  Specifically, the following co-authored publications form the basis of this work:
\begin{itemize}
\item ``{\bf The automation of next-to-leading order electroweak calculations}''  
  \\{}R.~Frederix, S.~Frixione, V.~Hirschi, D.~Pagani, {\bf H.-S.~Shao} and M.~Zaro
  \\{}JHEP {\bf 1807}, 185 (2018), [arXiv:1804.10017 [hep-ph]].
 \item  ``{\bf RIP $H b \bar b$: How other Higgs production modes conspire to kill a rare signal at the LHC}'' 
  \\{}D.~Pagani, {\bf H.-S.~Shao} and M.~Zaro
  \\{}JHEP, {\bf 2011}, 036 (2020), [arXiv:2005.10277 [hep-ph]].
 \item ``{\bf Automated EW corrections with isolated photons: $t \bar t \gamma$, $t \bar t \gamma\gamma$ and $t \gamma j$ as case studies}''
  \\{}D.~Pagani, {\bf H.-S.~Shao}, I.~Tsinikos and M.~Zaro
  \\{}JHEP, {\bf 2109}, 155 (2021), [arXiv:2106.02059 [hep-ph]].
 \item ``{\bf The complete NLO corrections to dijet hadroproduction}'' 
  \\{}R.~Frederix, S.~Frixione, V.~Hirschi, D.~Pagani, {\bf H.-S.~Shao} and M.~Zaro
  \\{}JHEP {\bf 1704}, 076 (2017) [arXiv:1612.06548 [hep-ph]].
 \item ``{\bf Electroweak and QCD corrections to top-pair hadroproduction in association with heavy bosons}''
  \\{}S.~Frixione, V.~Hirschi, D.~Pagani, {\bf H.-S.~Shao} and M.~Zaro
  \\{}JHEP {\bf 1506}, 184 (2015), [arXiv:1504.03446[hep-ph]].
% \item ``{\bf UFO 2.0 -- The Universal Feynman Output format}''
%\\{}L.~Darm\'e, \ldots, {\bf H.-S.~Shao} \textit{et al.}
%\\{}Eur.\ Phys.\ J.\ {\bf C83}, no.07, 631 (2023), [arXiv:2304.09883 [hep-ph]].
\item ``{\bf The automated computation of tree-level and next-to-leading order differential cross sections, and their matching to parton shower simulations}''
  \\{}J.~Alwall, R.~Frederix, S.~Frixione, V.~Hirschi, F.~Maltoni, O.~Mattelaer, {\bf H.-S.~Shao}, T.~Stelzer, P.~Torrielli and M.~Zaro
  \\{}JHEP {\bf 1407}, 079 (2014), [arXiv:1405.0301[hep-ph]].
\end{itemize}
The last paper serves as the standard reference for the \mgamcshort\ framework, while the first paper is the main reference for the development of NLO EW corrections within the same framework.

As a disclaimer, this dissertation does not aim to provide a comprehensive overview of the topic, particularly given the existence of a recent, excellent review on EW corrections~\cite{Denner:2019vbn} and a standard reference~\cite{Denner:1991kt}. Due to the vastness of the background material, I will briefly outline the main characteristics and primarily focus on the techniques pertinent to the realization of NLO EW automation within the \mgamcshort\ framework.

The remaining context is organized as follows. Chapter~\ref{SEC:NLO} introduces the basic concepts and techniques for NLO QCD computations within the \mgamcshort\ framework. Chapter~\ref{SEC:NLOEW} presents the key features of the automated computations of NLO short-distance cross sections in the context of mixed QCD and EW coupling expansions, highlighting some remaining issues. To exemplify the relevance of NLO EW corrections for collider physics, Chapter~\ref{SEC:PHENO} examines NLO EW corrections across a wide range of processes at the LHC and explores subleading NLO terms with varying impacts through representative examples. Finally, Chapter \ref{SEC:outlook} provides an outlook for future studies.

%It bridges the gap between theoretical physics and computational techniques, making high-precision predictions more accessible to researchers and paving the way for deeper insights into particle interactions at the quantum level.

%\newpage
%asdasdasd

\chapter{Next-to-Leading Order QCD Calculations in a Nutshell}
\label{SEC:NLO}

\section{General structure}

Following the notation used in refs.~\cite{Frederix:2009yq,Alwall:2014hca,Frederix:2018nkq,AH:2024ueu}, let us consider a generic $2\to n$ process, with its partonic subprocess denoted as $\Ione\Itwo\to \ident_3\cdots\ident_{n+2}$, where the identity of the $i$th particle is represented by $\ident_i$. The subprocess can be expressed as an ordered list $r=\left(\ident_1,\ldots,\ident_{n+2}\right)$. The 4-momentum of the external particle $\ident_i$ is denoted as $k_i$, with momentum conservation expressed as $k_1+k_2=\sum_{i=3}^{n+2}{k_i}$. The set of all possible subprocesses forms a space $\allprocn$, with $r\in \allprocn$. We denote the ultraviolet (UV) renormalized $\ell$-loop helicity amplitude for the subprocess $r$ as $\ampnll(r)$. For simplicity, we restrict ourselves to the special case where the LO features only one coupling order and only QCD corrections are considered in this chapter. For a more general case, we refer interested readers to Chapter~\ref{SEC:NLOEW}. 

At NLO, in addition to the $2\to n$ subprocesses, we also need to consider the $2\to n+1$ real-emission partonic processes, forming a set denoted as $\allprocnpo$. In the context of QCD corrections, this involves adding a light parton in the final states of the subprocesses in $\allprocn$. Without loss of generality, we can assume there are $\nlightB$ massless, strongly interacting particles in the Born-like processes, which implies there can be at most $\nlightR=\nlightB+1$ light jets in the real contribution. In this scenario, it is well known that the NLO QCD accurate cross section at a $P_1P_2$ collider in collinear factorization can be written as
\begin{eqnarray}
\sigma^{(\mathrm{NLO~QCD})}&=&\sum_{r\in\mathcal{R}_n}{\int{\mathrm{d}x_1\mathrm{d}x_2 f_{\Ione}^{(P_1)}(x_1,\mu_F^2) f_{\Itwo}^{(P_2)}(x_2,\mu_F^2)\left[\hat{\sigma}^{(B)}(r)+\hat{\sigma}^{(V)}(r)\right]}}\nonumber\\
&&+\sum_{r\in\mathcal{R}_{n+1}}{\int{\mathrm{d}x_1\mathrm{d}x_2 f_{\Ione}^{(P_1)}(x_1,\mu_F^2) f_{\Itwo}^{(P_2)}(x_2,\mu_F^2)\left[\hat{\sigma}^{(R)}(r)+\hat{\sigma}^{(\mathrm{PDF})}(r)\right]}},\label{eq:NLOxs}
\end{eqnarray}
where $x_1$ ($x_2$) is the longitudinal momentum fraction of the parton $\Ione$ ($\Itwo$) inside the beam particle $P_1$ ($P_2$), $f_{\ident}^{(P)}(x,\mu_F^2)$ is the parton distribution function (PDF) for the parton $\ident$ inside the initial particle $P$,  and $\mu_F$ is the factorization scale.  In the case of a fixed-energy lepton beam, without considering initial photon radiation, $f_{\ident}^{(P)}(x,\mu_F^2)$ can also simply be equal to $\delta(1-x)$. The partonic cross sections $\hat{\sigma}^{(X)}(r)$ for Born ($X=B$), virtual ($X=V$), and real ($X=R$) contributions can be obtained from the phase space integrated amplitude squared (also known as matrix elements):
\begin{eqnarray}
\hat{\sigma}^{(B)}(r)&=&\int{\phspn \frac{\JetsB}{\avg(r)}\ampsqnt(r)},\label{eq:bornxs}\\
\hat{\sigma}^{(V)}(r)&=&\int{\phspn \frac{\JetsB}{\avg(r)}\ampsqnl(r)},\label{eq:virtualxs}\\
\hat{\sigma}^{(R)}(r)&=&\int{\phspnpo \frac{\JetsB}{\avg(r)}\ampsqnpot(r)},\label{eq:realxs}
\end{eqnarray}
where $\avg(r)$ is the final state symmetry factor, $\mathrm{d}\phi_n$ is the $n$-body phase space measure, and $\JetsB$ is the measurement function that ensures the presence of at least $\nlightB$ light jets. The amplitude squared can be obtained from~\footnote{We have assumed that the initial particles are massless.}
\begin{eqnarray}
\ampsqnt(r)&=&\frac{1}{2s}\frac{1}{\omega(\Ione)\omega(\Itwo)}\mathop{\sum_{\rm color}}_{\rm spin}\abs{\ampnt(r)}^2,\label{eq:BornME}\\
\ampsqnl(r)&=&\frac{1}{2s}\frac{1}{\omega(\Ione)\omega(\Itwo)}\mathop{\sum_{\rm color}}_{\rm spin}2\Re\left\{\ampnt(r){\ampnl(r)}^{\star}\right\},\label{eq:virtualME}\\
\ampsqnpot(r)&=&\frac{1}{2s}\frac{1}{\omega(\Ione)\omega(\Itwo)}\mathop{\sum_{\rm color}}_{\rm spin}\abs{\ampnpot(r)}^2,\label{eq:RealME}
\end{eqnarray}
where the Mandelstam variable $s=(k_1+k_2)^2=2k_1\cdot k_2$, and $\omega(\ident)$ represents the product of spin and color degrees of 
freedom for particle $\ident$. While the presence of the jet function $\JetsB$ ensures that the $n$-body phase space integration in the Born and virtual processes is finite, infrared (IR) singularities--commonly referred to as soft and/or collinear divergences--still arise in both the loop integration of the one-loop amplitude $\ampnl(r)$ and the $(n+1)$-body phase space integration in the real cross section $\hat{\sigma}^{(R)}(r)$, necessitating regularization. To simultaneously regularize the UV and IR divergences--where the former appears only in the unrenormalized one-loop amplitude--a customary choice is to use dimensional regularization~\cite{tHooft:1972tcz,Bollini:1972ui}, which respects Lorentz and gauge invariance. However, several variants of dimensional regularization schemes exist (see table I in ref.~\cite{Shao:2011zza} for instance). In the conventional dimensional regularization (CDR) scheme with $d=4-2\epsilon$ dimensions, the average factors are given by $\omega(q)=\omega(\bar{q})=2\NC=6$ and $\omega(g)=2(1-\ep)(\NC^2-1)=16(1-\ep)$. If initial collinear QCD radiation is allowed, we need to renormalize the PDF in the QCD-improved parton model using collinear factorization, denoted as the partonic cross section $\hat{\sigma}^{(\mathrm{PDF})}(r)$ in eq.\eqref{eq:NLOxs}, with its concrete expression provided later in eq.\eqref{eq:PDFxs}. Thanks to the validity of Kinoshita-Lee-Nauenberg (KLN) theorem~\cite{Kinoshita:1962ur,Lee:1964is} and factorization, the combination of the virtual, real, and PDF contributions in eq.\eqref{eq:NLOxs} is free of any IR divergences. In fact, this statement can be generalized to any IR-safe observables~\cite{CTEQ:1993hwr}.

Since the computations of Born cross sections are widely acknowledged to be well established, we will introduce the general concepts for calculating real and virtual corrections. We follow the standard paradigm of treating the real and virtual parts separately. However, it is worth mentioning that some novel approaches within a non-standard paradigm have been proposed in the literature, where real and virtual contributions are treated on equal footing. One such innovative idea is the local unitarity approach~\cite{Capatti:2020xjc,Capatti:2022tit}. The code based on the local unitarity construction has successfully produced the first previously unknown cross section, namely the two-loop light-by-light scattering~\cite{AH:2023kor}.

\section{Real corrections}

Analytical calculations of real radiative corrections are generally infeasible unless one focuses on fully inclusive quantities, such as total cross sections or decay widths of simple processes. For fully differential results, numerical approaches must be employed. In particular, multi-dimensional phase space integration typically relies on adaptive Monte Carlo importance sampling methods. This necessitates flexible, process-independent procedures that facilitate numerical evaluations of real-emission contributions without any regulators in the regular phase-space regions. Two fundamentally different solutions have been proposed in the literature: the phase-space slicing and IR subtraction approaches. Both methods leverage the universal factorization properties of squared amplitudes and the phase space measure in the soft or collinear regions of gauge theories.

In slicing approaches, the $(n+1)$-body phase space integral in eq.\eqref{eq:realxs} are divided into singular and regular regions using technical slicing cut parameters. The real cross section in the regular regions can be numerically integrated in $d=4$ dimensions, while the one-body $d$-dimensional phase space integral in the soft or collinear regions must be carried out analytically. This is feasible only by taking the slicing cut parameters to be very small in both the phase space measure and the real matrix element, in conjunction with an $n$-body phase space Monte Carlo integration. Each phase space region is integrated individually, and only their sum reproduces the full phase-space integral in the asymptotic limit as the slicing cut parameters approach zero. However, there is a trade-off between residual dependencies on the cut parameters and increasing numerical integration errors as the slicing cuts decrease. Therefore, in practice, one must search for a plateau in the integrated results while accounting for numerical integration errors by adjusting the slicing cut parameters. The typical slicing approaches include one-cutoff~\cite{Giele:1991vf,Giele:1993dj,Keller:1998tf} and two-cutoff~\cite{Harris:2001sx} methods. We will not introduce them here, but we refer interested readers to the corresponding references.

On the other hand, the IR subtraction method relies on constructing simple counterterms that allow for analytic integration over one-body unresolved phase space in $d$ dimensions.  Ideally, these counterterms should locally match the IR behavior of the real matrix element in the $(n+1)$-body phase space integral, ensuring that the numerical integration over $\phspnpo$ of the difference between the real matrix element and the subtraction counterterms in $d=4$ is manageable. These are referred to as local counterterms. To recover the original real contributions, we must add back the phase-space integrated counterterms. Thus, the original local counterterms must be simple enough to allow for analytical integration over the singular regions within dimensional regularization, at least for the one-body unresolved phase space of a real emission particle. This singular analytical integration can be performed once and for all in a process-independent manner. The two widely adopted NLO subtraction methods were originally proposed by Frixione, Kunszt, and Signer (FKS)~\cite{Frixione:1995ms,Frixione:1997np}, and by Catani and Seymour~\cite{Catani:1996jh,Catani:1996vz}, respectively. They are commonly referred to as the FKS and dipole subtraction schemes. The former is inspired by the cancellation of collinear divergences, while the latter originates from the cancellation of soft divergences. The recoil schemes, which determine the mapping between the original $(n+1)$-body phase space and the phase space for counterterms, involve all external particles (\ie, globally) in the FKS scheme and one parton (\ie, locally) in the dipole method. Both approaches are effective for processes involving both simple and complicated elementary particles. Initially devised for massless colored particles, they have been generalized to include massive colored elementary particles~\cite{Phaf:2001gc,Catani:2002hc,Frederix:2009yq} and quarkonia~\cite{Butenschoen:2019lef,Butenschoen:2020mzi,AH:2024ueu}. These methods have also been implemented in various public computer codes~\cite{Gleisberg:2007md,Frederix:2008hu,Frederix:2009yq,Czakon:2009ss,Hasegawa:2009tx,Alioli:2010xd}.

IR subtraction methods outperform slicing approaches for at least two reasons. First, the former do not involve any approximations, whereas the phase-space integrals in the singular regions of slicing approaches are typically approximated using the leading terms in the expansions of the small cut parameters. Second, sliced phase-space integrals suffer from large logarithms of small cutoff parameters, which are expected to numerically cancel when combining all integration regions. On the other hand, these cancellations are significantly mitigated in subtraction methods, as they occur only between the local counterterms and the real matrix element at the level of individual phase space points. Consequently, subtraction techniques serve as the backbone of contemporary NLO automation codes. Since real emission computations in the \mgamcshort\ framework are based on the FKS subtraction, we will describe the method in the following section, largely following the discussions in refs.~\cite{Frederix:2009yq,AH:2024ueu}.

\subsection{FKS subtraction\label{sec:FKSsub}}

For any given $r\in\allprocnpo$, we can always reorder the final-state particles so that the final massless quarks, antiquarks, and gluons are indexed by $3\leq i\leq \nlightR+2$, while the strongly interacting particles are indexed by $\nini\leq j \leq \nlightR+\nheavy+2$, where $\nini=1, 2, 3$ corresponds to hadron-hadron, lepton-hadron, and lepton-lepton collisions, respectively. $\nheavy$ represents the number of massive colored partons in the final state. The key idea in the FKS subtraction is to introduce a set of ordered pairs, referred to as the set of FKS pairs, via
\begin{eqnarray}
\FKSpairs(r)&=&\Big\{(i,j)\;\Big|\;3\le i\le\nlightR+2\,, 
\nini\le j\le\nlightR+\nheavy+2\,, 
i\ne j\,,
\nonumber\\*&&\phantom{aaa}
\ampsqnpot(r)\JetsB\to\infty~~{\rm if}~~k_i^0\to 0~~
{\rm or}~~k_j^0\to 0~~{\rm or}~~\vec{k}_i\parallel \vec{k}_j\Big\}.
\phantom{aaaaa}
\label{PFKSdef}
\end{eqnarray}
This implies that a pair of particles belongs to the set of FKS pairs if they induce soft or collinear singularities (or both) in the $(n+1)$-body real matrix elements. It is important to note that $k_j^0\to 0$ (where $k_j^0$ is the energy component of the four-momentum $k_j$) is irrelevant when $j=1,2$. In the calculation of an NLO cross section within the FKS formalism, each pair belonging to $\FKSpairs(r)$ corresponds to a set of subtractions for these soft and collinear singularities.

In the FKS formalism, we multiply the real emission matrix elements by the partition function $\Sfunij$, which partitions the phase space into different kinematic regions, each containing at most one soft and one collinear singularity. The partitioning is achieved through the introduction of a set of positive-definite functions
\begin{eqnarray}
\Sfunij(r)\,,\;\;\;\;\;\;(i,j)\in\FKSpairs(r)\,,
\end{eqnarray}
where the argument $r\in\allprocnpo$ indicates that we can choose different $\Sfun$ functions for different subprocesses. As described in ref.~\cite{Frederix:2009yq}, $\Sfunij$ is defined in various regions as follows:
\begin{eqnarray}
\!\!\!\sum_{(i,j)\in \FKSpairs(r)}{\Sfunij(r)} &=&1, \nonumber \\
\lim_{\vec{k}_i\parallel\vec{k}_j}{\Sfunij(r)} &=& h_{ij}\left(\frac{k_i^0}{k_i^0+k_j^0}\right),\,~~~~{\rm if}~~m_i=\sqrt{k_i^2}=m_j=\sqrt{k_j^2}=0\,,\nonumber \\
\lim_{k_i^0 \to 0}{\Sfunij(r)} &=& c_{ij}\,,\phantom{}\;
{\rm if}~~~\ident_i=g\,,\,~~~~{\rm with}~~0<c_{ij}\le 1\qquad{\rm and}\,\,\,\,\,
\mathop{\sum_{j}}_{(i,j)\in\FKSpairs(r)}\!\!\!\!\!c_{ij}=1\,,\nonumber\\
\lim_{\vec{k}_k\parallel\vec{k}_l}{\Sfunij(r)}&=&0~~\forall\,
\{k,l\}\ne\{i,j\}~~{\rm with}~~(k,l)\in\FKSpairs(r)~~{\land}~~
m_k=m_l=0\,,\nonumber\\
\lim_{k_k^0\to 0}{\Sfunij(r)}&=&0~~\forall k~~{\rm with}~~\ident_k=g
~~{\rm and}~~\exists l~~{\rm with}~~(k,l)\in\FKSpairs(r)~{\lor}~
(l,k)\in\FKSpairs(r).
\nonumber\\&&
\label{eq:SfunC}
\end{eqnarray}
In other words, $\Sfunij$ approaches zero in all regions of the phase space where the real emission matrix elements diverge, except in cases where particle $i$ is soft, or particles $i$ and $j$ are collinear. The functions $h_{ij}(z)$ introduced in eq.\eqref{eq:SfunC} 
are defined for \mbox{$0\le z\le 1$} and possess the following properties:
\begin{eqnarray}
h_{ij}(z)&=&1\,,~~~~~~~~~{\rm if}~~\nini\le j\le 2\,,
\label{eq:hdef1}
\\
h_{ij}(z)&=&h(z)\,,~~~~~{\rm if}~~3\le j\le \nlightR+\nheavy+2\,,
\label{eq:hdef2}
\end{eqnarray}
with $h(z)$ being a positive-definite function such that
\begin{equation}
\lim_{z\to 0}{h(z)}=1\,,\;\;\;\;\;\;
\lim_{z\to 1}{h(z)}=0\,,\;\;\;\;\;\;
h(z)+h(1-z)=1\,.
\label{eq:hdef3}
\end{equation}
The specific construction of the partition function $\Sfunij$ is not crucial for our discussion. Its implementation in \mfks, a module in \mgamcshort, can be found in section 5.2 of ref.~\cite{Frederix:2009yq}. It is worth mentioning that the idea of using partition functions to separate IR singularities in real matrix elements has been extended in several subtraction methods~\cite{Frixione:2004is,Czakon:2010td,Czakon:2011ve,Czakon:2014oma,Caola:2017dug,Magnea:2018hab} for NNLO computations.

The IR-divergence-subtracted real cross sections can be formulated in the center-of-mass frame of the incoming partons:
\begin{eqnarray}
k_1=\frac{\sqrt{s}}{2}(1,0,0,1)\,,\;\;\;\;\;
k_2=\frac{\sqrt{s}}{2}(1,0,0,-1)\,.
\end{eqnarray}
In this frame, for each pair $(i,j)\in \FKSpairs(r)$, we can introduce the variables $\xii$ and $\yij$, where
\begin{eqnarray}
k_i^0&=&\frac{\sqrt{s}}{2}\xii\,,
\label{eq:xiidef}
\\
\vec{k}_i\mydot\vec{k}_j&=&\abs{\vec{k}_i}\abs{\vec{k}_j}\yij\,.
\label{eq:yijdef}
\end{eqnarray}
Thus, $\xii$ represents the rescaled energy of the FKS parton $i$, while $\yij$ denotes the cosine of the angle between the FKS parton $i$ and its sister $j$. The soft and collinear singularities of $\Sfunij(r)\ampsqnpot(r)\JetsB$ correspond to $\xii=0$ and $\yij=1$, respectively. The IR-divergence locally subtracted partonic cross section is
\begin{eqnarray}
\hat{\sigma}^{(R)}_{{\rm FKS}}(r)&=&\sum_{(i,j)\in\FKSpairs(r)}{\hat{\sigma}_{ij,{\rm FKS}}^{(R)}(r)},\label{eq:FKSRinFKSsectors}
\end{eqnarray}
where $\hat{\sigma}_{ij,{\rm FKS}}^{(R)}(r)$ is
\begin{equation}
\hat{\sigma}_{ij,{\rm FKS}}^{(R)}(r)=\int{\mathrm{d}\xii \mathrm{d}\yij \mathrm{d}\phii\tphspnij(r)\xic\omyijd
\Big((1-\yij)\xii^2\ampsqnpot(r)\Big)
\Sfunij(r)\frac{\JetsB}{\avg(r)}}\,.
\label{eq:dsigijnpo}
\end{equation}
The variable $\phii$ is the azimuthal direction of the FKS parton. The quantity $\tphspnij(r)$ is the reduced $n$-body phase space obtained via the following relation:
\begin{eqnarray}
\phspnpo(r)&=&\xii^{1-2\epsilon} \mathrm{d}\xii (1-\yij^2)^{-\epsilon}\mathrm{d}\yij \mathrm{d}\Omega_{i}^{(2-2\epsilon)} \frac{s^{1-\epsilon}}{(4\pi)^{3-2\epsilon}}\times 
\left\{\begin{array}{ll}
\phspn(r^{j\oplus \bar{i},\isubrmv}), & {\rm if}\quad j\leq 2\\
\frac{k_i^0+k_j^0}{k_j^0}\phspn(r^{j\oplus i,\isubrmv}), & {\rm if}\quad j\geq 3, j\neq i\end{array}\right.\nonumber\\
&=&\xii^{1-2\epsilon} \mathrm{d}\xii (1-\yij^2)^{-\epsilon}\mathrm{d}\yij \mathrm{d}\Omega_{i}^{(2-2\epsilon)} \tphspnij(r)\,.
\label{eq:tphspdef}
\end{eqnarray}
The reduced phase space measure has the following limits:
\begin{eqnarray}
\lim_{\xii\to 0}{\tphspnij(r)}&=&\frac{s^{1-\epsilon}}{(4\pi)^{3-2\epsilon}}\phspn(r^{\isubrmv})\,,\,~~~~~~~~~{\rm if}~~m_i=0\,,
\label{eq:tphisoft}
\\
\lim_{\yij\to 1}{\tphspnij(r)}&=&\frac{s^{1-\epsilon}}{(4\pi)^{3-2\epsilon}}\times\left\{\begin{array}{ll}
\phspn(r^{j\oplus \bar{i},\isubrmv}), & {\rm if}\quad j\leq 2\\
\frac{k_i^0+k_j^0}{k_j^0}\phspn(r^{j\oplus i,\isubrmv}), & {\rm if}\quad j\geq 3\end{array}\right.\,,~~~~~{\rm if}~~m_i=m_j=0\,.
\label{eq:tphicoll}
\end{eqnarray}
where
\begin{eqnarray}
r^{\isubrmv}&=&\left(\ident_1,\ldots,\remove{\ident}{0.2}_i,\ldots,\ident_j,\ldots,\ident_{n+3}\right),\label{eq:risubrmvdef}\\
r^{j\oplus i, \isubrmv}&=&\left(\ident_1,\ldots,\remove{\ident}{0.2}_i,\ldots,\ident_{j\oplus i},\ldots,\ident_{n+3}\right),\label{eq:rjplusiisubrmvdef}\\
r^{1\oplus \bar{i}, \isubrmv}&=&\left(\ident_{1\oplus \bar{i}},\ident_2,\ldots,\remove{\ident}{0.2}_i,\ldots,\ident_{n+3}\right),\label{eq:r1plusiisubrmvdef}\\
r^{2\oplus \bar{i}, \isubrmv}&=&\left(\ident_1,\ident_{2\oplus \bar{i}},\ldots,\remove{\ident}{0.2}_i,\ldots,\ident_{n+3}\right),\label{eq:r2plusiisubrmvdef}
\end{eqnarray}
such that $r^{\isubrmv},r^{j\oplus i, \isubrmv}, r^{j\oplus\bar{i},\isubrmv}\in \allprocn$. $r^{\isubrmv}$ in eq.\eqref{eq:risubrmvdef} represents a reduced partonic process obtained by simply removing parton $i$ from the original list $r$. Similarly, $r^{j\oplus i, \isubrmv}$ in eq.\eqref{eq:rjplusiisubrmvdef}  is constructed by removing parton $i$ and replacing parton $j$ with one whose
identity is $\ident_{j\oplus i}$. The definitions of the processes $r^{1\oplus \bar{i}, \isubrmv}$ and $r^{2\oplus \bar{i}, \isubrmv}$ in eqs.\eqref{eq:r1plusiisubrmvdef} and \eqref{eq:r2plusiisubrmvdef}, respectively, are similar to $r^{j\oplus i, \isubrmv}$, but in these cases, the anti-identity of parton $i$, $\ident_{\bar{i}}$, is used instead. In $d=4$ dimensions, we can simply set $\epsilon=0$ and $\mathrm{d}\Omega_i^{(2)}=\mathrm{d}\phii$. The distributions entering eq.\eqref{eq:dsigijnpo} are defined as 
follows, for any test functions $f()$ and $g()$:
\begin{eqnarray}
\int_0^{\ximax}{d\xii f(\xii)\xic}&=&
\int_0^{\ximax}{d\xii \frac{f(\xii)-f(0)\stepf(\xicut-\xii)}{\xii}}\,,
\label{eq:distrxii}
\\
\int_{-1}^1{d\yij g(\yij)\omyijd}&=&
\int_{-1}^1{d\yij \frac{g(\yij)-g(1)\stepf(\yij-1+\delta)}{1-\yij}}\,,
\label{eq:distryij}
\end{eqnarray}
where
\begin{equation}
\ximax=1-\frac{1}{s}\left(\sum_{k=3}^{n+3}m_k\right)^2\,,
\end{equation}
and $\stepf()$ is the Heaviside theta function.
In eqs.\eqref{eq:distrxii} and \eqref{eq:distryij}, $\xicut$ and $\delta$ are 
free parameters, that can be chosen in the ranges
\begin{equation}
0<\xicut\le\ximax\,,\;\;\;\;\;
0<\delta\le 2\,.
\end{equation}
Different values of $\xicut$ and $\delta$ could be chosen for each FKS sector $(i,j)\in\FKSpairs(r)$ when computing $\hat{\sigma}_{ij,{\rm FKS}}^{(R)}(r)$. In \mfks~\cite{Frederix:2009yq}, a common value $\xicut$ is used for all $(i,j)$ pairs. The same value of $\delta$ ($\delta=\deltaI$) is applied to the initial state collinear singularities $(i,j)\in \FKSpairs(r), j\leq 2$, while another value of $\delta$ ($\delta=\deltaO$) is used for the final state collinear singularities $(i,j)\in \FKSpairs(r) , j\geq 3$. Now, we can introduce the quantity in $d=4$ dimensions
\begin{eqnarray}
\Sigma_{ij}(r; \xii, \yij)&=&\Big((1-\yij)\xii^2\ampsqnpot(r)\Big)
\Sfunij(r)\frac{\JetsB}{\avg(r)}\, 
\mathrm{d}\phii\tphspnij(r),\label{eq:Sigmaijdef}
\end{eqnarray}
so that
\begin{eqnarray}
\hat{\sigma}_{ij,{\rm FKS}}^{(R)}(r)=\int{\xic\omyijd \Sigma_{ij}(r; \xii, \yij) \mathrm{d}\xii \mathrm{d}\yij}.\label{eq:sigmaFKSijdef1}
\end{eqnarray}
If we expand the plus distributions, we have
\begin{eqnarray}
\hat{\sigma}_{ij,{\rm FKS}}^{(R)}(r)&=&
\int{\frac{1}{\xii(1-\yij)}
\Big[\Sigma_{ij}(r; \xii,\yij)-\Sigma_{ij}(r;\xii,1)\stepf(\yij-1+\delta)}
\nonumber\\*&&-\Sigma_{ij}(r; 0,\yij)\stepf(\xicut-\xii)
+\Sigma_{ij}(r; 0,1)\stepf(\xicut-\xii)\stepf(\yij-1+\delta)
\Big]\mathrm{d}\xii \mathrm{d}\yij
\nonumber\\*&=&\hat{\sigma}_{ij}^{(R)}(r)-\hat{\sigma}_{ij}^{(C)}(r)-\hat{\sigma}_{ij}^{(S)}(r)+\hat{\sigma}_{ij}^{(SC)}(r).
\label{eq:dsigijnpoE}
\end{eqnarray}
The first term in the integrand, called ``events", contributes to the original partonic real cross section 
\begin{equation}
\hat{\sigma}^{(R)}(r)=\sum_{(i,j)\in \FKSpairs(r)}{\hat{\sigma}^{(R)}_{ij}(r)},\label{eq:sigmaRdef}
\end{equation}
 while the remaining three terms are local subtraction counterterms, which are called ``collinear counterevent" ($X=C$), ``soft counterevent" ($X=S$), and ``soft-collinear counterevent" ($X=CS$), respectively. The sums of the counterterms over the FKS pairs can be similarly defined as
 \begin{equation}
\hat{\sigma}^{(X)}(r)=\sum_{(i,j)\in \FKSpairs(r)}{\hat{\sigma}^{(X)}_{ij}(r)},\quad X=C,S,CS.\label{eq:sigmaXdef}
\end{equation}
 
 The collinear and soft-collinear local counterterms can be derived from the collinear limit of the real-emission matrix elements
\begin{eqnarray}
\lim_{\vec{k}_i\parallel \vec{k}_j}{(1-\yij)\xii^2\ampsqnpot(r)\Sfunij(r)}&=&\frac{4}{s}g_s^2\mu^{2\epsilon}\xii P_{\ident_{j\oplus \bar{i}}\ident_j}^{(\mathrm{QCD}),<}(1-\xii,\epsilon)\ampsqnt(r^{j\oplus \bar{i},\isubrmv})\nonumber\\
&&+\frac{4}{s}g_s^2\mu^{2\epsilon}\xi_i\underbrace{Q^{(\mathrm{QCD})}_{\ident_{j\oplus \bar{i}}^\star\ident_j}(1-\xii)\ampsqnttilde_{ij}(r^{j\oplus \bar{i},\isubrmv})}_{\equiv \Delta^{(\mathrm{QCD})}_{ij}},\label{eq:localinitialcoll0}\\
&&~~~~~~~~~~~~~~~~~~~~~~~~~~~~~~~~j=\nini,\ldots,2,\nonumber\\
\lim_{\vec{k}_i\parallel \vec{k}_j}{(1-\yij)\xii^2\ampsqnpot(r)\Sfunij(r)}&=&\frac{4}{s}g_s^2\mu^{2\epsilon}\frac{1-z}{z}h(1-z)P_{\ident_j\ident_{j\oplus i}}^{(\mathrm{QCD}),<}(z,\epsilon)\ampsqnt(r^{j\oplus i,\isubrmv})\nonumber\\
&&+\frac{4}{s}g_s^2\mu^{2\epsilon}\frac{1-z}{z}h(1-z)\underbrace{Q^{(\mathrm{QCD})}_{\ident_j\ident_{j\oplus i}^\star}(z)\ampsqnttilde_{ij}(r^{j\oplus i,\isubrmv})}_{\equiv \Delta^{(\mathrm{QCD})}_{ij}},\label{eq:localfinalcoll0}\\
&&~~~~~~~~~~~~~~~~~~~~~~~~~~~~~~~~j=3,\ldots, \nlightR+2, j\neq i,\nonumber
\end{eqnarray}
where $\bar{i}$ denotes the antiparticle of the particle $i$ and $g_s=\sqrt{4\pi\alpha_s}$. $P_{\ident_k\ident_l}^{(\mathrm{QCD}),<}(z,\epsilon)$ is the unregularized Altarelli-Parisi kernel~\cite{Altarelli:1977zs} in QCD for $z<1$ in $d=4-2\epsilon$ dimensions that can be found in the literature (see, \eg, eqs.(D.15-D.18) in ref.~\cite{Frederix:2009yq}). The quantities $Q^{(\mathrm{QCD})}_{\ident_{j\oplus \bar{i}}^\star\ident_j}$ and $Q^{(\mathrm{QCD})}_{\ident_j\ident_{j\oplus i}^\star}$ are given in eqs.(D.3-D.10) of ref.~\cite{Frederix:2009yq}. The reduced matrix element is defined as~\cite{Frixione:1995ms,Frederix:2009yq}
\begin{eqnarray}
\ampsqnttilde_{ij}(r^{j\oplus i,\isubrmv})&=&\frac{1}{2s}
\frac{1}{\omega(\Ione)\omega(\Itwo)}
\Re{\left\{\frac{\langle ij\rangle}{[ij]}\tilde{\mathop{\sum_{\rm color}}_{\rm spin}}\!
\ampnt_+(r^{j\oplus i,\isubrmv})
{\ampnt_-(r^{j\oplus i,\isubrmv})}^{\star}\right\}}\,,~~~~~\label{eq:ampsqtilde0}
\end{eqnarray}
where the spinor-helicity formalism takes the conventions of ref.~\cite{Mangano:1990by}. The notation $\ampnt_{\pm}$ represents the helicity amplitude with the helicity of the parton $j\oplus i$ being $\pm$. The sum with a tilde indicates a summation over the color and spin of the external states, excluding the spin of the particle $j\oplus i$.
The $\Delta^{(\mathrm{QCD})}_{ij}$ term vanishes upon integrating the azimuthal variable $\mathrm{d}\phii$ of the parton $i$. In eq.\eqref{eq:localfinalcoll0}, we assume $k_i=(1-z)\left(k_i+k_j\right)$ and $k_j=z\left(k_i+k_j\right)$. This assumption applies specifically to the initial-state collinear singularities ($\nini\leq j\leq 2$)
\begin{eqnarray}
\Sigma_{ij}(r; \xi_i, 1)&=&\frac{4}{\left(4\pi\right)^3}g_s^2\xii \left[P_{\ident_{j\oplus \bar{i}}\ident_j}^{(\mathrm{QCD}),<}(1-\xii,0)\ampsqnt(r^{j\oplus \bar{i},\isubrmv})+\Delta^{(\mathrm{QCD})}_{ij}\right]\nonumber\\
&&\times\frac{\JetsB}{\avg(r)}\, 
\mathrm{d}\phii\phspn(r^{j\oplus \bar{i},\isubrmv}),\label{eq:Sigmaijinitialcollinear}
\end{eqnarray}
and to the final state collinear singularities ($3\leq j \leq \nlightR+2, j\neq i$)
\begin{eqnarray}
\Sigma_{ij}(r; \xi_i, 1)&=&\frac{4}{\left(4\pi\right)^3}g_s^2\frac{1-z}{z}\frac{h(1-z)}{z}\left[P_{\ident_j\ident_{j\oplus i}}^{(\mathrm{QCD}),<}(z,0)\ampsqnt(r^{j\oplus i,\isubrmv})+\Delta^{(\mathrm{QCD})}_{ij}\right]\nonumber\\
&&\times\frac{\JetsB}{\avg(r)}\, 
\mathrm{d}\phii\phspn(r^{j\oplus i,\isubrmv})\,.\label{eq:Sigmaijfinalcollinear}
\end{eqnarray}
For the soft-collinear counterparts, we need to take $\xi_i\to 0$ and $z\to 1$ on the right-hand side (r.h.s) of the above two equations. The soft local counterterm is
\begin{eqnarray}
\Sigma_{ij}(r; 0, \yij)&=&\frac{s}{\left(4\pi\right)^3}(1-\yij)\Big(\lim_{\xii\to 0}{\xii^2\ampsqnpot(r)}\Big)
c_{ij}\frac{\JetsB}{\avg(r)}\, 
\mathrm{d}\phii\phspn(r^{\isubrmv}).\label{eq:Sigmaijinsoft}
\end{eqnarray}
The soft limit of the real matrix element has the eikonal form
\begin{eqnarray}
\lim_{\xii\to 0}{\xii^2\ampsqnpot(r)}&=&g_s^2\mathop{\sum_{k,l=\nini}}_{k,l\neq i, k\leq l}^{\nlightR+\nheavy+2}{\left(\lim_{\xii\to 0}{\frac{\xii^2 k_k\cdot k_l}{k_k\cdot k_i k_l\cdot k_i}}\right)\ampsqnt_{\mathrm{QCD},kl}(r^{\isubrmv})},\label{eq:softME}
\end{eqnarray}
where the color-linked Born matrix element is defined as
\begin{eqnarray}
\ampsqnt_{\mathrm{QCD},kl}(r)&=&-\frac{1}{2s}
\frac{2-\delta_{kl}}{\omega(\Ione)\omega(\Itwo)}\mathop{\sum_{\rm color}}_{\rm spin}
\ampnt(r) \Qop_{\mathrm{QCD}}(\ident_k)\mydot\Qop_{\mathrm{QCD}}(\ident_l)
{\ampnt(r)}^{\star}.
\label{eq:colorlinkedBornME}
\end{eqnarray}
In the above equation, $\delta_{kl}$ is the Kronecker delta function, $\Qop_{\mathrm{QCD}}(\ident) $ represents the QCD color generator associated with the particle $\ident$
\begin{align}\label{SUN}
\Qop_{\mathrm{QCD}}(\ident) = \{t^a\}_{a=1}^{8} ,\quad  \{-t^{aT}\}_{a=1}^8, \quad \{T^a\}_{a=1}^8 \quad \ident\in {\irrep{3},\irrepbar{3},\irrep{8}}  
\end{align}
with $t^a$ and $T^a$ being the SU(3) generators in the fundamental and adjoint representations, respectively. We take the final quark or initial antiquark as $\irrep{3}$, while the initial quark and final antiquark are taken as $\irrepbar{3}$. The matrix element of the adjoint representation is $T^a_{bc}=-if_{abc}$ with $f_{abc}$ being the anti-symmetric structure constants of SU(3).

In order to recover the original NLO cross section, we must reintroduce the elements that were subtracted from the real emission contribution; these are our integrated counterterms. The soft integrated counterterm is given by
\begin{equation}
\hat{\sigma}^{(S)}(r)=\int{\phspn(r^{\isubrmv})\frac{\JetsB}{\avg(r^{\isubrmv})}\frac{\alpha_s}{2\pi}\left[\sum_{k=\nini}^{\nlightB+\nheavy+2}{\sum_{l= k}^{\nlightB+\nheavy+2}{\left(\hat{\mathcal{E}}_{kl}^{(m_k,m_l)}+\mathcal{E}_{kl}^{(m_k,m_l)}\right)\ampsqnt_{\mathrm{QCD},kl}(r^{\isubrmv})}}\right]}\,,\label{eq:integratedsoftCT1}
\end{equation}
where the pole and finite parts of the eikonal integrals $\hat{\mathcal{E}}_{kl}^{(m_k,m_l)}$ and $\mathcal{E}_{kl}^{(m_k,m_l)}$ can be found in appendix A of ref.~\cite{Frederix:2009yq}. The collinear renormalization of initial PDFs leads to
\begin{equation}
\hat{\sigma}^{(\mathrm{PDF})}(r)=\sum_{k=\nini}^{2}\int_{1-\ximax}^1 \mathrm{d}z \left(\frac{1}{\bar{\ep}}P^{(\mathrm{QCD})}_{\ident_{k\oplus \bar{i}}\ident_k}(z,0)-K^{(\mathrm{QCD})}_{\ident_{k\oplus \bar{i}}\ident_k}(z)\right)\int{\phspn(r^{k\oplus \bar{i},\isubrmv})\frac{\JetsB}{\avg(r^{\isubrmv})}\frac{\alpha_s}{2\pi}\ampsqnt(r^{k\oplus \bar{i},\isubrmv})}\label{eq:PDFxs}
\end{equation}
with $1/\bar{\ep}=1/\ep+\log{\left(4\pi\right)}-\gamma_E$ and $\gamma_E$ being the Euler-Mascheroni constant. The four dimensional regularized Altarelli-Parisi splitting kernels in QCD are
\begin{equation}
    P^{(\mathrm{QCD})}_{ab}(z,0) = \frac{(1-z)P^{(\mathrm{QCD}),<}_{ab}(z,0)}{(1-z)_{+}} + \gamma_{\mathrm{QCD}}(a)\delta_{ab}\,\delta(1-z).\label{eq:QCDAPregkernal}
\end{equation}
The subtraction scheme-dependent function $K^{(\mathrm{QCD})}_{ab}(z)$ is zero for $\overline{\rm MS}$ PDFs. It is important to note that eq.\eqref{eq:PDFxs} implies $r^{k\oplus \bar{i},\isubrmv}\in\allprocn$; otherwise, the matrix element of the reduced Born process is zero. The sum of the collinear, soft-collinear, and PDF counterterms, all associated with collinear divergences, yields
\begin{equation}
\hat{\sigma}^{(C)}(r)-\hat{\sigma}^{(SC)}(r)+\hat{\sigma}^{(\mathrm{PDF})}(r)=\hat{\sigma}^{(C)}_{\rm sing.}(r)+\hat{\sigma}^{(C,n)}_{\rm FIN}(r)+\hat{\sigma}^{(C,n+1)}_{\rm FIN}(r)\,,\label{eq:sumofcollinearintCT1}
\end{equation}
where we have decomposed into a singular term
\begin{equation}
\hat{\sigma}^{(C)}_{\rm sing.}(r)=\int{\phspn(r^{\isubrmv})\frac{\alpha_s}{2\pi}\frac{\JetsB}{\avg(r^{\isubrmv})}\sum_{k=\nini}^{\nlightB+2}\frac{(4\pi)^\ep}{\Gamma(1-\ep)}
\left(\frac{\mu^2}{Q_{\rm ES}^2}\right)^\ep\frac{1}{\ep}}\left[\gamma_{\mathrm{QCD}}(\ident_k)+C_{\mathrm{QCD}}(\ident_k)\log\!\left(\frac{\xicut^2s}{4(k_k^0)^2}\right)\right]\ampsqnt(r^{\isubrmv})\,,
\end{equation}
a finite $n$-body contribution
\begin{eqnarray}
\hat{\sigma}^{(C,n)}_{\rm FIN}(r)&=&\int{\phspn(r^{\isubrmv})\frac{\alpha_s}{2\pi}\frac{\JetsB}{\avg(r^{\isubrmv})}\Bigg\lbrace-\log\!\left(\dfrac{\mu^2}{Q_{\rm ES}^2}\right)\sum_{k=\nini}^{2}\Big(\gamma_{\mathrm{QCD}}(\ident_k)+2C_{\mathrm{QCD}}(\ident_k)\log(\xicut)\Big)}\nonumber\\
&&+\sum_{k= 3}^{\nlightB+2}\left[\gamma_{\mathrm{QCD}}^\prime(\ident_k)-\log\!\left(\dfrac{s\deltaO}{2Q_{\rm ES}^2}\right)\left(\gamma_{\mathrm{QCD}}(\ident_k)-2C_{\mathrm{QCD}}(\ident_k)\log\!\left(\dfrac{2k^0_k}{\xicut\sqrt{s}}\right)\right)\right.\nonumber\\
&&\left.+2C_{\mathrm{QCD}}(\ident_k)\left(\log^2\!\left(\dfrac{2k^0_k}{\sqrt{s}}\right)-\log^2(\xicut)\right)-2\gamma_{\mathrm{QCD}}(\ident_k)\log\!\left(\dfrac{2k^0_k}{\sqrt{s}}\right)\right]\Bigg\rbrace\ampsqnt(r^{\isubrmv})\,,\nonumber\\
\end{eqnarray}
and a finite contribution with a degenerated $(n+1)$-body phase space
\begin{eqnarray}
\hat{\sigma}^{(C,n+1)}_{\rm FIN}(r)&=&\sum_{k=\nini}^{2}\int_0^{\ximax} d\xii\left\lbrace\left[\left(\frac{1}{\xii}\right)_c\log\!\left(\frac{s\deltaI}{2\mu^2}\right)+2\left(\frac{\log(\xii)}{\xii}\right)_c\right]\right.\nonumber\\
&&\left.\times\xii P_{\ident_{k\oplus \bar{i}}\ident_k}^{(\mathrm{QCD}),<}(1-\xii,0)-\left(\frac{1}{\xii}\right)_c\xii P_{\ident_{k\oplus \bar{i}}\ident_k}^{(\mathrm{QCD}),\prime<}(1-\xii,0)-K^{(\mathrm{QCD})}_{\ident_{k\oplus \bar{i}}\ident_k}(1-\xii)\right\rbrace\nonumber\\
&&\times\int{\phspn(r^{k\oplus \bar{i},\isubrmv})\frac{\JetsB}{\avg(r^{\isubrmv})}\frac{\alpha_s}{2\pi}\ampsqnt(r^{k\oplus \bar{i},\isubrmv})}\,.
\end{eqnarray}
We have introduced the dimensional regularization scale $\mu$ and the Ellis-Sexton scale $Q_{\rm ES}$~\cite{Ellis:1985er}. The QCD Casimir factors are
\begin{equation}
\begin{aligned}
C_{\mathrm{QCD}}(\ident)&=\left\{\begin{array}{ll}
C_F, &~~{\rm if}~~\ident\in \irrep{3},\irrepbar{3}\\
C_A, &~~{\rm if}~~\ident\in \irrep{8}\\
\end{array}\right..
\end{aligned}
\end{equation}
The QCD collinear anomalous dimensions are
\begin{equation}
\begin{aligned}
\gamma_{\mathrm{QCD}}(\ident)&=\left\{\begin{array}{ll}
\frac{3}{2}C_F\,,\hfill&~~{\rm if}~~\ident=q,\bar{q}\\
\frac{11}{6}C_A-\frac{2}{3}T_Fn_q\,,&~~{\rm if}~~\ident=g\\
\end{array}\right.,
\end{aligned}
\end{equation}
with $T_F=\frac{1}{2}$ and $n_q$ being the number of massless quark flavors, and
\begin{equation}
\begin{aligned}
\gamma_{\mathrm{QCD}}^\prime(\ident)&=\left\{\begin{array}{ll}
\left(\frac{13}{2}-\frac{2\pi^2}{3}\right)C_F\,,\hfill&~~{\rm if}~~\ident=q,\bar{q}\\
\left(\frac{67}{9}-\frac{2\pi^2}{3}\right)C_A-\frac{23}{9}T_Fn_q\,,&~~{\rm if}~~\ident=g\\
\end{array}\right..
\end{aligned}
\end{equation}
$P_{ab}^{(\mathrm{QCD}),\prime<}(z,0)$ is the $\epsilon$ part of the $d=4-2\epsilon$ dimensional unregularized Altarelli-Parisi splitting kernels
\begin{eqnarray}
P_{ab}^{(\mathrm{QCD}),<}(z,\epsilon)&=&P_{ab}^{(\mathrm{QCD}),<}(z,0)+\epsilon P_{ab}^{(\mathrm{QCD}),\prime<}(z,0).
\end{eqnarray}
Finally, the total sum of the integrated FKS and PDF counterterms is given by
\begin{eqnarray}
\hat{\sigma}^{(I)}_{\mathrm{FKS}}(r)&=&\hat{\sigma}^{(S)}(r)+\hat{\sigma}^{(C)}(r)-\hat{\sigma}^{(SC)}(r)+\hat{\sigma}^{(\mathrm{PDF})}(r).
\end{eqnarray}

The original NLO cross section in the FKS subtraction scheme, as given in eq.\eqref{eq:NLOxs}, can be rewritten as
\begin{eqnarray}
\sigma^{(\mathrm{NLO~QCD})}&=&\sum_{r\in\mathcal{R}_n}{\int{\mathrm{d}x_1\mathrm{d}x_2 f_{\Ione}^{(P_1)}(x_1,\mu_F^2) f_{\Itwo}^{(P_2)}(x_2,\mu_F^2)\left[\hat{\sigma}^{(B)}(r)+\hat{\sigma}^{(V)}(r)\right]}}\nonumber\\
&&+\sum_{r\in\mathcal{R}_{n+1}}{\int{\mathrm{d}x_1\mathrm{d}x_2 f_{\Ione}^{(P_1)}(x_1,\mu_F^2) f_{\Itwo}^{(P_2)}(x_2,\mu_F^2)\left[\hat{\sigma}^{(R)}_{\mathrm{FKS}}(r)+\hat{\sigma}^{(I)}_{\mathrm{FKS}}(r)\right]}}.\label{eq:NLOxsFKS}
\end{eqnarray}
Although the Born, virtual, FKS counterevents, and events in eq.\eqref{eq:NLOxsFKS} can be integrated separately after pole cancellation, it is preferable to integrate them together, as noted in ref.~\cite{Alwall:2014hca}, in order to reduce the probability of mis-binning and thereby enhances the numerical stability of the final result. Additionally, the number of FKS pairs defined in eq.\eqref{PFKSdef}, which scales as $\nlightR(\nlightR+\nheavy)$, can often be significantly reduced by exploiting the symmetries arising from identical final-state particles~\cite{Frederix:2009yq}.

\section{Virtual corrections}

The second essential ingredient in an NLO calculation is the UV renormalized one-loop amplitude $\ampnl(r)$, or its interference with the Born amplitude $\ampsqnl(r)$ in the Born kinematics $r\in\allprocn$. The key quantities that need to be computed are the one-loop UV-unrenormalized amplitude $\ampnl_{\mathrm{U}}(r)$ and the tree-like UV renormalization amplitude $\ampnUV(r)$, \ie,
\begin{equation}
\ampnl(r)=\ampnl_{\mathrm{U}}(r)+\ampnUV(r).
\end{equation}
While the computation of the latter can be performed similarly to that of the Born amplitude $\ampnt(r)$ using the usual UV renormalization Feynman rules, the former requires addressing one-loop Feynman integrals, which can be both UV and IR divergent and thus necessitate regulators.  In dimensional regularization, loop calculations are conducted in $d=4-2\epsilon$ dimensions of spacetime. The convergence of loop integrals depends on the choice of $d$: UV-divergent integrals are manageable only when $\Re{d}<4$, while IR-divergent integrals are well defined for $\Re{d}>4$. The structure of loop integrals permits analytic continuation to arbitrary complex values of $d$, enabling regularization of integrals with UV and IR singularities, with the limit $d\to 4$ (or $\epsilon\to 0$) limit taken safely. However, for loop integrals that are both UV and IR divergent, this continuation becomes more subtle, as no value of $d$ exists for which those integrals are finite. In such cases, ref.~\cite{Leibbrandt:1975dj} provides arguments for their proper analytic continuation.

For the one-loop UV-unrenormalized amplitude, it is a simple collection of one-loop Feynman diagrams
\begin{eqnarray}
\ampnl_{\mathrm{U}}(r)&=&\sum_{\mathrm{diagrams}}{\mathcal{C}(r)},
\end{eqnarray}
where $\mathcal{C}(r)$ denotes the contribution of a single one-loop Feynman diagram after loop integration, with implicit dependence of spin, color, and Lorentz indices. A standard technique for evaluating $\mathcal{C}(r)$ is the Feynman integral reduction procedure (such as the Passarino-Veltman reduction~\cite{Passarino:1978jh}). This means that all one-loop integrals can be algebraically reduced to a set of standard scalar integrals with at most four loop propagators in the limit as $d$ approaches $4$. In other words, we have
\begin{eqnarray}
\mathcal{C}(r)&=&\mathrm{Red}\left[\mathcal{C}(r)\right]=\sum_{i}{c_i(\mathcal{C})\mathcal{J}_i^{(\mathrm{Red})}}+R(\mathcal{C}),
\end{eqnarray}
where the one-loop scalar integrals $\mathcal{J}_i^{(\mathrm{Red})}$ are independent of $\mathcal{C}$. The function $R$ depends on external momenta and internal/external masses and may equal zero. Importantly, as its name suggests, unlike one-loop scalar integrals that generally depend on logarithms or dilogarithms, $R$ depends solely on a rational function of the Lorentz invariant scalar products of external momenta and masses. The essence of any reduction procedure is that it is an algebraic operation determining the coefficients $c_i$ and $R$. The remaining task is to evaluate the basis integrals $\mathcal{J}_i^{(\mathrm{Red})}$, which are significantly simpler than any $\mathcal{C}(r)$ and can be expressed in terms of logarithms and dilogarithms, as pioneered by 't Hooft and Veltman~\cite{tHooft:1978jhc}. As the
notation $\{\mathcal{J}_i^{(\mathrm{Red})}\}$ suggests, different reduction procedures may utilize different sets of one-loop scalar integrals.

It is not restrictive to express $\mathcal{C}(r)$ in the following form:
\begin{eqnarray}
\mathcal{C}(r)&=&\int{\mathrm{d}^d\bar{\ell}\frac{\bar{N}(\bar{\ell})}{\bar{D}_0\bar{D}_1\cdots \bar{D}_{N-1}}},\label{eq:generic1loop}
\end{eqnarray}
where the denominators of the loop propagators are defined as
\begin{eqnarray}
\bar{D}_i&=&\left(\bar{\ell}+p_i\right)^2-M_i^2,\quad  0\leq i\leq N-1.\label{eq:loopdendef}
\end{eqnarray}
Here, $p_i$ represents a linear combination of the external momenta flowing into the loop, and $M_i$ denotes the mass of the particle in the $i$th loop line. We use the capital $M$ to differentiate between the masses of loop and external particles. Furthermore, for any $4$-dimensional quantity $X$, its $d$-dimensional counterpart is denoted by $\bar{X}$, while its $(-2\epsilon)$-dimensional part is denoted by $\tilde{X}$. In this context, the $d$-dimensional loop momentum $\bar{\ell}$ can be decomposed into a $4$-dimensional part $\ell$ and a $(-2\epsilon)$-dimensional part $\tilde{\ell}$,\ie,
\begin{eqnarray}
\bar{\ell}&=&\ell+\tilde{\ell}
\end{eqnarray}
with $\ell\cdot\tilde{\ell}=0$. Similar decompositions apply to the Dirac matrices $\bar{\gamma}^\mu$ and the metric tensor $\bar{g}^{\mu\nu}$. In general, $p_i$ can either be a $d$-dimensional quantity when working in CDR or a $4$-dimensional one in the 't Hooft-Veltman regularization scheme (see the difference in table I of ref.~\cite{Shao:2011zza}). For numerical computations, it is more convenient to work in the latter, so we use $p_i$ instead of $\bar{p}_i$ in eq.\eqref{eq:loopdendef}. 

The numerator function $\bar{N}(\bar{\ell})$ in eq.\eqref{eq:generic1loop} is process-dependent and can be quite complex. However, it can always be expressed as a polynomial in the loop momentum $\bar{\ell}$. In automated approaches that rely on numerical methods, handling the non-integer dimensions required by dimensional regularization poses obvious challenges. Thus, it is convenient to define the purely $4$-dimensional part of the numerator function as:
\begin{eqnarray}
N(\ell)&=&\lim_{\epsilon\to 0}{\left.\bar{N}(\bar{\ell})\right|_{\bar{\ell}\to \ell, \bar{\gamma}^\mu \to \gamma^\mu, \bar{g}^{\mu\nu}\to g^{\mu\nu}}}.\label{eq:N4dim}
\end{eqnarray}
The quantity $N(\ell)$ does not involve non-integer dimensions and can therefore be treated using the standard computational techniques. The $(-2\epsilon)$-dimensional counterpart is simply given by the difference
\begin{eqnarray}
\tilde{N}(\ell,\tilde{\ell})&=&\bar{N}(\bar{\ell})-N(\ell).
\end{eqnarray}
This implies that $\mathcal{C}(r)$ can be decomposed into two components:
\begin{eqnarray}
\mathcal{C}(r)&=&\mathcal{C}_{\mathrm{non-}R_2}(r)+\mathcal{C}_{R_2}(r),
\end{eqnarray}
where
\begin{eqnarray}
\mathcal{C}_{\mathrm{non-}R_2}(r)&=&\int{\mathrm{d}^d\bar{\ell}\frac{N(\ell)}{\bar{D}_0\bar{D}_1\cdots \bar{D}_{N-1}}},\\
\mathcal{C}_{R_2}(r)&=&\int{\mathrm{d}^d\bar{\ell}\frac{\tilde{N}(\ell,\tilde{\ell})}{\bar{D}_0\bar{D}_1\cdots \bar{D}_{N-1}}}.
\end{eqnarray}
The second term, $\mathcal{C}_{R_2}(r)$, represents the rational term of type $R_2$ following the nomenclature used in ref.~\cite{Ossola:2008xq}.  Refs.~\cite{Binoth:2006hk,Bredenstein:2008zb} demonstrate that $R_2$ rational terms of IR origin cancel in the UV-unrenormalized scattering amplitude $\ampnl_{\mathrm{U}}(r)$ and arise only from wave-function renormalization constants. Consequently, only the $R_2$ rational terms of UV origin need to be calculated. Based on this observation, it was shown in  ref.~\cite{Ossola:2008xq} that the $R_2$ rational terms can be derived from a tree-level amplitude using a finite and universal set of theory-dependent Feynman rules, which can be established once and for all in each given theory. For instance, the $R_2$ Feynman rules in QCD~\cite{Draggiotis:2009yb},  the SM~\cite{Garzelli:2009is,Garzelli:2010qm,Shao:2011tg},  Higgs effective field theory~\cite{Page:2013xla,Chen:2019fhs}, and minimal supersymmetric SM~\cite{Shao:2012ja} are well documented in the literature. Nowadays, for any given model, the Feynman rules for $R_2$ terms can be automatically derived using \nloct~\cite{Degrande:2014vpa}, a module of the {\tt Mathematica} package \feynrules~\cite{Christensen:2008py,Alloul:2013bka}. This approach introduces no additional complexities and maintains a computational workload that is negligible compared to that required for the 4-dimensional part $\mathcal{C}_{\mathrm{non-}R_2}$. Therefore, it is the most widely used method in modern tools. This results in the computation of a tree-level $R_2$ amplitude $\ampnRT(r)$:
\begin{eqnarray}
\ampnRT(r)&=&\sum_{\mathrm{diagrams}}{\mathcal{C}_{R_2}(r)},
\end{eqnarray}
whose complexity is anyhow comparable to that of the UV renormalization amplitude $\ampnUV(r)$. The UV-unrenormalized one-loop amplitude can also be expressed as:
\begin{eqnarray}
\ampnl_{\mathrm{U}}(r)&=&\ampnl_{\mathrm{U,non-}R_2}(r)+\ampnRT(r)
\end{eqnarray}
with
\begin{eqnarray}
\ampnl_{\mathrm{U,non-}R_2}(r)&=&\sum_{\mathrm{diagrams}}{\mathcal{C}_{\mathrm{non-}R_2}(r)}.
\end{eqnarray}

After separating out the $R_2$ term, we still need to evaluate the non-$R_2$ part $\mathcal{C}_{\mathrm{non-}R_2}(r)$ using an integral reduction procedure:
\begin{eqnarray}
\mathcal{C}(r)&=&\mathrm{Red}[\mathcal{C}_{\mathrm{non-}R_2}(r)]+\mathcal{C}_{R_2}(r).
\end{eqnarray}
Since the pioneer work by Passarino and Veltman~\cite{Passarino:1978jh}, numerous one-loop reduction methods have been proposed in the literature. These methods can be broadly classified into three categories: tensor-integral reduction (TIR)~\cite{Passarino:1978jh,Stuart:1987tt,Stuart:1989de,Davydychev:1991va,Bern:1993kr,Campbell:1996zw,Tarasov:1996br,Fleischer:1999hq,Binoth:1999sp,Giele:2004iy,Denner:2005nn,Binoth:2005ff,Diakonidis:2008ij,Fleischer:2010sq,Shao:2011wx}, integrand-level reduction~\cite{Ossola:2006us,delAguila:2004nf,Mastrolia:2008jb,Mastrolia:2010nb,Mastrolia:2012bu}, and generalized unitarity reduction~\cite{Bern:1994zx,Britto:2004nc,Ellis:2007br,Ellis:2008ir}. While it is not necessary to discuss each method in detail here, it is worthwhile to provide brief comments on each approach.

As the name suggests, TIR methods perform reductions for tensor integrals, since the numerator can be expressed in the following form:
\begin{eqnarray}
N(\ell)&=&\sum_{r=0}^{r_{\mathrm{max}}}{C^{(r)}_{\mu_1\cdots\mu_r}\prod_{i=1}^{r}{\ell^{\mu_i}}}.
\end{eqnarray}
The tensor integrals in $\ampnl_{\mathrm{U,non-}R_2}$ take the form:
\begin{eqnarray}
\int{\mathrm{d}^d\bar{\ell}\frac{\ell^{\mu_1}\cdots \ell^{\mu_r}}{\bar{D}_0\bar{D}_1\cdots \bar{D}_{N-1}}},
\end{eqnarray}
where the rank $r$ should not be confused with a partonic process $r$. In TIR, one employs a set of identities derived from Lorentz variance, translation invariance, integration by parts, and other principles, which allow us to reduce the tensor integrals to those with fewer denominators or lower rank. However, solving these identities requires the inversion of certain matrices, such as Gram or (modified) Cayley matrices. The numerically stability of the system can become an issue if the determinants of these matrices approach zero. In such cases, different iterative solutions to the reduction equations that avoid these instabilities are available, as implemented in \collier~\cite{Denner:2016kdg} based on the work of ref.~\cite{Denner:2005nn}.

The best-known integrand-level reduction method is the OPP method, named after its authors Ossola, Papadopoulos, and Pittau~\cite{Ossola:2006us}. The main idea of the OPP method is to expand the numerator function $N(\ell)$ in terms of the denominators $\bar{D}_0,\ldots,\bar{D}_{N-1}$. Due to the dimensionality of $\ell$, $N(\ell)$ can be decomposed into $4$-dimensional denominators $D_i=(\ell+p_i)^2-M_i^2=\bar{D}_i-\tilde{\ell}^2$ as follows:
\begin{eqnarray}
N(\ell)&=&\sum_{0\leq i_0<i_1<i_2<i_3\leq N-1}{\left[d_{i_0i_1i_2i_3}+\hat{d}_{i_0i_1i_2i_3}(\ell)\right]\mathop{\prod_{i=0}^{N-1}}_{i\neq i_0,i_1,i_2,i_3}{D_i}}\nonumber\\
&&+\sum_{0\leq i_0<i_1<i_2\leq N-1}{\left[c_{i_0i_1i_2}+\hat{c}_{i_0i_1i_2}(\ell)\right]\mathop{\prod_{i=0}^{N-1}}_{i\neq i_0,i_1,i_2}{D_i}}\nonumber\\
&&+\sum_{0\leq i_0<i_1\leq N-1}{\left[b_{i_0i_1}+\hat{b}_{i_0i_1}(\ell)\right]\mathop{\prod_{i=0}^{N-1}}_{i\neq i_0,i_1}{D_i}}\nonumber\\
&&+\sum_{0\leq i_0\leq N-1}{\left[a_{i_0}+\hat{a}_{i_0}(\ell)\right]\mathop{\prod_{i=0}^{N-1}}_{i\neq i_0}{D_i}}\nonumber\\
&&+\hat{P}(\ell)\prod_{i=0}^{N-1}{D_i}.\label{eq:Nldecomp}
\end{eqnarray}
The terms proportional to the coefficients $\hat{d}$, $\hat{c}$, $\hat{b}$, $\hat{a}$, and $\hat{P}$ vanish upon integration over $\mathrm{d}^d\bar{\ell}$; these are referred to as spurious terms. The functional forms of $\ell$ in these spurious coefficients are established and utilized in ref.~\cite{Ossola:2006us} for any renormalizable theory. In principle, one can derive these spurious coefficients as well as the $\ell$-independent coefficients $d$, $c$, $b$, $a$ by solving the master equation \eqref{eq:Nldecomp} using a sufficient number of numerical complex values of $\ell$ for each phase space point. Notably, this operation can be conducted at the amplitude or amplitude squared level instead of on a diagram-by-diagram basis. However, the resulting system of algebraic equations can be quite large, prompting various strategies for efficient resolution in the literature. The original approach proposed in ref.~\cite{Ossola:2006us} involves selecting specific values of $\ell$ to systematically set 4, 3, 2, or 1 of the possible denominators $D_i$ to zero. This transforms the system of equations into a block-triangular form, facilitating a sequential solution: first solving for the coefficients of all possible 4-point functions, followed by the coefficients for all 3-point functions, and so on. This method can be further optimized~\cite{Mastrolia:2008jb,Mastrolia:2012bu}. To match the dimensionality of denominators, one must replace all $4$-dimensional denominators $D_i$ with their $d$-dimensional counterparts $\bar{D}_i$ on the r.h.s of eq.\eqref{eq:Nldecomp}. This yields the cut-constructive (cc) numerator:
\begin{eqnarray}
N_{\mathrm{cc}}(\ell,\tilde{\ell}^2)&=&\left.N(\ell)\right|_{\mathrm{r.h.s.~of~eq}.\eqref{eq:Nldecomp},D_i\to \bar{D}_i}.
\end{eqnarray}
Here, $N(\ell)$ on the r.h.s refers to the expression replaced from the r.h.s of eq.\eqref{eq:Nldecomp}. The difference between the two numerators gives rise to the rational terms of type $R_1$~\cite{Ossola:2008xq}:
\begin{eqnarray}
\mathcal{C}_{R_1}(r)&=&\int{\mathrm{d}^d\bar{\ell}\frac{N_{R_1}(\ell,\tilde{\ell}^2)}{\bar{D}_0\bar{D}_1\cdots \bar{D}_{N-1}}},
\end{eqnarray}
where
\begin{eqnarray}
N_{R_1}(\ell,\tilde{\ell}^2)&=&N(\ell)-N_{\mathrm{cc}}(\ell,\tilde{\ell}^2).
\end{eqnarray}
The cc term is given by:
\begin{eqnarray}
\mathcal{C}_{\mathrm{cc}}(r)&=&\int{\mathrm{d}^d\bar{\ell}\frac{N_{\mathrm{cc}}(\ell,\tilde{\ell}^2)}{\bar{D}_0\bar{D}_1\cdots \bar{D}_{N-1}}}\nonumber\\
&=&\sum_{0\leq i_0<i_1<i_2<i_3\leq N-1}{d_{i_0i_1i_2i_3}\int{\mathrm{d}^d\bar{\ell}\frac{1}{\bar{D}_{i_0}\bar{D}_{i_1}\bar{D}_{i_2}\bar{D}_{i_3}}}}\nonumber\\
&&+\sum_{0\leq i_0<i_1<i_2\leq N-1}{c_{i_0i_1i_2}\int{\mathrm{d}^d\bar{\ell}\frac{1}{\bar{D}_{i_0}\bar{D}_{i_1}\bar{D}_{i_2}}}}\nonumber\\
&&+\sum_{0\leq i_0<i_1\leq N-1}{b_{i_0i_1}\int{\mathrm{d}^d\bar{\ell}\frac{1}{\bar{D}_{i_0}\bar{D}_{i_1}}}}\nonumber\\
&&+\sum_{0\leq i_0\leq N-1}{a_{i_0}\int{\mathrm{d}^d\bar{\ell}\frac{1}{\bar{D}_{i_0}}}}.\label{eq:Ccutcon}
\end{eqnarray}
The non-$R_2$ term is then:
\begin{eqnarray}
\mathcal{C}_{\mathrm{non-}R_2}(r)&=&\mathcal{C}_{\mathrm{cc}}(r)+\mathcal{C}_{R_1}(r).
\end{eqnarray}
Similar to $R_2$, $R_1$ rational terms arise only in the presence of $1/\epsilon$ poles of UV origin, which can be understood from the TIR reduction of one-loop integrals~\cite{Denner:2005nn}. Consequently, these terms can be easily identified through simple power counting. The $R_1$ terms can always be computed alongside the cc terms. The OPP reduction algorithm has been implemented in various numerical tools, including \cuttools~\cite{Ossola:2007ax}, \samurai~\cite{Mastrolia:2010nb}, and \ninja~\cite{Peraro:2014cba}.

The generalized unitarity methods are extensions of tree-level on-shell techniques to the one-loop case. They exploit multiple branch cuts to place one or more internal loop particles on-shell. Consequently, the one-loop amplitude can be fully determined by these on-shell conditions. Generalized unitarity methods are closely related to the OPP reduction; for instance, ref.~\cite{Ellis:2007br} demonstrated that the coefficients of the scalar one-loop integrals in the OPP reduction can be entirely determined in terms of on-shell tree-level amplitudes with complex external momenta. The generalized unitarity based efforts have primarily focus on studies of processes involving high-multiplicity final states at NLO in QCD. However, they have yet to be generalized to scenarios involving massive loop particles.

After reduction, the final task is to evaluate the one-loop scalar integrals $\mathcal{J}_i^{(\mathrm{Red})}$.  In the $d\to 4$ limit, these integrals can take a special form, as seen on the r.h.s of eq.\eqref{eq:generic1loop}, with $1\leq N\leq 4$ and $\bar{N}(\bar{\ell})=1$ (also see eq.\eqref{eq:Ccutcon}). Depending on the value of $N$, they are typically referred to as scalar $N$-point functions/integrals, where the term ``scalar" indicates that the numerator function is unity. The integral $\mathcal{J}_i^{(\mathrm{Red})}$ generally exhibits a Laurent expansion in the dimensional regulator $\epsilon$:
\begin{eqnarray}
\mathcal{J}_i^{(\mathrm{Red})}&=&\sum_{l=-2}^{+\infty}{\epsilon^{l}\mathcal{J}_{i,l}^{(\mathrm{Red})}}.
\end{eqnarray}
For all possible cases, the analytic expressions for $\mathcal{J}_{i,l}^{(\mathrm{Red})}$ with $l\leq 0$ are known in the literature. These expressions can be written in terms of logarithms and dilogarithms, regardless of whether the masses of the loop particles are real or complex.  They have been implemented in several public numerical codes~\cite{vanOldenborgh:1990yc,Hahn:1998yk,Ellis:2007qk,Binoth:2008uq,Nhung:2009pm,vanHameren:2010cp,Cullen:2011kv,Carrazza:2016gav,Denner:2016kdg}.

\section{The \mgamclong\ framework\label{sec:mgamc}}

\mgamcshort~\cite{Alwall:2014hca}, the successor to \mg~\cite{Alwall:2011uj} and \amcatnlo~\cite{Frederix:2011zi,Frederix:2011qg,Frederix:2011ss}, is a single comprehensive framework that automates the computation of both LO, either tree-level or loop-induced~\cite{Hirschi:2015iia}, and NLO accurate (differential) cross sections, including their matching to parton shower programs via the MC@NLO method~\cite{Frixione:2002ik}. The code is designed to be flexible, enabling calculations that can be performed even by those who are not familiar with quantum field theory. In this sense, it provides all the necessary elements for studies in the SM and BSM phenomenology. In particular, \mgamcshort\ employs the FKS subtraction method, which is automated in the \mfks\ module~\cite{Frederix:2009yq}~\footnote{The resonance treatment in FKS subtraction has been updated in \mfks\ since the study of ref.~\cite{Frederix:2016rdc}.} for addressing IR singularities in real emission contributions (\cf\ eq.\eqref{eq:NLOxsFKS}). The computations of one-loop amplitudes are conducted in the \ml\ module~\cite{Hirschi:2011pa,Alwall:2014hca}, which dynamically switches between two integral-reduction techniques: TIR and integrand-level reduction (such as OPP). For matching NLO QCD computations with parton showers, the MC@NLO formalism is utilized.~\footnote{The MC@NLO implementation yields both positive and negative weights. The fraction of negative weights determines the effective statistics. To reduce the number of negative weight events, two variants of the original MC@NLO formalism, known as  MC@NLO-$\Delta$ and \textit{Born spreading}, have been proposed in ref.~\cite{Frederix:2020trv} and ref.~\cite{Frederix:2023hom}, respectively.} Finally, inclusive samples that are accurate at the NLO QCD level but span many jet multiplicities can be obtained using the Frederix-Frixione (FxFx) merging method~\cite{Frederix:2012ps}. Let us describe the \ml\ and \mfks\ modules in more detail.

\ml\ employs an original method for generating one-loop Feynman diagrams by leveraging the existing tree-level diagram generation algorithm in \mgamcshort. This approach directly generates the so-called {\it L-cut diagrams}, which represent loop diagrams with exactly one loop propagator cut open. It effectively transforms the one-loop diagrams into tree-level diagrams but with two additional final-state particles compared to the original loop process. Significant improvements have been made in \ml5~\cite{Alwall:2011uj} compared to its predecessor \ml4~\cite{Hirschi:2011pa}. The efficiency of loop numerator generation has been enhanced through an in-house implementation of the \openloops\ idea~\cite{Cascioli:2011va}, enabling for dynamic switching between interfaces for three different integrand-level reduction programs: \cuttools~\cite{Ossola:2007ax}, \samurai~\cite{Mastrolia:2010nb}, and \ninja~\cite{Peraro:2014cba,Hirschi:2016mdz}, and four distinct TIR codes: \pjfry~\cite{Fleischer:2011zz}, \iregi~\cite{Alwall:2014hca,Shao:2016knn}, \golem~\cite{Binoth:2008uq}, and \collier~\cite{Denner:2016kdg}. This is critical for improving the procedure for fixing numerically unstable loop-integral reductions, which are detected by newly designed self-diagnostic numerical stability tests.  Another key feature of \ml5 is its flexibility. The introduction of the \aloha\ module~\cite{deAquino:2011ub}, which allows for the on-the-fly generation of \helas\ routines~\cite{Murayama:1992gi} within any model in the Universal Feynman Output (UFO) format~\cite{Degrande:2011ua,Darme:2023jdn}, enables broad applicability to BSM physics. Furthermore, \ml5 allows users to select specific contributions from the full one-loop matrix element, such as restricting the allowed particle content, requiring certain propagators, or choosing any arbitrary gauge-invariant subset of diagrams. As long as the input UFO model contains the relevant information, the corresponding UV and $R_2$ counterterm contributions will be accurately accounted for. Communication between \ml\ and \mfks\ is handled via the Binoth Les Houches accord~\cite{Binoth:2010xt,Alioli:2013nda}.

The FKS subtraction formalism in \mfks\ has evolved from its original version~\cite{Frederix:2009yq} to the current implementation in \mgamcshort. These improvements are summarized in section 2.4.1 of ref.~\cite{Alwall:2014hca}. Briefly, the complete set of the bookkeeping real emission processes, $\allprocnpo$, is now generated more efficiently by considering all possible $\ident_{j\oplus i}\to \ident_{i}\ident_{j}$ branchings, based on the underlying Born processes $\allprocn$.~\footnote{In this way, certain IR-finite real partonic processes might not be included in an NLO run in \mgamcshort\ but can be straightforwardly incorporated as a separate LO run. An example of this is the NLO QCD calculation of inclusive Higgs boson hadroproduction ($pp\to h+X$) in the infinite top-quark mass limit, where the real partonic process $q\bar{q}\to h g$ is IR finite.} In \mgamcshort, \mfks\ maps real emission processes $\allprocnpo$ to Born processes $\allprocn$, in contrast to the original implementation. This allows adaptive importance sampling techniques to randomly select a real-emission process and an FKS pair, enhancing efficiency in complex processes and offering advantages when matching to parton showers. Since evaluating virtual matrix elements is usually the slowest part, \mfks\ uses a ``virtual trick" when integrating the one-loop contribution (see section 2.4.3 of ref.~\cite{Alwall:2014hca}). This method reduces the fraction of phase space points when \ml\ must be called by approximating the virtual matrix element as a constant times the Born matrix element over a phase-space grid. The difference between the exact virtual matrix element and this approximation is added back in, but the number of phase space points needed for this integration is significantly smaller. The constant and the fraction of \ml\ calls are dynamically updated using the information gathered by the numerical integrator during previous iterations.

The building blocks of \mgamcshort\ parametrize cross sections in a theory- and process-independent way, with theory- and process-specific information provided at runtime, fully under the user's control. Specifically, information about the theory, such as particle content and interactions, is provided in the form of a UFO model. Most models can be automatically constructed from the Lagrangian using tools like \feynrules\ and \nloct, the latter being crucial for embedding the UV and $R_2$ counterterms necessary for one-loop computations. Over the last decade, numerous efforts have extended the \mgamcshort\ framework to perform various BSM studies at NLO QCD accuracy~\cite{Artoisenet:2013puc,Maltoni:2013sma,Degrande:2014tta,Demartin:2014fia,Degrande:2014sta,Durieux:2014xla,Demartin:2015uha,Degrande:2015vaa,Backovic:2015soa,Degrande:2015vpa,Degrande:2015xnm,Mattelaer:2015haa,Neubert:2015fka,BuarqueFranzosi:2015jrv,Das:2016pbk,Fuks:2016ftf,Degrande:2016dqg,Degrande:2016hyf,Maltoni:2016yxb,Arina:2016cqj,BessidskaiaBylund:2016jvp,Zhang:2016omx,Degrande:2016aje,Bell:2016ekl,Mattelaer:2016ynf,Bell:2017rgi,BuarqueFranzosi:2017jrj,Fuks:2017vtl,Cacciapaglia:2017gzh,Hirschi:2018etq,Cacciapaglia:2018rqf,Cacciapaglia:2018qep,Frixione:2019fxg,Fuks:2019clu,Degrande:2020evl,Arina:2020udz,Fuks:2020zbm,Deandrea:2021vje,AH:2023hft,Javurkova:2024bwa}. These studies showcase the flexibility of the framework, because they necessitate the support of many novel aspects absent in the SM, such as non-renormalizable operators~\cite{Degrande:2016dqg,Maltoni:2016yxb,Hirschi:2018etq,Degrande:2020evl}, Majorana fermions~\cite{Degrande:2015vaa,Frixione:2019fxg}, and spin-$2$ particles~\cite{Das:2016pbk}, and of infrequently occurring phenomena in the SM like  the strategies for overcoming problems posed to NLO computations by the presence of narrow resonance(s) in a theory with a sufficiently rich particle spectrum~\cite{Frixione:2019fxg}.

\chapter{Automation of Electroweak Radiative Corrections}
\label{SEC:NLOEW}

The automation of the calculation of EW radiative corrections is still ongoing. The methods are fundamentally similar to those used for QCD corrections, with some necessary modifications. However, the complexity increases due to the mixing of QCD and EW corrections, the larger number of contributing terms, the presence of multiple and diverse mass scales, the chiral structure of EW interactions, and the instability of many SM particles.

\section{Generalities\label{eq:EWgeneral}}

When considering perturbative corrections, it is important to recognize that we typically consider an expansion based on a single quantity, such as a coupling constant in a fixed-order computation. However, this is just a specific case of a broader scenario where the expansion occurs simultaneously in two or more couplings, all treated as small parameters. This is referred to as a mixed-coupling expansion. While there is often a clear numerical hierarchy among these couplings, such as $\alpha_s\gg \alpha$, the mixed-coupling scenario is not merely academic. As we will demonstrate, there are cases where it becomes necessary to address it. 

\begin{figure}[htbp!]
\centering
\includegraphics[width=0.8\textwidth]{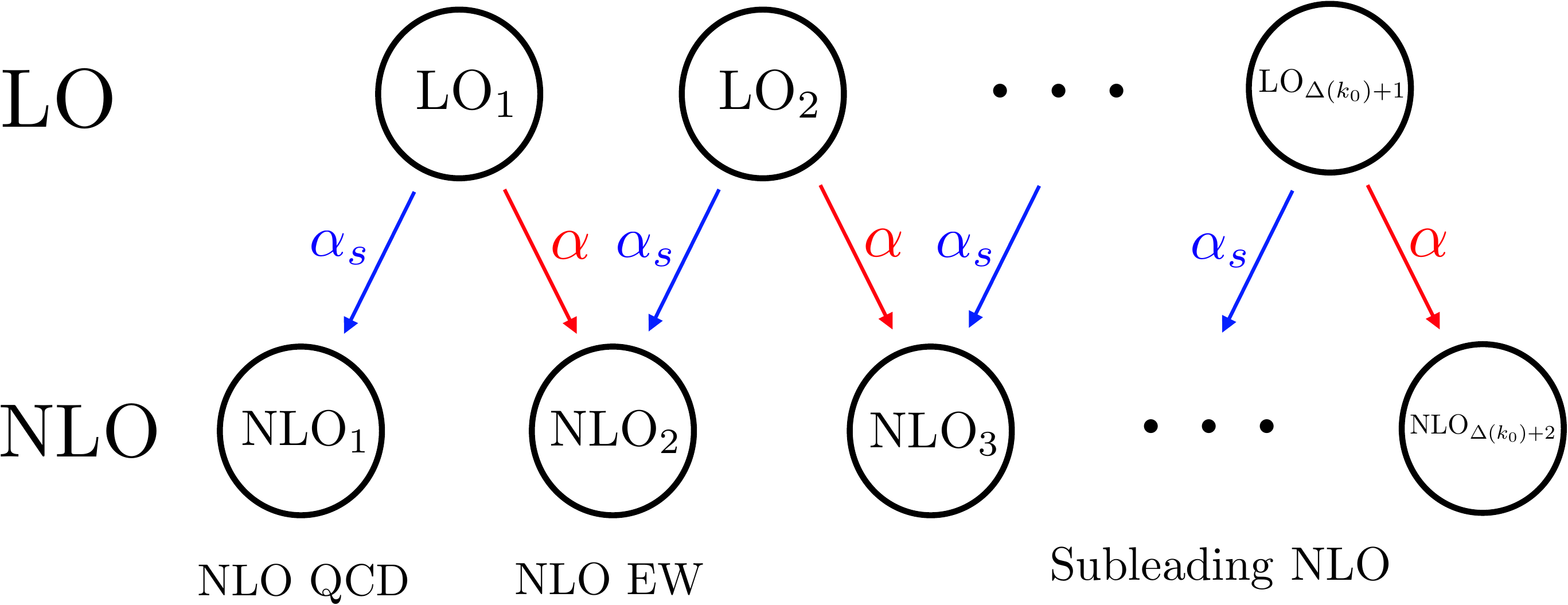}
\caption{A graphical illustration of QCD (blue, right-to-left arrows) and EW (red, left-to-right arrows) corrections to a generic process. Each circle represents a specific contribution at a fixed order in terms of the QCD and EW coupling constants, while the black ellipses indicate the potential presence of additional cross section contributions. See the text for details.\label{fig:mixedorderexpand}}
\end{figure}

To explore a generic mixed-coupling expansion in a concrete way, let us consider a generic observable $\Sigma$ (\eg, a cross section for a given scattering process) that receives contributions from processes involving both QCD and EW interactions. While the specific nature of these interactions is not crucial, what matters for our discussion is that $\Sigma$ generally depends on multiple coupling constants. Let us assume the observable $\Sigma(\alpha_s,\alpha)$ admits the following Taylor expansion in the couplings~\cite{Alwall:2014hca,Frederix:2018nkq}:
\begin{eqnarray}
\Sigma(\alpha_s,\alpha)&=&\alpha_s^{c_s(k_0)}\alpha^{c(k_0)}\sum_{p=0}^{+\infty}{\sum_{q=0}^{\Delta(k_0)+p}{\Sigma_{k_0+p,q}\alpha_s^{\Delta(k_0)+p-q}\alpha^q}}\nonumber\\
&=&\Sigma^{({\rm LO})}(\alpha_s,\alpha)+\Sigma^{({\rm NLO})}(\alpha_s,\alpha)+\ldots,\label{eq:Sigmaexp0}
\end{eqnarray}
where we have identified the (N)LO contribution $\Sigma^{({\rm (N)LO})}$ with the $p=0$ ($1$) terms, \ie,
\begin{eqnarray}
\Sigma^{({\rm LO})}(\alpha_s,\alpha)&=&\alpha_s^{c_s(k_0)}\alpha^{c(k_0)}\sum_{q=0}^{\Delta(k_0)}{\Sigma_{k_0,q}\alpha_s^{\Delta(k_0)-q}\alpha^q}\nonumber\\
&=&\Sigma_{\mathrm{LO}_1}+\ldots+\Sigma_{\mathrm{LO}_{\Delta(k_0)+1}},\\
\Sigma^{({\rm NLO})}(\alpha_s,\alpha)&=&\alpha_s^{c_s(k_0)}\alpha^{c(k_0)}\sum_{q=0}^{\Delta(k_0)+1}{\Sigma_{k_0+1,q}\alpha_s^{\Delta(k_0)+1-q}\alpha^q}\nonumber\\
&=&\Sigma_{\mathrm{NLO}_1}+\ldots+\Sigma_{\mathrm{NLO}_{\Delta(k_0)+2}}.
\end{eqnarray}
The non-negative integers $k_0$, $c_s(k_0)$, $c(k_0)$, and $\Delta(k_0)$ are observable-dependent quantities, with the constraint $k_0=c_s(k_0)+c(k_0)+\Delta(k_0)$. A particular example is the cross section for dijet hadroproduction given in ref.~\cite{Frederix:2016ost}, where $k_0=\Delta(k_0)=2$ and $c_s(k_0)=c(k_0)=0$. In a well-behaved perturbative series, where the coefficients $\Sigma_{k_0+p,q}$ are of the same order, we observe the hierarchies
\begin{equation}
\Sigma_{{\mathrm{N}^p\mathrm{LO}}_i}\gg \Sigma_{{\mathrm{N}^p\mathrm{LO}}_{i+1}},\quad \forall\ i, p,\label{eq:hierachy0}
\end{equation}
due to $\alpha_s\gg \alpha$, and 
\begin{equation}
\Sigma_{{\mathrm{N}^{p}\mathrm{LO}}_i}\gtrsim \Sigma_{{\mathrm{N}^{p+1}\mathrm{LO}}_i},\quad \forall\ i, p,\label{eq:hierachy1}
\end{equation}
due to the perturbativity $\alpha_s\lesssim 1$. In real situations, however, such hierarchies may be violated, and I will give a few examples in Chapter \ref{SEC:PHENO}. The hierarchy eq.\eqref{eq:hierachy0} suggests that we identify $\Sigma_{{\mathrm{N}^p\mathrm{LO}}_i}$ as the leading ($i = 1$) or $i$th-leading ($i > 1$) term of the N$^p$LO contribution to the observable $\Sigma$.  It is customary to refer to $\Sigma_{\mathrm{NLO}_1}$ and $\Sigma_{\mathrm{NLO}_2}$ as the NLO QCD and NLO EW corrections, respectively, while $\Sigma_{\mathrm{NLO}_i}$ with $i\geq 3$ are called subleading NLO corrections, as they are typically expected to be numerically subdominant compared to the first, based on the considerations of eq.\eqref{eq:hierachy0}. Figure~\ref{fig:mixedorderexpand} summarizes our discussion thus far. These discussions are not limited to QCD and EW corrections and can be easily generalized to other interactions and more than two couplings.

\begin{figure}[htbp!]
\centering
\includegraphics[width=0.32\textwidth]{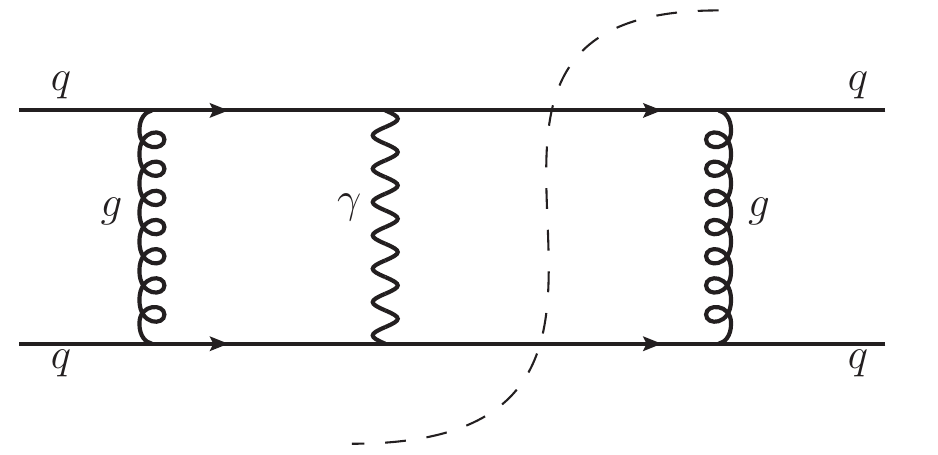}
\includegraphics[width=0.32\textwidth]{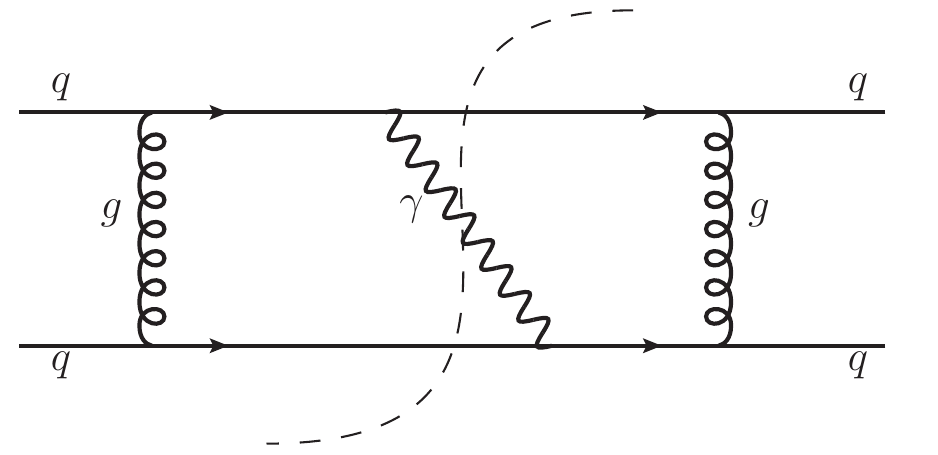}
\includegraphics[width=0.32\textwidth]{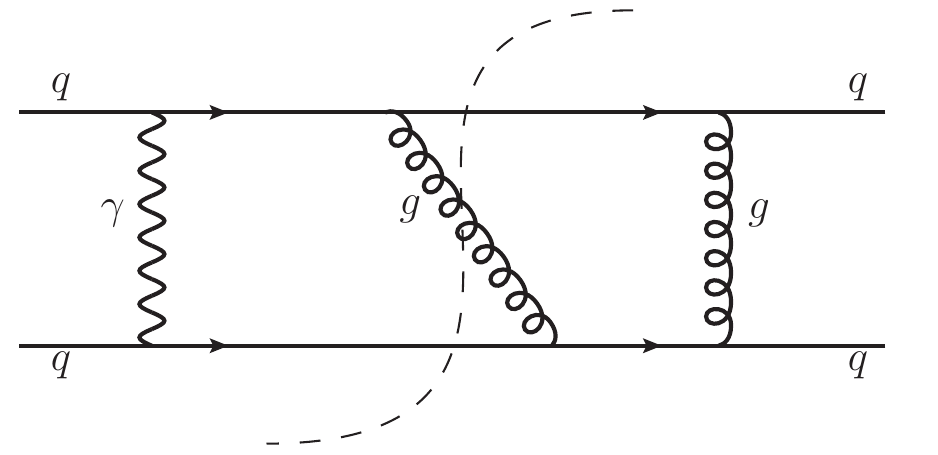}
\caption{Forward scattering graphs for one-loop (left) and real-emission (middle and right) $qq\to qq$ contributions to dijet production at NLO EW ($\mathrm{NLO}_2$, $\mathcal{O}(\alpha_s^2\alpha)$). The dashed lines represent Cutkosky cuts. \label{fig:qq2qqcut}}
\end{figure}

While the interpretation of $\Sigma_{\mathrm{NLO}_1}$ is generally straightforward, it is somehow misleading when talking about NLO EW and subleading NLO contributions. At one loop, there
is no clear-cut way to define purely EW contributions on a diagrammatic basis, aside from the last term $\Sigma_{\mathrm{NLO}_{\Delta(k_0)+2}}$. For example, in the case of $qq\to qq$ contributing to the dijet process at NLO EW $\mathcal{O}(\alpha_s^2\alpha)$, both one-loop virtual and real emission contributions exist, as shown in figure~\ref{fig:qq2qqcut}. Broadly, the middle diagram of that figure can be identified as representing EW real-emission corrections to the $\mathcal{O}(\alpha_s^2)$ Born contribution $\mathrm{LO}_1$ (\ie, the red arrow from $\mathrm{LO}_1$ to $\mathrm{NLO}_2$ in figure~\ref{fig:mixedorderexpand}), because the photon is cut. On the other hand, the diagram on the right shows QCD real-emission corrections to the $\mathrm{LO}_2$ ($\mathcal{O}(\alpha_s\alpha)$) contribution, corresponding to the blue arrow from $\mathrm{LO}_2$ to $\mathrm{NLO}_2$ in figure~\ref{fig:mixedorderexpand}), as indicated by the gluon being cut. Additionally, the one-loop virtual diagram on the left of figure~\ref{fig:qq2qqcut} cannot be interpreted as either arrow in figure~\ref{fig:mixedorderexpand} since its underlying Born process is ambiguous. This suggests that, although being useful in a technical sense, separating QCD and EW corrections are not always physically meaningful.  In general, both must be considered simultaneously to obtain a sensible and NLO-accurate result. Moreover, $\mathrm{NLO}_i$ ($i\geq 2$) can receive contributions from heavy-boson radiation (HBR), such as in the second diagram in figure~\ref{fig:qq2qqcut}, where the photon $\gamma$ is replaced by a $Z$ boson. Although heavy-boson virtual diagrams (\eg, replacing $\gamma$ with $Z$ in the first diagram of figure~\ref{fig:qq2qqcut}) are always included in NLO EW and subleading NLO corrections, HBR in real emission is  typically excluded from conventional NLO EW corrections since it contributes at LO to a different final state. In other words, HBR itself is IR finite. For example, in the dijet case, HBR corresponds to LO dijet plus heavy-boson associated production, rather than NLO EW radiative corrections to the dijet process. In the following,  to be precise, we exclude HBR from NLO corrections unless explicitly stated otherwise.

%\section{Techniques}

\section{FKS subtraction in mixed QCD and EW coupling expansions}

From the perspective of IR divergences, only photon emission in NLO EW calculations can lead to singularities in phase space, as other bosons (Higgs, $W^\pm$, $Z$) are massive.  In this regard, the IR singularities in NLO QCD corrections already represent the worst-case scenario due to the non-Abelian nature of QCD. Thus, the FKS subtraction for EW corrections can largely follow a literal translation of its formulation in QCD, by taking the Abelian limits. At NLO and in the soft or collinear limit, this generally amounts to the following substitutions:
\begin{equation}
g_s\to e,\quad \alpha_s\to \alpha, \quad \Qop_{\mathrm{QCD}}(\ident)\to \Qop_{\mathrm{EW}}(\ident),\quad C_A\to 0,\quad C_F\to \left(e(\ident)\right)^2,\quad T_F\to N_c(\ident)\left(e(\ident)\right)^2,\label{eq:QCD2QEDrep}
\end{equation}
where $e=\sqrt{4\pi\alpha}$, $e(\ident)$ is the electric charge of particle $\ident$ in units of the positron charge, $N_c(\ident)$ is the number of colors of particle $\ident$ (for instance, $N_c(\ident)=1$ if $\ident\in \irrep{1}$, and $N_c(\ident)=3$ if $\ident\in \irrep{3},\irrepbar{3}$). The EW charge generator $\Qop_{\mathrm{EW}}(\ident)$, the counterpart of the QCD color generator $\Qop_{\mathrm{QCD}}(\ident)$, is a scalar operator defined as
\begin{eqnarray}
\Qop_{\mathrm{EW}}(\ident)&=&(-1)^{s(\ident)}e(\ident),
\end{eqnarray}
where
\begin{eqnarray}
s(\ident)&=&\left\{\begin{array}{ll} 0, & \ident~\mathrm{is~a~final~particle}\\
1, & \ident~\mathrm{is~an~initial~particle}\\
\end{array}\right..\label{eq:sIdef}
\end{eqnarray}
Now, we will explicitly formulate the FKS subtraction in the mixed QCD and EW coupling expansion,  largely following section 3 in ref.~\cite{Frederix:2018nkq}.

Let us first explicitly introduce the powers of coupling dependence in amplitudes and matrix elements. For any process $r\in\allprocn$, we write the tree-level amplitudes and matrix elements as:
\begin{equation}
\ampnt_{(p,q)}(r)\propto g_s^p e^q,\quad \ampsqnt_{(p,q)}(r)\propto \alpha_s^p\alpha^q,\quad \forall p,q\in \mathbb{N}.
\end{equation}
We can define the similar expressions for $r\in\allprocnpo$ and for the one-loop case. At N$^l$LO, the matrix elements always satisfy $p+q=k_0+l$. The matrix elements defined in eqs.\eqref{eq:BornME}, \eqref{eq:virtualME}, and \eqref{eq:RealME} can be generalized accordingly to
\begin{eqnarray}
\ampsqnt_{(p,q)}(r)&=&\frac{1}{2s}\frac{1}{\omega(\Ione)\omega(\Itwo)}\mathop{\sum_{\rm color}}_{\rm spin}\mathop{\sum_{p_1,p_2}}_{q_1,q_2}{\delta_{(\frac{p_1+p_2}{2})p}\delta_{(\frac{q_1+q_2}{2})q}\delta_{\mathrm{mod}(p_1+p_2,2)0}\delta_{\mathrm{mod}(q_1+q_2,2)0}}\nonumber\\
&&\times\left(2-\delta_{p_1p_2}\delta_{q_1q_2}\right)\Re\left\{\ampnt_{(p_1,q_1)}(r)\ampnt_{(p_2,q_2)}(r)^{\star}\right\},\quad \quad \quad \quad p+q=k_0,\label{eq:BornME2}\\
\ampsqnl_{(p,q)}(r)&=&\frac{1}{2s}\frac{1}{\omega(\Ione)\omega(\Itwo)}\mathop{\sum_{\rm color}}_{\rm spin}\mathop{\sum_{p_1,p_2}}_{q_1,q_2}{\delta_{(\frac{p_1+p_2}{2})p}\delta_{(\frac{q_1+q_2}{2})q}\delta_{\mathrm{mod}(p_1+p_2,2)0}\delta_{\mathrm{mod}(q_1+q_2,2)0}}\nonumber\\
&&\times \delta_{(p_1+q_1+2)(p_2+q_2)}2\Re\left\{\ampnt_{(p_1,q_1)}(r){\ampnl_{(p_2,q_2)}(r)}^{\star}\right\},\quad \quad \quad p+q=k_0+1,\label{eq:virtualME2}\\
\ampsqnpot_{(p,q)}(r)&=&\frac{1}{2s}\frac{1}{\omega(\Ione)\omega(\Itwo)}\mathop{\sum_{\rm color}}_{\rm spin}\mathop{\sum_{p_1,p_2}}_{q_1,q_2}{\delta_{(\frac{p_1+p_2}{2})p}\delta_{(\frac{q_1+q_2}{2})q}\delta_{\mathrm{mod}(p_1+p_2,2)0}\delta_{\mathrm{mod}(q_1+q_2,2)0}}\nonumber\\
&&\times \left(2-\delta_{p_1p_2}\delta_{q_1q_2}\right)\Re\left\{\ampnpot_{(p_1,q_1)}(r)\ampnpot_{(p_2,q_2)}(r)^\star\right\}, \quad \quad \quad p+q=k_0+1.\label{eq:RealME2}
\end{eqnarray}
The additional average factors in CDR are $\omega(\ell^\pm)=2N_c(\ell^\pm)=2$ for charged leptons $\ell^\pm$, and $\omega(\gamma)=2(1-\epsilon)N_c(\gamma)=2(1-\epsilon)$ for photons. Similarly, the color-linked Born matrix element in eq.\eqref{eq:colorlinkedBornME} can be written as
\begin{eqnarray}
\ampsqnt_{(p,q),\mathrm{QCD},kl}(r)&=&-\frac{1}{2s}
\frac{2-\delta_{kl}}{\omega(\Ione)\omega(\Itwo)}\mathop{\sum_{\rm color}}_{\rm spin}
\mathop{\sum_{p_1,p_2}}_{q_1,q_2}\delta_{(\frac{p_1+p_2}{2})p}\delta_{(\frac{q_1+q_2}{2})q}\delta_{\mathrm{mod}(p_1+p_2,2)0}\delta_{\mathrm{mod}(q_1+q_2,2)0}\nonumber\\
&&\times \left(2-\delta_{p_1p_2}\delta_{q_1q_2}\right)\Re\left\{\ampnt_{(p_1,q_1)}(r) \Qop_{\mathrm{QCD}}(\ident_k)\mydot\Qop_{\mathrm{QCD}}(\ident_l)
{\ampnt_{(p_2,q_2)}(r)}^{\star}\right\},
\label{eq:colorlinkedBornME2}
\end{eqnarray}
where we also have $p+q=k_0$. Obviously, if $\ident\in \irrep{1}$, we have $\Qop_{\mathrm{QCD}}(\ident)=\vec{0}$. The charge-linked Born matrix element can then be expressed as
\begin{eqnarray}
\ampsqnt_{(p,q),\mathrm{EW},kl}(r)&=&-\frac{1}{2s}
\frac{2-\delta_{kl}}{\omega(\Ione)\omega(\Itwo)}\mathop{\sum_{\rm color}}_{\rm spin}
\mathop{\sum_{p_1,p_2}}_{q_1,q_2}\delta_{(\frac{p_1+p_2}{2})p}\delta_{(\frac{q_1+q_2}{2})q}\delta_{\mathrm{mod}(p_1+p_2,2)0}\delta_{\mathrm{mod}(q_1+q_2,2)0}\nonumber\\
&&\times \left(2-\delta_{p_1p_2}\delta_{q_1q_2}\right)\Re\left\{\ampnt_{(p_1,q_1)}(r) \Qop_{\mathrm{EW}}(\ident_k)\mydot\Qop_{\mathrm{EW}}(\ident_l)
{\ampnt_{(p_2,q_2)}(r)}^{\star}\right\}\nonumber\\
&=&(-1)^{1+s(\ident_k)+s(\ident_l)}(2-\delta_{kl})e(\ident_k)e(\ident_l)\ampsqnt_{(p,q)}(r), \quad \quad \quad p+q=k_0.
\label{eq:chargelinkedBornME2}
\end{eqnarray}
The counterpart of eq.\eqref{eq:ampsqtilde0} can now be read as
\begin{eqnarray}
\ampsqnttilde_{(p,q),ij}(r^{j\oplus i,\isubrmv})&=&\frac{1}{2s}
\frac{1}{\omega(\Ione)\omega(\Itwo)}
\Re{\Bigg\{\frac{\langle ij\rangle}{[ij]}\tilde{\mathop{\sum_{\rm color}}_{\rm spin}}\mathop{\sum_{p_1,p_2}}_{q_1,q_2}\!
\delta_{(\frac{p_1+p_2}{2})p}\delta_{(\frac{q_1+q_2}{2})q}\delta_{\mathrm{mod}(p_1+p_2,2)0}\delta_{\mathrm{mod}(q_1+q_2,2)0}}\nonumber\\
&&\times\left(2-\delta_{p_1p_2}\delta_{q_1q_2}\right)\ampnt_{(p_1,q_1),+}(r^{j\oplus i,\isubrmv})
{\ampnt_{(p_2,q_2),-}(r^{j\oplus i,\isubrmv})}^{\star}\Bigg\}\,.~~~~~\label{eq:ampsqtilde1}
\end{eqnarray}

For any given process $r\in\allprocnpo$, the final particles are reordered in such a way that the final massless charged or colored fermions, photons, and gluons are indexed as $3\leq i\leq \nlightR+2$, while the charged or colored particles are indexed as $\nini\leq j\leq \nlightR+\nheavy+2$. In this context, we typically have $\nini=1$ for hadron-hadron, lepton-hadron, and lepton-lepton collisions, unless the initial partons are both color- and charge-neutral, such as neutrinos and photons.~\footnote{For instance, $\nini=2$ for neutrino deep inelastic scattering in neutrino-hadron collisions, and $\nini=3$ for coherent photon-photon scattering processes in ultraperipheral high-energy collisions of protons and/or nuclei~\cite{Shao:2022cly,Shao:2024dmk}.} The definitions of FKS pairs $\FKSpairs(r)$ and the partition function $\Sfunij$ remain unchanged, where, in particular, the jet measurement function $\JetsB$ treats all colored or charged fermions, photons, and gluons in a democratic manner. 

The entire discussion in section \ref{sec:FKSsub} remains valid up to the explicit expressions for the local counterterms, if we apply the following substitutions
\begin{eqnarray}
&&\Sigma_{ij}(r;\xii,\yij)\to \Sigma_{(p,q),ij}(r;\xii,\yij),\quad \ampsqnpot(r)\to\ampsqnpot_{(p,q)}(r),\nonumber\\
&&\hat{\sigma}^{(R)}_{ij,\mathrm{FKS}}(r)\to \hat{\sigma}^{(R)}_{(p,q),ij,\mathrm{FKS}}(r),\quad \hat{\sigma}_{ij}^{(X)}(r)\to \hat{\sigma}_{(p,q),ij}^{(X)}(r),\quad \hat{\sigma}^{(X)}(r)\to\hat{\sigma}^{(X)}_{(p,q)}(r)
\end{eqnarray}
in eqs.\eqref{eq:Sigmaijdef}, \eqref{eq:sigmaFKSijdef1}, \eqref{eq:dsigijnpoE}, \eqref{eq:sigmaRdef}, \eqref{eq:sigmaXdef}, and \eqref{eq:Sigmaijinsoft}. In particular, the collinear and soft-collinear local counterterms in eqs.\eqref{eq:localinitialcoll0} and \eqref{eq:localfinalcoll0} become
\begin{eqnarray}
\lim_{\vec{k}_i\parallel \vec{k}_j}{(1-\yij)\xii^2\ampsqnpot_{(p,q)}(r)\Sfunij(r)}&=&\frac{4}{s}g_s^2\mu^{2\epsilon}\xii P_{\ident_{j\oplus \bar{i}}\ident_j}^{(\mathrm{QCD}),<}(1-\xii,\epsilon)\ampsqnt_{(p-1,q)}(r^{j\oplus \bar{i},\isubrmv})\nonumber\\
&&+\frac{4}{s}e^2\mu^{2\epsilon}\xii P_{\ident_{j\oplus \bar{i}}\ident_j}^{(\mathrm{EW}),<}(1-\xii,\epsilon)\ampsqnt_{(p,q-1)}(r^{j\oplus \bar{i},\isubrmv})\nonumber\\
&&+\frac{4}{s}g_s^2\mu^{2\epsilon}\xi_i\underbrace{Q^{(\mathrm{QCD})}_{\ident_{j\oplus \bar{i}}^\star\ident_j}(1-\xii)\ampsqnttilde_{(p-1,q),ij}(r^{j\oplus \bar{i},\isubrmv})}_{\equiv \Delta^{(\mathrm{QCD})}_{(p-1,q),ij}}\nonumber\\
&&+\frac{4}{s}e^2\mu^{2\epsilon}\xi_i\underbrace{Q^{(\mathrm{EW})}_{\ident_{j\oplus \bar{i}}^\star\ident_j}(1-\xii)\ampsqnttilde_{(p,q-1),ij}(r^{j\oplus \bar{i},\isubrmv})}_{\equiv \Delta^{(\mathrm{EW})}_{(p,q-1),ij}},\label{eq:localinitialcoll1}\\
&&~~~~~~~~~~~~~~~~~~~~~~~~~~~~~~~~j=\nini,\ldots,2,\nonumber\\
\lim_{\vec{k}_i\parallel \vec{k}_j}{(1-\yij)\xii^2\ampsqnpot_{(p,q)}(r)\Sfunij(r)}&=&\frac{4}{s}g_s^2\mu^{2\epsilon}\frac{1-z}{z}h(1-z)P_{\ident_j\ident_{j\oplus i}}^{(\mathrm{QCD}),<}(z,\epsilon)\ampsqnt_{(p-1,q)}(r^{j\oplus i,\isubrmv})\nonumber\\
&&+\frac{4}{s}e^2\mu^{2\epsilon}\frac{1-z}{z}h(1-z)P_{\ident_j\ident_{j\oplus i}}^{(\mathrm{EW}),<}(z,\epsilon)\ampsqnt_{(p,q-1)}(r^{j\oplus i,\isubrmv})\nonumber\\
&&+\frac{4}{s}g_s^2\mu^{2\epsilon}\frac{1-z}{z}h(1-z)\underbrace{Q^{(\mathrm{QCD})}_{\ident_j\ident_{j\oplus i}^\star}(z)\ampsqnttilde_{(p-1,q),ij}(r^{j\oplus i,\isubrmv})}_{\equiv \Delta^{(\mathrm{QCD})}_{(p-1,q),ij}}\nonumber\\
&&+\frac{4}{s}e^2\mu^{2\epsilon}\frac{1-z}{z}h(1-z)\underbrace{Q^{(\mathrm{EW})}_{\ident_j\ident_{j\oplus i}^\star}(z)\ampsqnttilde_{(p,q-1),ij}(r^{j\oplus i,\isubrmv})}_{\equiv \Delta^{(\mathrm{EW})}_{(p,q-1),ij}},\label{eq:localfinalcoll1}\\
&&~~~~~~~~~~~~~~~~~~~~~~~~~~~~~~~~j=3,\ldots, \nlightR+2, j\neq i.\nonumber
\end{eqnarray}
The EW/QED Altarelli-Parisi splitting functions $P_{\ident_k\ident_l}^{(\mathrm{EW}),<}(z,\epsilon)$ and the EW/QED $Q$ kernels $Q^{(\mathrm{EW})}_{\ident_{j\oplus \bar{i}}^\star\ident_j}$ and $Q^{(\mathrm{EW})}_{\ident_j\ident_{j\oplus i}^\star}$ are obtained by substituting their QCD counterparts using the replacement rules in eq.\eqref{eq:QCD2QEDrep}. The non-zero unregularized EW splitting functions are given by:
\begin{eqnarray}
P_{\ident\gamma}^{(\mathrm{EW}),<}(z,\epsilon)&=&N_c(\ident)\left(e(\ident)\right)^2\left(z^2+(1-z)^2-2\epsilon z(1-z)\right),\label{eq:EWunregAP1}\\
P_{\ident\ident}^{(\mathrm{EW}),<}(z,\epsilon)&=&\left(e(\ident)\right)^2\left(\frac{1+z^2}{1-z}-\epsilon(1-z)\right),\label{eq:EWunregAP2}\\
P_{\gamma\ident}^{(\mathrm{EW}),<}(z,\epsilon)&=&\left(e(\ident)\right)^2\left(\frac{1+(1-z)^2}{z}-\epsilon z\right).\label{eq:EWunregAP3}
\end{eqnarray}
The non-vanishing EW $Q$ kernels are:
\begin{eqnarray}
Q_{\ident \gamma^\star}^{(\mathrm{EW})}(z)&=&4N_c(\ident)\left(e(\ident)\right)^2z(1-z),\\
Q_{\gamma^\star \ident}^{(\mathrm{EW})}(z)&=&-4\left(e(\ident)\right)^2\frac{1-z}{z}.
\end{eqnarray}
The spin-correlated terms $\Delta_{(p-1,q),ij}^{(\mathrm{QCD})}$ and $\Delta_{(p,q-1),ij}^{(\mathrm{EW})}$ vanish after integration over the azimuthal variable $\mathrm{d}\varphi_i$. The collinear counterterms in eqs.\eqref{eq:Sigmaijinitialcollinear} and \eqref{eq:Sigmaijfinalcollinear} need to be amended as 
\begin{eqnarray}
\Sigma_{(p,q),ij}(r; \xi_i, 1)&=&\frac{4}{\left(4\pi\right)^3}\xii\left\{g_s^2 \left[P_{\ident_{j\oplus \bar{i}}\ident_j}^{(\mathrm{QCD}),<}(1-\xii,0)\ampsqnt_{(p-1,q)}(r^{j\oplus \bar{i},\isubrmv})+\Delta^{(\mathrm{QCD})}_{(p-1,q),ij}\right]\right.\nonumber\\
&&\left.+e^2 \left[P_{\ident_{j\oplus \bar{i}}\ident_j}^{(\mathrm{EW}),<}(1-\xii,0)\ampsqnt_{(p,q-1)}(r^{j\oplus \bar{i},\isubrmv})+\Delta^{(\mathrm{EW})}_{(p,q-1),ij}\right]\right\}\nonumber\\
&&\times\frac{\JetsB}{\avg(r)}\, 
\mathrm{d}\phii\phspn(r^{j\oplus \bar{i},\isubrmv}),\quad\quad\quad\quad\quad\mathrm{if}~\nini\leq j\leq 2,\label{eq:Sigmaijinitialcollinear2}\\
\Sigma_{(p,q),ij}(r; \xi_i, 1)&=&\frac{4}{\left(4\pi\right)^3}\frac{1-z}{z}\frac{h(1-z)}{z}\left\{g_s^2\left[P_{\ident_j\ident_{j\oplus i}}^{(\mathrm{QCD}),<}(z,0)\ampsqnt_{(p-1,q)}(r^{j\oplus i,\isubrmv})+\Delta^{(\mathrm{QCD})}_{(p-1,q),ij}\right]\right.\nonumber\\
&&\left.+e^2\left[P_{\ident_j\ident_{j\oplus i}}^{(\mathrm{EW}),<}(z,0)\ampsqnt_{(p,q-1)}(r^{j\oplus i,\isubrmv})+\Delta^{(\mathrm{EW})}_{(p,q-1),ij}\right]\right\}\nonumber\\
&&\times\frac{\JetsB}{\avg(r)}\, 
\mathrm{d}\phii\phspn(r^{j\oplus i,\isubrmv}), \quad\quad\quad\quad\quad\mathrm{if}~3\leq j\leq \nlightR+2,j\neq i\,.\label{eq:Sigmaijfinalcollinear2}
\end{eqnarray}
The soft limit of the real matrix element (analogous to eq.\eqref{eq:softME}) is given by:
\begin{equation}
\lim_{\xii\to 0}{\xii^2\ampsqnpot_{(p,q)}(r)}=\mathop{\sum_{k,l=\nini}}_{k,l\neq i, k\leq l}^{\nlightR+\nheavy+2}{\left(\lim_{\xii\to 0}{\frac{\xii^2 k_k\cdot k_l}{k_k\cdot k_i k_l\cdot k_i}}\right)\left[g_s^2\ampsqnt_{(p-1,q),\mathrm{QCD},kl}(r^{\isubrmv})+e^2\ampsqnt_{(p,q-1),\mathrm{EW},kl}(r^{\isubrmv})\right]},\label{eq:softME2}
\end{equation}

Regarding the integrated counterterms, we can write the new forms explicitly. The soft integrated counterterm (\cf\ eq.\eqref{eq:integratedsoftCT1}) is given by
\begin{equation}
\begin{aligned}
\hat{\sigma}_{(p,q)}^{(S)}(r)&=\int{\phspn(r^{\isubrmv})\frac{\JetsB}{\avg(r^{\isubrmv})}\Bigg[\sum_{k=\nini}^{\nlightB+\nheavy+2}{\sum_{l= k}^{\nlightB+\nheavy+2}{\left(\hat{\mathcal{E}}_{kl}^{(m_k,m_l)}+\mathcal{E}_{kl}^{(m_k,m_l)}\right)}}}\nonumber\\
&\times\left(\frac{\alpha_s}{2\pi}\ampsqnt_{(p-1,q),\mathrm{QCD},kl}(r^{\isubrmv})+\frac{\alpha}{2\pi}\ampsqnt_{(p,q-1),\mathrm{EW},kl}(r^{\isubrmv})\right)\Bigg]\,.\label{eq:integratedsoftCT2}
\end{aligned}
\end{equation}
The PDF counterterm (see  eq.\eqref{eq:PDFxs}) becomes
\begin{eqnarray}
\hat{\sigma}_{(p,q)}^{(\mathrm{PDF})}(r)&=&\sum_{k=\nini}^{2}\int_{1-\ximax}^1 \mathrm{d}z \left(\frac{1}{\bar{\ep}}P^{(\mathrm{QCD})}_{\ident_{k\oplus \bar{i}}\ident_k}(z,0)-K^{(\mathrm{QCD})}_{\ident_{k\oplus \bar{i}}\ident_k}(z)\right)\int{\phspn(r^{k\oplus \bar{i},\isubrmv})\frac{\JetsB}{\avg(r^{\isubrmv})}\frac{\alpha_s}{2\pi}\ampsqnt_{(p-1,q)}(r^{k\oplus \bar{i},\isubrmv})}\nonumber\\
&&+\sum_{k=\nini}^{2}\int_{1-\ximax}^1 \mathrm{d}z \left(\frac{1}{\bar{\ep}}P^{(\mathrm{EW})}_{\ident_{k\oplus \bar{i}}\ident_k}(z,0)-K^{(\mathrm{EW})}_{\ident_{k\oplus \bar{i}}\ident_k}(z)\right)\int{\phspn(r^{k\oplus \bar{i},\isubrmv})\frac{\JetsB}{\avg(r^{\isubrmv})}\frac{\alpha}{2\pi}\ampsqnt_{(p,q-1)}(r^{k\oplus \bar{i},\isubrmv})}.\nonumber\\
&&~\label{eq:PDFxs2}
\end{eqnarray}
The EW regularized splitting kernels are obtained from the unregularized ones via (the counterpart of eq.\eqref{eq:QCDAPregkernal})
\begin{equation}
    P^{(\mathrm{EW})}_{ab}(z,0) = \frac{(1-z)P^{(\mathrm{EW}),<}_{ab}(z,0)}{(1-z)_{+}} + \gamma_{\mathrm{EW}}(a)\delta_{ab}\,\delta(1-z).\label{eq:EWAPregkernal}
\end{equation}
The sum of collinear, soft-collinear, and PDF counterterms is (\cf\ eq.\eqref{eq:sumofcollinearintCT1})
\begin{equation}
\hat{\sigma}_{(p,q)}^{(C)}(r)-\hat{\sigma}_{(p,q)}^{(SC)}(r)+\hat{\sigma}_{(p,q)}^{(\mathrm{PDF})}(r)=\hat{\sigma}^{(C)}_{(p,q),{\rm sing.}}(r)+\hat{\sigma}^{(C,n)}_{(p,q),{\rm FIN}}(r)+\hat{\sigma}^{(C,n+1)}_{(p,q),{\rm FIN}}(r)\,,\label{eq:sumofcollinearintCT2}
\end{equation}
where 
\begin{eqnarray}
\hat{\sigma}^{(C)}_{(p,q),{\rm sing.}}(r)&=&\int{\phspn(r^{\isubrmv})\frac{\JetsB}{\avg(r^{\isubrmv})}\sum_{k=\nini}^{\nlightB+2}\frac{(4\pi)^\ep}{\Gamma(1-\ep)}
\left(\frac{\mu^2}{Q_{\rm ES}^2}\right)^\ep\frac{1}{\ep}}\nonumber\\
&&\times\left\{\frac{\alpha_s}{2\pi}\left[\gamma_{\mathrm{QCD}}(\ident_k)+C_{\mathrm{QCD}}(\ident_k)\log\!\left(\frac{\xicut^2s}{4(k_k^0)^2}\right)\right]\ampsqnt_{(p-1,q)}(r^{\isubrmv})\right.\nonumber\\
&&\left.+\frac{\alpha}{2\pi}\left[\gamma_{\mathrm{EW}}(\ident_k)+C_{\mathrm{EW}}(\ident_k)\log\!\left(\frac{\xicut^2s}{4(k_k^0)^2}\right)\right]\ampsqnt_{(p,q-1)}(r^{\isubrmv})\right\}\,,
\end{eqnarray}
\begin{eqnarray}
\hat{\sigma}^{(C,n)}_{(p,q),{\rm FIN}}(r)&=&\int{\phspn(r^{\isubrmv})\frac{\alpha_s}{2\pi}\frac{\JetsB}{\avg(r^{\isubrmv})}\Bigg\lbrace-\log\!\left(\dfrac{\mu^2}{Q_{\rm ES}^2}\right)\sum_{k=\nini}^{2}\Big(\gamma_{\mathrm{QCD}}(\ident_k)+2C_{\mathrm{QCD}}(\ident_k)\log(\xicut)\Big)}\nonumber\\
&&+\sum_{k= 3}^{\nlightB+2}\left[\gamma_{\mathrm{QCD}}^\prime(\ident_k)-\log\!\left(\dfrac{s\deltaO}{2Q_{\rm ES}^2}\right)\left(\gamma_{\mathrm{QCD}}(\ident_k)-2C_{\mathrm{QCD}}(\ident_k)\log\!\left(\dfrac{2k^0_k}{\xicut\sqrt{s}}\right)\right)\right.\nonumber\\
&&\left.+2C_{\mathrm{QCD}}(\ident_k)\left(\log^2\!\left(\dfrac{2k^0_k}{\sqrt{s}}\right)-\log^2(\xicut)\right)-2\gamma_{\mathrm{QCD}}(\ident_k)\log\!\left(\dfrac{2k^0_k}{\sqrt{s}}\right)\right]\Bigg\rbrace\ampsqnt_{(p-1,q)}(r^{\isubrmv})\nonumber\\
&&+\int{\phspn(r^{\isubrmv})\frac{\alpha}{2\pi}\frac{\JetsB}{\avg(r^{\isubrmv})}\Bigg\lbrace-\log\!\left(\dfrac{\mu^2}{Q_{\rm ES}^2}\right)\sum_{k=\nini}^{2}\Big(\gamma_{\mathrm{EW}}(\ident_k)+2C_{\mathrm{EW}}(\ident_k)\log(\xicut)\Big)}\nonumber\\
&&+\sum_{k= 3}^{\nlightB+2}\left[\gamma_{\mathrm{EW}}^\prime(\ident_k)-\log\!\left(\dfrac{s\deltaO}{2Q_{\rm ES}^2}\right)\left(\gamma_{\mathrm{EW}}(\ident_k)-2C_{\mathrm{EW}}(\ident_k)\log\!\left(\dfrac{2k^0_k}{\xicut\sqrt{s}}\right)\right)\right.\nonumber\\
&&\left.+2C_{\mathrm{EW}}(\ident_k)\left(\log^2\!\left(\dfrac{2k^0_k}{\sqrt{s}}\right)-\log^2(\xicut)\right)-2\gamma_{\mathrm{EW}}(\ident_k)\log\!\left(\dfrac{2k^0_k}{\sqrt{s}}\right)\right]\Bigg\rbrace\ampsqnt_{(p,q-1)}(r^{\isubrmv})\,,\nonumber\\
&&~\label{eq:hatsigmaCnpqFIN}
\end{eqnarray}
and
\begin{eqnarray}
\hat{\sigma}^{(C,n+1)}_{(p,q),{\rm FIN}}(r)&=&\sum_{k=\nini}^{2}\int_0^{\ximax} d\xii\left\lbrace\left[\left(\frac{1}{\xii}\right)_c\log\!\left(\frac{s\deltaI}{2\mu^2}\right)+2\left(\frac{\log(\xii)}{\xii}\right)_c\right]\right.\nonumber\\
&&\left.\times\xii P_{\ident_{k\oplus \bar{i}}\ident_k}^{(\mathrm{QCD}),<}(1-\xii,0)-\left(\frac{1}{\xii}\right)_c\xii P_{\ident_{k\oplus \bar{i}}\ident_k}^{(\mathrm{QCD}),\prime<}(1-\xii,0)-K^{(\mathrm{QCD})}_{\ident_{k\oplus \bar{i}}\ident_k}(1-\xii)\right\rbrace\nonumber\\
&&\times\int{\phspn(r^{k\oplus \bar{i},\isubrmv})\frac{\JetsB}{\avg(r^{\isubrmv})}\frac{\alpha_s}{2\pi}\ampsqnt_{(p-1,q)}(r^{k\oplus \bar{i},\isubrmv})}\nonumber\\
&&+\sum_{k=\nini}^{2}\int_0^{\ximax} d\xii\left\lbrace\left[\left(\frac{1}{\xii}\right)_c\log\!\left(\frac{s\deltaI}{2\mu^2}\right)+2\left(\frac{\log(\xii)}{\xii}\right)_c\right]\right.\nonumber\\
&&\left.\times\xii P_{\ident_{k\oplus \bar{i}}\ident_k}^{(\mathrm{EW}),<}(1-\xii,0)-\left(\frac{1}{\xii}\right)_c\xii P_{\ident_{k\oplus \bar{i}}\ident_k}^{(\mathrm{EW}),\prime<}(1-\xii,0)-K^{(\mathrm{EW})}_{\ident_{k\oplus \bar{i}}\ident_k}(1-\xii)\right\rbrace\nonumber\\
&&\times\int{\phspn(r^{k\oplus \bar{i},\isubrmv})\frac{\JetsB}{\avg(r^{\isubrmv})}\frac{\alpha}{2\pi}\ampsqnt_{(p,q-1)}(r^{k\oplus \bar{i},\isubrmv})}\,.
\end{eqnarray}
The EW Casimir factors and anomalous dimensions are
\begin{eqnarray}
C_{\mathrm{EW}}(\ident)&=&\left(e(\ident)\right)^2,\\
\gamma_{\mathrm{EW}}(\ident)&=&\left\{\begin{array}{ll}
\frac{3}{2}\left(e(\ident)\right)^2\,,\hfill&~~{\rm if}~~\ident\neq \gamma\\
-\frac{2}{3}\sum_{\ident^\prime=\ell^-,q}{N_c(\ident^\prime)\left(e(\ident^\prime)\right)^2}\,,&~~{\rm if}~~\ident=\gamma\\
\end{array}\right.,\\
\gamma_{\mathrm{EW}}^\prime(\ident)&=&\left\{\begin{array}{ll}
\left(\frac{13}{2}-\frac{2\pi^2}{3}\right)\left(e(\ident)\right)^2\,,\hfill&~~{\rm if}~~\ident\neq \gamma\\
-\frac{23}{9}\sum_{\ident^\prime=\ell^-,q}{N_c(\ident^\prime)\left(e(\ident^\prime)\right)^2}\,,&~~{\rm if}~~\ident=\gamma\\
\end{array}\right..
\end{eqnarray}
In the collinear anomalous dimensions, the sum over $\ident^\prime$ accounts for all massless charged leptons and massless quarks, which can indeed be generalized to include all charged massless fermions.~\footnote{Note that the introduction of new massless bosonic degrees of freedom, such as a massless scalar, in the theory may exhibit different IR structures (see,\eg, ref.~\cite{Campiglia:2017dpg}). Hence, we only consider gluons and photons are massless bosons in this context.}
Therefore, the sum of the integrated FKS and PDF counterterms is given by
\begin{eqnarray}
\hat{\sigma}^{(I)}_{(p,q),\mathrm{FKS}}(r)&=&\hat{\sigma}_{(p,q)}^{(S)}(r)+\hat{\sigma}_{(p,q)}^{(C)}(r)-\hat{\sigma}_{(p,q)}^{(SC)}(r)+\hat{\sigma}_{(p,q)}^{(\mathrm{PDF})}(r).
\end{eqnarray}

The NLO correction term at $\mathcal{O}(\alpha_s^p\alpha^q)$ is given by
\begin{eqnarray}
\sigma^{(\mathrm{NLO})}_{(p,q)}&=&\sum_{r\in\mathcal{R}_n}{\int{\mathrm{d}x_1\mathrm{d}x_2 f_{\Ione}^{(P_1)}(x_1,\mu_F^2) f_{\Itwo}^{(P_2)}(x_2,\mu_F^2)\hat{\sigma}^{(V)}_{(p,q)}(r)}}\nonumber\\
&&+\sum_{r\in\mathcal{R}_{n+1}}{\int{\mathrm{d}x_1\mathrm{d}x_2 f_{\Ione}^{(P_1)}(x_1,\mu_F^2) f_{\Itwo}^{(P_2)}(x_2,\mu_F^2)\left[\hat{\sigma}^{(R)}_{(p,q),\mathrm{FKS}}(r)+\hat{\sigma}^{(I)}_{(p,q),\mathrm{FKS}}(r)\right]}},\label{eq:NLOpqxsFKS}
\end{eqnarray}
where $\hat{\sigma}^{(V)}_{(p,q)}(r)$ is defined in eq.\eqref{eq:virtualxs}, by substituting $\ampsqnl(r)$ with $\ampsqnl_{(p,q)}(r)$ on the r.h.s.

\section{Renormalization and input-parameter schemes\label{sec:renormalization}}

The UV renormalization is essential for canceling UV singularities in higher-order calculations. The renormalizability of non-Abelian gauge theories with spontaneous symmetry breaking--and thus of the SM--was first proven by 't Hooft and Veltman~\cite{tHooft:1971akt,tHooft:1971qjg,tHooft:1972tcz,tHooft:1972qbu}. To define a renormalization scheme, it is necessary to select a set of independent input parameters. While there is significant freedom in choosing these parameters, it is convenient to fix the renormalization constants by imposing renormalization conditions directly on the parameters in the physical basis. This typically involves fixing the renormalized masses of the gauge bosons $W^\pm, Z$, the Higgs boson $h$, and the fermions $f$ to the locations of the poles of their propagators. The UV renormalization counterterms are chosen such that the finite renormalized parameters equal the physical parameters in all orders of perturbation theory. This approach is known as the on-shell (OS) renormalization scheme in the literature~\cite{Ross:1973fp}. 

In order to illustrate this further, let us consider the propagator, or two-point Green function, of a particle with mass $M$ and virtuality $p^2$, which diverges when $p^2\to M^2$:
\begin{equation}
G_D(p^2)=-i\left[p^2-M_0^2+\Sigma(p^2)\right]^{-1},\label{eq:GD1}
\end{equation}
where $M_0$ is the bare mass of the particle, and $\Sigma(p^2)$ represents one or more loops of the two-point one-particle-irreducible (1PI) contribution. The subscript ``$D$" indicates that the propagator results from the Dyson summation of the geometric series of multiple insertions of $\Sigma(p^2)$ in the two-point Green function. In general, there is an imaginary part in the denominator of eq.\eqref{eq:GD1} stemming from $\Sigma(p^2)$, which prevents the propagator from diverging in the limit $p^2\to M^2$, as the virtuality must remain real. However, by effectively setting the pole mass of the propagator to be different from the corresponding Lagrangian mass parameter, one might violate gauge invariance~\cite{Stuart:1991xk,Stuart:1991cc,Sirlin:1991rt,Sirlin:1991fd}. Equation \eqref{eq:GD1} suggests an alternative (potentially gauge-violating) way for the propagator:
\begin{equation}
G_R(p^2)=-i\left[p^2-M^2+i\Gamma M\right]^{-1},\label{eq:GD2}
\end{equation}
where $M$ is the OS mass, and $\Gamma$ is the total decay width of the particle. Equations \eqref{eq:GD1} and \eqref{eq:GD2} can be related via the OS mass renormalization condition:
\begin{equation}
M_0^2-\underbrace{\Re{\Sigma_{\mathrm{U}}(M^2)}}_{\equiv\delta M^2}=M^2,\label{eq:OSmassrenorm}
\end{equation}
and the optical theorem:
\begin{equation}
\Im{\Sigma_{\mathrm{U}}(M^2)}=\Gamma M.\label{eq:WidthFromOT}
\end{equation}
Here, we identify the one-loop 1PI $\Sigma(p^2)$ in eq.\eqref{eq:GD1} as the unrenormalized self-energy function (hence the subscript ``$\mathrm{U}$"). For the OS wavefunction renormalization, we have:
\begin{equation}
Z=1+\underbrace{\left.\Re{\left(\frac{\partial}{\partial p^2}\Sigma_{\mathrm{U}}(p^2)\right)}\right|_{p^2=M^2}}_{\equiv-\delta Z}.\label{eq:OSwaverenorm}
\end{equation}
This admits the following Taylor expansion around $p^2=M^2$:
\begin{equation}
\Re{\Sigma_{\mathrm{U}}(p^2)}=\delta M^2-(p^2-M^2)\delta Z+\mathcal{O}\left(\left(p^2-M^2\right)^2\right).
\end{equation}
It yields the renormalized two-point self energy function:
\begin{equation}
\Sigma_{\mathrm{R}}^{(\mathrm{OS})}(p^2)=\Sigma_{\mathrm{U}}(p^2)-\delta M^2+(p^2-M^2)\delta Z,\label{eq:renormalizedSigma}
\end{equation}
which satisfies the following renormalization conditions:
\begin{eqnarray}
\Re{\Sigma_{\mathrm{R}}^{(\mathrm{OS})}(M^2)}&=&0,\label{eq:sigmaROSrenorm1}\\
\lim_{p^2\to M^2}{\left[\frac{1}{p^2-M^2}\Re{\Sigma_{\mathrm{R}}^{(\mathrm{OS})}(p^2)}\right]}&=&1.\label{eq:sigmaROSrenorm2}
\end{eqnarray}
Thus, we can rewrite eq.\eqref{eq:GD1} as:
\begin{equation}
G_D(p^2)=-i\left[Z(p^2-M^2)+\Sigma_{\mathrm{R}}^{(\mathrm{OS})}(p^2)\right]^{-1}.\label{eq:GD3}
\end{equation}
In the SM, their generalization to vectors and fermions can be found in eq.(3.7) of ref.~\cite{Denner:1991kt}, for instance. The OS condition simplifies practical calculations since we do not need to consider one-loop diagrams with loops attached to an external leg only. We will revisit these equations when discussing complex renormalization in the complex-mass scheme in section \ref{sec:CMS}. An important point to stress is that in the OS scheme, all mass and wavefunction renormalization constants are real.~\footnote{To be more precise, due to the possible presence of an imaginary part from interactions, such as the complex phase in the CKM matrix, what we really need to apply to the two-point self-energy functions in the OS scheme is the operator $\tilde{\Re}$, which only removes the imaginary absorptive part of the loop function~\cite{Denner:1991kt}. An even more precise definition of $\tilde{\Re}$ can be found in footnote 29 of ref.~\cite{Frederix:2018nkq}.}

Regarding the electroweak coupling $\alpha=e^2/(4\pi)$, it is mainly renormalized in three different ways in the literature:
\begin{itemize}
\item {\it $\alpha(0)$ scheme}: In this scheme, the renormalized $\alpha$ is fixed in such a way that it coincides with the value measured in the Thomson (\ie, zero photon momentum) limit.  In this limit, all higher-order corrections to Compton scattering vanish, making the lowest-order QED cross section, known as the Thomson cross section, exact in all orders of perturbation theory. Here, the electric charge $e$ renormalization constant (recalling $e_0=Z_ee=(1+\delta Z_e)e$, where $e_0$ is the bare electric charge) is given by
\begin{equation}
\delta Z_e^{(\alpha(0))}=\frac{1}{2}\left.\frac{\partial \Sigma_{\mathrm{U},T}^{(\gamma\gamma)}(p^2)}{\partial p^2}\right|_{p^2=0}-\frac{s_W}{c_W}\frac{\Sigma^{(\gamma Z)}_{\mathrm{U},T}(0)}{M_Z^2},\label{eq:Zeinalpha0}
\end{equation}
where $M_Z$ ($M_W$) is the OS mass of the $Z$ ($W^\pm$) boson, $s_W$ and $c_W$ are sine and cosine of the Weinberg (or weak mixing) angle, related by $c_W^2=M_W^2/M_Z^2$, and $\Sigma_{\mathrm{U},T}^{(V_1V_2)}(p^2)$ is the unrenormalized transverse part of the two-point $V_1V_2$ 1PI function. $\delta Z_e^{(\alpha(0))}$ depends on the masses of all charged fermions, particularly the perturbatively problematic light quark masses. In the Thomson limit, the masses of all fermion cannot be neglected. The input value of $\alpha$ in the scheme is approximately $\alpha(0)\approx 1/137.035 999 084(21)$~\cite{ParticleDataGroup:2024cfk}.
\item {\it $\alpha(M_Z^2)$ scheme}: This alternative scheme was firstly introduced in refs.~\cite{Burkhardt:2001xp,Eidelman:1995ny}, where the value of $\alpha$ is chosen as $\alpha(M_Z^2)$. The effective value of  $\alpha(M_Z^2)$ is obtained from the renormalization-group running from the Thomson limit to the $Z$ pole:
\begin{equation}
\alpha(M_Z^2)=\frac{\alpha(0)}{1-\Delta \alpha(M_Z^2)},
\end{equation}
where~\cite{Dittmaier:2001ay}
\begin{equation}
\Delta \alpha(M_Z^2)=\mathop{\sum_{\ident=\ell^-,q}}_{\ident\neq t}{\left[\left.\frac{\partial \Sigma_{\mathrm{U},T}^{(\gamma\gamma),\ident}(p^2)}{\partial p^2}\right|_{p^2=0}-\frac{\Re{\Sigma_{\mathrm{U},T}^{(\gamma\gamma),\ident}(M_Z^2)}}{M_Z^2}\right]}.\label{eq:DeltaalphaMZ2}
\end{equation}
Here, $\Sigma_{\mathrm{U},T}^{(\gamma\gamma),\ident}(p^2)$ represents the unrenormalized one-loop transverse photon self-energy function with $\ident$ as the loop particle. The sum in eq.\eqref{eq:DeltaalphaMZ2} runs over all charged leptons and quarks except the top quark. The electric charge renormalization constant becomes
\begin{equation}
\delta Z_e^{(\alpha(M_Z^2))}=\delta Z_e^{(\alpha(0))}-\frac{1}{2}\Delta \alpha(M_Z^2).
\end{equation}
In contrast to the $\alpha(0)$ scheme, the counterterm $\delta Z_e^{(\alpha(M_Z^2))}$ does not involve light quark and lepton mass, allowing all quark and lepton masses except the top quark to be safely set to zero. The quark or hadron contribution in $\Delta \alpha(M_Z^2)$ is essentially non-perturbative and should be determined either from experimental data using dispersion relations~\cite{Eidelman:1995ny} or from lattice calculations. The numerical value of $\alpha(M_Z^2)$ is around $1/128.94$. Using a similar equation as eq.\eqref{eq:DeltaalphaMZ2}, one can define $\alpha(\mu_R^2)$ at any scale $\mu_R$ other than $\mu_R=0$ and $\mu_R=M_Z$. It is important not to confuse $\alpha(\mu_R^2)$ with $\alpha$ in the $\msbar$ scheme.
\item {\it $G_\mu$ scheme}: This scheme offers the possibility to absorb the universal top-mass enhanced correction to the EW $\rho$ parameter~\cite{Ross:1975fq}. The transition from $\alpha(0)$ to $G_\mu$ (the Fermi constant deduced from muon decay) is governed by the quantity $\Delta r$~\cite{Sirlin:1980nh,Marciano:1980pb,Sirlin:1981yz,Denner:1991kt} via
\begin{equation}
\alpha_{G_\mu}=\frac{\sqrt{2}G_\mu M_W^2(M_Z^2-M_W^2)}{\pi M_Z^2}=\alpha(0)\left(1+\Delta r\right)+\mathcal{O}(\alpha^3),
\end{equation}
where 
\begin{eqnarray}
\Delta r&=&\left.\frac{\partial \Sigma_{\mathrm{U},T}^{(\gamma\gamma)}(p^2)}{\partial p^2}\right|_{p^2=0}-\frac{c_W^2}{s_W^2}\left(\frac{\Sigma_{\mathrm{U},T}^{(ZZ)}(M_Z^2)}{M_Z^2}-\frac{\Sigma_{\mathrm{U},T}^{(WW)}(M_W^2)}{M_W^2}\right)+\frac{\Sigma_{\mathrm{U},T}^{(WW)}(0)-\Sigma_{\mathrm{U},T}^{(WW)}(M_W^2)}{M_W^2}\nonumber\\
&&+2\frac{c_W}{s_W}\frac{\Sigma_{\mathrm{U},T}^{(\gamma Z)}(0)}{M_Z^2}+\frac{\alpha(0)}{4\pi s_W^2}\left(6+\frac{7-4s_W^2}{2s_W^2}\log{c_W^2}\right).
\end{eqnarray}
The charge renormalization constant in this scheme is
\begin{equation}
\delta Z_e^{(G_\mu)}=\delta Z_e^{(\alpha(0))}-\frac{1}{2}\Delta r.
\end{equation}
The $G_\mu$ scheme corresponds to an $\alpha$ value at the EW scale, similar to the $\alpha(M_Z^2)$ scheme, where $\alpha_{G_\mu}\approx 1/132.183$. This scheme is also insensitive to the light quark and lepton masses.
\end{itemize}
Among these three schemes, it is advisable to use the $\alpha(0)$ scheme for on-shell external photons, the $\alpha(M_Z^2)$ scheme for off-shell internal photons, and the $G_\mu$ scheme for couplings involving $W^\pm$ and $Z$ bosons. Both $\alpha(M_Z^2)$ and $G_\mu$ schemes are appropriate for other cases. For incoherent initial photons in hadronic collisions or final-state photons as parts of electromagnetic showers,  it has been suggested to use either the $\alpha(M_Z^2)$ or $G_\mu$ scheme (see, \eg, section 4.3.3 in ref.~\cite{Denner:2019vbn} or ref.~\cite{Harland-Lang:2016lhw}). While one scheme may be better than another, they formally yield the same NLO expression; their only difference lies in higher-order corrections. Although still under debate, the difference between schemes can be viewed as a means to estimate missing higher order in $\alpha$ when there is no clear preference among $\alpha$ schemes. Since we typically do not perform scale evolution for $\alpha$ as we do for $\alpha_s$, we cannot use the traditional approach of varying the renormalization scale in $\alpha$ to estimate the missing higher order. Finally, a hybrid scheme that combines two or more $\alpha$ renormalization schemes is certainly possible, as demonstrated in ref.~\cite{Pagani:2021iwa} for processes involving isolated final photon(s). 

Analogous to the strong coupling $\alpha_s$, one can also define $\alpha$ in the $\msbar$ scheme (see, \eg\, section 5.1.2 in ref.~\cite{Denner:2019vbn} or section 5 in ref.~\cite{Degrassi:2003rw}). The charge renormalization constant in the $\msbar$ scheme is
\begin{equation}
\begin{aligned}
\delta Z_e^{(\msbar)}=&-\frac{\alpha^{(\msbar)}(\mu_R^2)}{4\pi}\Bigg[\left(\frac{7}{2}-\frac{2}{3}\sum_{\ident=\ell^-,q}{N_c(\ident)\left(e(\ident)\right)^2}\right)\frac{1}{\bar{\epsilon}}\\
&+\Theta(M_W^2-\mu_R^2)\frac{7}{2}\log{\frac{\mu_R^2}{M_W^2}}+\sum_{\ident=\ell^-,q}{\Theta(M_{\ident}^2-\mu_R^2)\left(-\frac{2}{3}\right)N_c(\ident)\left(e(\ident)\right)^2\log{\frac{\mu_R^2}{M_\ident^2}}}\Bigg],
\end{aligned}
\end{equation}
where the sum runs over all charged leptons and quarks, including the top quark. Similar to $\alpha_s$, the light leptons and quarks are renormalized in the $\msbar$ scheme, while the heavy degrees of freedom are subtracted at the zero momentum transfer to ensure decoupling. We can relate the $\msbar$ coupling to $\alpha(0)$ via
\begin{equation}
\alpha^{(\msbar)}(\mu_R^2)=\frac{\alpha(0)}{1-\Delta \alpha^{(\msbar)}(\mu_R^2)}
\end{equation}
with
\begin{equation}
\Delta \alpha^{(\msbar)}(\mu_R^2)=2\left(\delta Z_e^{(\alpha(0))}-\delta Z_e^{(\msbar)}\right).
\end{equation}
At one-loop level, the difference for $\mu_R=M_Z$ is
\begin{equation}
\Delta \alpha^{(\msbar)}(M_Z^2)-\Delta \alpha(M_Z^2)=\frac{\alpha(0)}{\pi}\left(\frac{5}{9}\mathop{\sum_{\ident=\ell^-,q}}_{\ident\neq t}{N_c(\ident)\left(e(\ident)\right)^2}-\frac{1}{6}+\frac{7}{4}\log{c_W^2}\right)=\frac{\alpha(0)}{\pi}\left(\frac{191}{54}+\frac{7}{4}\log{c_W^2}\right).
\end{equation}
This gives us $\alpha^{(\msbar)}(M_Z^2)\approx 1/127.955$. By construction, the $\msbar$ scheme is purely a short-distance scheme, making it insensitive to the masses of light fermions. However, this scheme is rarely used in practical NLO EW computations.

In the SM, EW radiative corrections affect the Higgs potential in such a way that its minimum is shifted. To correct for this shift, one introduces a counterterm to the vacuum
expectation value (vev) of the Higgs field, determined in such a way that the renormalized vev $v$ corresponds to the actual minimum of the effective Higgs potential. The renormalization condition for the renormalized one-point function of the Higgs boson $h$ is
\begin{equation}
\Gamma_{\mathrm{R}}^{(h)}=\Gamma_{\mathrm{U}}^{(h)}+\delta t=0,
\end{equation}
where $\delta t$ is the renormalization constant for the tadpole, and $\Gamma_{\mathrm{U}}^{(h)}$ is the unrenormalized 1PI one-point function. This condition ensures that we do not need to calculate any one-particle reducible tadpole diagrams. Depending on how $\delta t$ enters the counterterm structures of the Lagrangian, several different tadpole schemes exist. Since the final physical predictions do not depend on the tadpole scheme, in \mgamcshort~\cite{Frederix:2018nkq},  we simply use the tadpole scheme initially proposed in ref.~\cite{Denner:1991kt}, referred to as the parameter-renormalized tadpole scheme in ref.~\cite{Denner:2019vbn}. In this scheme, all bare masses are gauge-dependent, as they are connected to the gauge-independent bare parameters of the Lagrangian through gauge-dependent tadpole terms. This results in a gauge-dependent representation of the $S$-matrix in terms of bare parameters. However, this gauge dependence cancels in physical quantities when all parameters are defined using OS renormalization conditions. The gauge dependence of the counterterms results arises from momentum-independent tadpole contributions, which cancel in the OS scheme. Conversely, when some parameters are renormalized using the $\msbar$ scheme, the extra tadpole contributions do not cancel, leading to potential gauge dependence in the $S$-matrix. This issue becomes relevant in extensions of the SM, such as the Two-Higgs-Doublet Model, where some parameters are typically renormalized within the $\msbar$ scheme (see, \eg\, ref.~\cite{Denner:2018opp}). An alternative tadpole scheme to address this issue has recently been proposed in refs.~\cite{Dittmaier:2022maf,Dittmaier:2022ivi}.

For scattering processes at high-energy colliders, the renormalization of the quark-mixing Cabibbo-Kobayashi-Maskawa (CKM) matrix~\cite{Cabibbo:1963yz,Kobayashi:1973fv} in the SM is practically irrelevant. In higher-order corrections, neglecting the quark masses other than the top quark and possibly the bottom quark is a good approximation. Therefore, I will refrain from discussing it here, and instead refer interested readers to refs.~\cite{Denner:1991kt,Denner:2019vbn} for details on the CKM matrix renormalization and related issues.

The renormalization constants of derived quantities, such as the Weinberg angle or Yukawa couplings, should be expressed in terms of the input parameters (like $M_W$ and $M_Z$ for the Weinberg angle and the fermion masses for the Yukawa couplings). As a result, they are no longer free parameters. For instance, the $\msbar$ renormalized Yukawa coupling would not be consistent with the OS renormalization of the corresponding fermion mass in a NLO EW calculation. Similarly, introducing $s_W$ as an independent parameter alongside $M_W$ and $M_Z$, either by assigning it an arbitrary value or by using the sine of the effective weak mixing angle measured at the $Z$ pole from LEP, generally breaks gauge invariance, disrupts gauge anomaly cancellations, and would yield completely incorrect results, even at LO.

Finally, as already pointed out in section \ref{sec:mgamc}, the UV renormalization counterterms should be contained in the UFO model in order to enable NLO computations in \mgamcshort.

\section{Complex-mass scheme\label{sec:CMS}}

The presence of unstable short-lived particles complicates perturbative computations in quantum field theory. This issue becomes particularly pronounced for NLO EW calculations in the SM,  as all short-lived particles--such as $W^\pm$ and $Z$ bosons, and the top quark--have their width-to-mass ratio $\Gamma/M\sim \mathcal{O}(\alpha)$~\footnote{The Higgs boson presents a more complex scenario; its dominant decay channel $h\to b\bar{b}$ also features $\mathcal{O}(\alpha)$ at the lowest order, but is further suppressed by $M_b^2/M_h^2$. This makes other decay channels, like $h\to g g$, also significant.}. This effectively leads to $\mathcal{O}(\alpha)$ corrections in scattering amplitudes. The situation is further complicated by the fact that contributions from such particles can potentially spoil gauge invariance and unitarity; thus, their proper treatment may necessitate relaxing strict fixed-order accuracy. Among various solutions to this problem, the so-called complex-mass (CM) scheme~\cite{Denner:1999gp,Denner:2005fg} stands out. In the CM scheme, NLO calculations deliver accurate results across the entire phase space, \ie, in both on-shell and off-shell regions, provided the decay widths are at least NLO-accurate. Moreover, the CM scheme preserves essential properties of the $S$-matrix, such as gauge invariance and perturbative unitarity~\cite{Bauer:2012gn,Denner:2014zga}. Therefore, the CM scheme is the strategy of choice in \mgamcshort\ for dealing with unstable particles, and I will limit the discussion to it in this dissertation.

As explained in section \ref{sec:renormalization}, the Dyson-summation in the propagator provides a regularization to allow us to study the kinematically dominant resonant or on-shell region, since otherwise scattering amplitudes diverge when $p^2\to M^2$. The core idea of the CM scheme stems from the observation that the form eq.\eqref{eq:GD1} implies a complex pole $\bar{p}^2$:
\begin{equation}
\bar{p}^2-M_0^2+\Sigma(\bar{p}^2)=0 \quad \Longrightarrow \quad \bar{p}^2=\bar{M}^2-i\bar{\Gamma}\bar{M},\label{eq:polemass1}
\end{equation}
where $\bar{M}$ and $\bar{\Gamma}$ are usually called the pole mass and width respectively, to differentiate them from the OS mass $M$ and width $\Gamma$ defined in section \ref{sec:renormalization}. The pole mass can be related to the OS mass~\cite{Stuart:1991xk,Sirlin:1991fd} as follows:
\begin{equation}
M^2=\bar{M}^2+\bar{\Gamma}^2+\mathcal{O}(\alpha^3)=\bar{M}^2+\mathcal{O}(\alpha^2)
\end{equation}
with $\bar{\Gamma}/\bar{M}\sim\mathcal{O}(\alpha)$. In particular, for $W^\pm$ and $Z$ bosons in the SM, the conversion typically takes the following form~\cite{Stuart:1991xk,Sirlin:1991fd,Kniehl:1998fn,Grassi:2001bz,Kniehl:2002wn,Passarino:2010qk}:
\begin{eqnarray}
\bar{M}=\frac{M}{\sqrt{1+\frac{\Gamma^2}{M^2}}},\quad \bar{\Gamma}=\frac{\Gamma}{\sqrt{1+\frac{\Gamma^2}{M^2}}}.\label{eq:polemasswidthfromOS}
\end{eqnarray}
An appealing feature of the pole position $\bar{p}^2$ is that it remains unchanged under renormalization. In contrast, the OS mass becomes gauge dependent at the two-loop order~\cite{Sirlin:1991rt,Stuart:1991xk,Gambino:1999ai,Grassi:2001bz}. The CM scheme provides a framework that does not alter the theory but rearranges the perturbative expansion. This fact has been throughly utilized in the implementation and validation of the CM scheme in \mgamcshort, which provides a systematic test suite for checking its implementation (see appendix E.1 in ref.~\cite{Frederix:2018nkq}). In the CM scheme, the renormalization conditions of the theory are modified to yield complex-valued renormalized parameters, including the masses of unstable particles and a subset of coupling constants. At the NLO, however, extending the OS renormalization condition (\cf\ section  \ref{sec:renormalization}) to complex renormalization in the CM scheme presents many subtleties (see,\eg, section 5 in ref.~\cite{Frederix:2018nkq}). In the following, I will limit myself to discussing the pragmatic concerns of the implementation in \mgamcshort, closely following section 5 of ref.~\cite{Frederix:2018nkq}.

The complex pole of the propagator in eq.\eqref{eq:polemass1} suggests that for any unstable particle field, we can define a complex mass $m$, whose square is simply the complex pole
\begin{equation}
m^2=\bar{M}^2-i\bar{\Gamma}\bar{M}.\label{eq:polemass2}
\end{equation}
All derived parameters must be expressed in terms of the complex $m$ and other independent input parameters, thereby potentially acquiring an imaginary part. In the SM, these parameters can include $s_W,c_W$, Yukawa couplings, and the Fermi constant $G_\mu$ in the $\alpha(M_Z^2)$ scheme or $\alpha_{G_\mu}$ in the $G_\mu$ scheme. At NLO, the OS renormalization condition eqs.\eqref{eq:OSmassrenorm}, \eqref{eq:OSwaverenorm}, and \eqref{eq:renormalizedSigma} should be replaced with the complex renormalization condition in the CM scheme:
\begin{eqnarray}
M_0^2-\underbrace{\Sigma_{\mathrm{U}}(m^2)}_{\equiv \delta m^2}&=&m^2=\bar{M}^2-i\bar{\Gamma}\bar{M},\\
z&=&1+\underbrace{\left.\frac{\partial\Sigma_{\mathrm{U}}(p^2)}{\partial p^2}\right|_{p^2=m^2}}_{\equiv-\delta z},\\
\Sigma_{\mathrm{R}}^{(\mathrm{CM})}(p^2)&=&\Sigma_{\mathrm{U}}(p^2)-\delta m^2+(p^2-m^2)\delta z.
\end{eqnarray}
Equations \eqref{eq:sigmaROSrenorm1} and \eqref{eq:sigmaROSrenorm2} generalize to:
\begin{eqnarray}
\Re{\Sigma_{\mathrm{R}}^{(\mathrm{CM})}(m^2)}&=&0,\label{eq:sigmaRCMrenorm1}\\
\lim_{p^2\to m^2}{\left[\frac{1}{p^2-m^2}\Re{\Sigma_{\mathrm{R}}^{(\mathrm{CM})}(p^2)}\right]}&=&1.\label{eq:sigmaRCMrenorm2}
\end{eqnarray}
We adhere to the convention that lowercase and uppercase symbols ($m$ and $z$ versus $M$ and $Z$) represent the (complex) mass and wavefunction renormalization in the CM and OS schemes, respectively. These definitions, along with those relevant to coupling renormalization (which mirror the OS renormalization but replace OS masses with complex masses and omit the $\tilde{\Re}$ operator), ensure that by working in the CM scheme, one can proceed analogously to other renormalization schemes. 

The complex renormalization condition implies, in particular, the necessity of evaluating massive-particle self-energies at $p^2=m^2=\bar{M}^2-i\bar{\Gamma}\bar{M}$. Due to the presence of the additional negative imaginary part, analytical continuation might lead one to compute logarithms or any other multi-valued function in Riemann sheets different from the first. It is important to note that the first Riemann sheet requires the continuation of $\Im{p^2}>0$. Such continuation, however, is rather non-trivial even for the one-loop two-point function at NLO. To circumvent this issue, what is customary done in the literature (see,\eg, the discussion in section 6.6.3 of ref.~\cite{Denner:2019vbn}) is to Taylor expand $\Sigma_{\mathrm{U}}(p^2=m^2)$ around $p^2=\bar{M}^2+i0^+$, allowing the new one-loop self-energy function and its derivative to be evaluated in the first Riemann sheet as usually done. However, this procedure poses three possible issues. First, $\Sigma_{\mathrm{U}}(p^2=m^2)$ is approximated using its Taylor expansion up to NLO, meaning that higher-order expansion terms formally contributing to NNLO and beyond are ignored. Second, the pure Taylor expansion is insufficient for contributions with a branch point (such as logarithms) at $p^2=\bar{M}^2+i0^+$; the missing term due to the branch point must be added back manually; otherwise, it yields incorrect result at NLO accuracy. Finally, in a general theory with large-width particles (which is not the case in the SM), the Taylor expansion approximation fails in general, as elucidated in section 5.3 of ref.~\cite{Frederix:2018nkq}. To tackle the problem, a general method for the analytic continuation of two-point functions for arbitrary complex momentum variables and masses, based on trajectories in the complex plane, called the {\it trajectory method},  was proposed in ref.~\cite{Frederix:2018nkq}. This is the strategy adopted in the \mgamcshort\ framework and now also constitutes an extension of the UFO format~\cite{Darme:2023jdn}. We refer interested readers to section 5.3 of ref.~\cite{Frederix:2018nkq} for the method and appendix E.2 of that paper for algorithms to implement the method in code.

As mentioned earlier, derived couplings can potentially acquire an imaginary part in the CM scheme due to the complex renormalization. This is particularly true for $\alpha$ or $e$. The imaginary part of $e$ is not a free input parameter but is determined by the charge renormalization constant in an iterative manner. However, the presence of complex phase in $\alpha$ spoils the IR cancellation in the context of NLO computations, as clearly illustrated in section 5.4 of ref.~\cite{Frederix:2018nkq}. Thus, we need to find a way to eliminate the complex phase of $\alpha$. It is important to point out that the imaginary part of $\alpha$ in the $\alpha(0)$ scheme is entirely due to the spurious terms, as the charge renormalization constant only involves self-energy functions at zero momentum transfer (\cf\ eq.\eqref{eq:Zeinalpha0}), which do not develop imaginary parts for real internal masses. Therefore, the imaginary parts in $\alpha$ are of formal two-loop order. Following this reasoning, we conclude that it is consistent within NLO accuracy to set the imaginary part in $\alpha$ to zero. The situation does not change in other $\alpha$ schemes, as described in section \ref{sec:renormalization}. In the case of the $G_\mu$ scheme, $\alpha$ should be calculated from the real values of $G_\mu$ and the masses of $W^\pm$ and $Z$ bosons. To avoid the spurious terms of $\mathcal{O}(\alpha)$ in $\alpha$, there are several solutions, such as taking the absolute value or the real part of $\alpha$ in terms of complex masses or using real masses directly. However, any such solution is not particularly appealing, as it might spoil a gauge relation through higher-order terms. Alternatively, one can also turn $G_\mu$ into a complex value in such a way that $\alpha_{G_\mu}$ is real by using complex masses directly. Nevertheless, all these variants differ only at the order beyond NLO. Therefore, the imaginary part of the renormalized charge is generally considered irrelevant in a one-loop calculation, but it must be taken into account at the two-loop level. Thus, a full-fledged CM scheme at NNLO (and beyond) is still absent.

Finally, before closing this section, I wish to emphasize that in the context of a CM scheme based calculation, it is always possible to impose OS renormalization conditions on potentially unstable
particles, provided that these particles only appear in the final state and not as intermediate resonances that may go on-shell. This seemingly trivial statement crucially depends on a correct interpretation of the operator $\tilde{\Re}$ and the complex conjugate. I will not repeat the arguments here but refer readers to section 5.5 in ref.~\cite{Frederix:2018nkq}. 

%\section{Issues}

\section{Photon and lepton parton distribution functions\label{sec:photonleptonPDF}}

The photon PDF of the proton is instrumental in making precise predictions of photon-initiated processes at the LHC. In NLO EW calculations, it becomes an indispensable ingredient, as the photon density in the proton is roughly at $\mathcal{O}(\alpha)$ of the quark densities. Although $\alpha\ll 1$, the photon cannot be considered as a pure perturbative object since it generally mixes with hadrons, such as vector mesons, a known phenomenon referred to as the resolved photon in $ep$ physics. Before 2016, the available photon PDF sets in global PDF fits were either plagued by large uncertainties or relied on phenomenological models. The situation regarding photon PDF determination was drastically improved with the advent of the photon PDF determination based on a novel approach dubbed \luxqed\ in 2016~\cite{Manohar:2016nzj}. This approach is based on the observation that the virtual photon exchanged between the electron and proton in $ep$ collisions can be viewed as being emitted by either the electron in the proton target frame or by the proton in the rest frame of the electron. This allows us to relate the photon PDF to the proton structure functions, which have been precisely measured in various nuclear physics experiments. In the usual $\msbar$ scheme, the photon PDF can be written as~\cite{Manohar:2016nzj}
\begin{equation}
\begin{aligned}
xf_{\gamma}^{(p)}(x,\mu_F^2)=&\frac{1}{2\pi\alpha(\mu_F^2)}\int_x^1{\frac{\mathrm{d}z}{z}\Bigg\{\int_{\frac{x^2m_p^2}{1-z}}^{\frac{\mu_F^2}{1-z}}{\frac{\mathrm{d}Q^2}{Q^2}\left(\alpha(Q^2)\right)^2\Bigg[\left(z\frac{P^{(\mathrm{EW})}_{\gamma q}(z,0)}{\left(e(q)\right)^2}+\frac{2x^2m_p^2}{Q^2}\right)F_2^{(p)}\left(\frac{x}{z},Q^2\right)}}\\
&-z^2F_L^{(p)}\left(\frac{x}{z},Q^2\right)\Bigg]-\left(\alpha(\mu_F^2)\right)^2 z^2F_2^{(p)}\left(\frac{x}{z},\mu_F^2\right)\Bigg\}+\mathcal{O}(\alpha\alpha_s,\alpha^2)\,,\label{eq:LUXqed}
\end{aligned}
\end{equation}
where $m_p$ is the mass of the proton, and $F_2^{(p)}$ and $F_L^{(p)}$ are the proton structure functions in $ep$ scattering. From eqs.\eqref{eq:EWunregAP3} and \eqref{eq:EWAPregkernal}, we have
\begin{equation}
\frac{P^{(\mathrm{EW})}_{\gamma q}(z,0)}{\left(e(q)\right)^2}=\frac{1+(1-z)^2}{z}.
\end{equation}
Equation \eqref{eq:LUXqed} is accurate up to $\mathcal{O}(\alpha)$, with the missing higher-order corrections at $\mathcal{O}(\alpha\alpha_s)$ and $\mathcal{O}(\alpha^2)$, the latter two corrections being given in ref.~\cite{Manohar:2017eqh}. To determine $f_{\gamma}^{(p)}(x,\mu_F^2)$, one must include the elastic component of the proton structure functions, which corresponds to the elastic scattering $ep\to ep$, where the Bjorken $x_{\mathrm{Bj}}$ is located at $1$. Furthermore, it is also important to include the low photon virtuality $Q^2$ or the photoproduction data. The \luxqed\ approach leads to a photon PDF determination with a relative error below $2\%$ for $10^{-4}<x<0.1$ and below $3\%$ for $10^{-5}<x<0.5$, with a precision approximately $40$ times better than previous determinations. Such an extremely accurate photon PDF is sufficient for the LHC physics. Nowadays, all major PDF sets that include photon PDFs have adopted the \luxqed\ approach.

Following a similar idea, the charged lepton PDFs of the proton can also be determined from the same proton structure functions~\cite{Buonocore:2020nai}.~\footnote{An earlier attempt on determining lepton PDFs can be found in ref.~\cite{Bertone:2015lqa}.} The relation between the $\msbar$ lepton PDFs and the structure functions is, however, more complex. I will not present the equation explicitly here but refer to eq.(2.25) in ref.~\cite{Buonocore:2020nai}. Compared to the photon PDF, the lepton PDFs are further suppressed by an additional $\mathcal{O}(\alpha)$ factor, leading to the common belief that their phenomenological importance is secondary. However, they turn out to be relevant in some BSM studies, such as resonant lepto-quark production~\cite{Haisch:2020xjd}. 

While the numerical impact is usually marginal, I want to stress that it is important to include $\mathcal{O}(\alpha)$ corrections in the Dokshitzer-Gribov-Lipatov-Altarelli-Parisi (DGLAP) evolution of PDFs~\cite{Gribov:1972ri,Lipatov:1974qm,Altarelli:1977zs,Dokshitzer:1977sg} to achieve full NLO EW accuracy. The inclusion of photon and $\mathcal{O}(\alpha)$ corrections in scale evolution, along with possibly charged lepton contributions in the proton, should impact all quark and antiquark PDFs, as well as the gluon PDF. Several global-fit PDF sets available through \lhapdf~\cite{Buckley:2014ana} have included the photon PDF and $\mathcal{O}(\alpha)$ QED corrections.

We conclude this section with an aside regarding lepton structure functions for precision physics programs at future lepton colliders. At lepton colliders,  a well known fact is that radiative corrections from collinear electromagnetic emission off leptons can be quite large due to the presence of large logarithms stemming from the hierarchy between the tiny lepton mass and the hard scale at high energies. Such collinear logarithms are ubiquitous at lepton colliders and must therefore be resummed. While there are no unique ways to perform the resummation~\footnote{Alternative resummation techniques, such as QED parton showers~\cite{Anlauf:1991wr,Fujimoto:1993qh,Fujimoto:1993ge,Fujimoto:1993ge,Munehisa:1995si,CarloniCalame:2000pz,Shao:2014rwa}, will not be discussed here.}, a conventional approach is to define the lepton structure functions, the analogue of proton PDFs, and to resum the collinear logarithms by solving the DGLAP evolution equations. This task involves determining the initial conditions and then solving the coupled evolution equations. To match NLO EW accuracy, LO plus leading logarithm (LO+LL) is insufficient. Therefore, the structure functions of the electron/positron $f^{(e^\pm)}_{\ident}(x,\mu_F^2)$ with $\ident=\ell^\pm,\gamma,q,\bar{q}$ ($q\neq t$) have been advanced from LO+LL~\cite{Skrzypek:1990qs,Skrzypek:1992vk,Cacciari:1992pz,Beenakker:1996kt} known in the 1990s to NLO plus next-to-leading logarithmic (NLO+NLL) accuracy recently~\cite{Frixione:2019lga,Bertone:2019hks,Frixione:2012wtz,Bertone:2022ktl}. Note that in the context of lepton structure functions, (N)LO refers to the order of the initial conditions at the initial scale (typically the lepton mass), while (N)LL indicates the accuracy of their scale evolution. The only non-null initial condition at LO for the electron structure functions is $e^-$, which has $f^{(e^-)}_{\ident}(x,m_e^2)=\delta_{\ident e^-}\delta(1-x)+\mathcal{O}(\alpha)$. The photon receives a non-zero contribution starting at $\mathcal{O}(\alpha)$, while other partons are non-zero only beyond NLO at the initial scale. However, all the partons can be generated by the evolution equations at sufficiently high scales. 
Automated computations of NLO EW corrections in the \mgamcshort\ for $e^-e^+$ colliders~\cite{Bertone:2022ktl} take into account effects from both the structure functions and the beamstrahlung, with the latter accounting for collective phenomena in the beam dynamics of accelerators. Unlike proton PDFs, the electron PDF peaks sharply at $x=1$ due to the Dirac delta function $f^{(e^-)}_{e^-}(x,m_e^2)$. Consequently, the numerical implementation must address at least two aspects to avoid degrading numerical accuracy. For the evaluation of the electron PDF, analytical ingredients in the region $x\simeq 1$ must be computed. Additionally, due to dominant contribution at $x\to 1$ in $e^-e^+$ collisions, phase space generation--a convolution of the matrix elements with the electron PDF--must be carefully redesigned. This issue exists already at LO~\cite{Frixione:2021zdp}, but becomes more severe at NLO, particularly in the context of FKS subtraction. For the latter, explained in appendix C of ref.~\cite{Bertone:2022ktl},  the issue arises from how the initial $x_1$ and $x_2$ are generated. The original treatment based on the so-called event projection method~\cite{Frixione:2002ik,Frederix:2011ss} works only if both incoming partons have distributions in $x_1$ and $x_2$ that are not Dirac delta function alike, which is suitable for matching NLO computations to the hadronic parton shower in cases of initial backward evolution. This has been remedied by abandoning event projection for the FKS sectors relevant to cases where $\nini\leq j\leq 2$ and the branching $\ident_j\to \ident_{j\oplus \bar{i}} \ident_i$ is a QED one. Finally, there has been progress in determining the muon PDFs, which target a future high-energy muon collider. In particular, $f^{(\mu^-)}_{\ident}(x,\mu_F^2)$ are known at NLO+NLL for partons being photons, charged (anti)leptons, (anti)quarks, and gluons~\cite{Frixione:2023gmf}, and for $\ident$ additionally being weak bosons at LO+LL~\cite{Han:2020uid,Han:2021kes,Garosi:2023bvq}. In ref.~\cite{Garosi:2023bvq}, polarization effects have also been studied.

\section{Isolated final photons and dressed leptons\label{sec:taggedphotonlepton}}

Photons and leptons can be regarded both as particles that enter the short-distance process and as observable (or taggable) objects, a point that has already been throughly discussed in ref.~\cite{Frederix:2016ost}. I will limit myself here to summarizing the conclusions of that paper. The key point is that short-distance photons and massless leptons can be
identified with the corresponding taggable objects only up to a certain $\mathrm{NLO}_{i_0}$ term, beyond which (\ie\, for $\mathrm{NLO}_i$, $i>i_0$) this identification leads to IR-unsafe observables. Typically, the value of $i_0$ is process dependent, resulting in IR unsafety manifesting in the third- or even the second-leading NLO contribution. In the case of the second-leading NLO term ($\mathrm{NLO}_2$), one can work around this issue using the $\alpha(0)$ scheme. The solution, applicable for the usual inclusive processes in hadronic collisions, proposed in ref.~\cite{Frederix:2016ost} is as follows:
\begin{enumerate}
\item Short-distance photons and massless charged leptons are not taggable objects.
\item A taggable photon is a photon that emerges from a fragmentation process.
\item A taggable massless lepton is either a lepton that emerges from a fragmentation process or a dressed lepton, defined as an object whose four-momentum matches that of a very narrow jet that containing the short-distance lepton.
\item These rules imply that photons and massless charged leptons must be treated on the same footing as gluons and quarks in short-distance computations (\ie, the democratic approach).
\item Short-distance computations should be performed in $\msbar$-like EW renormalization schemes, such as the $G_\mu$ or $\alpha(M_Z^2)$ ones, regardless of the initial- and final-state particle contents of the process of interest.
\end{enumerate}
We refer to this approach as the fragmentation function approach. In addition to being completely general and compatible with established procedures, working with fragmentation functions has the appealing feature of aligning QCD and QED within a similar framework, and of rendering conceptually alike treatments of initial- and final-state photons. The necessary theoretical framework for implementing fragmentation functions in the \mgamcshort\ framework has been presented in ref.~\cite{Frederix:2018nkq}. In particular, the FKS subtraction formalism with fragmentation has been derived in that paper. Unfortunately, the determinations of non-perturbative fragmentation functions for tagged photons is plagued by large uncertainties (if known), and they can only be measured (if at all) for sufficiently large momentum fractions. 

While the solution based on the concept of fragmentation functions has not been fully exploited yet, alternative but simpler solutions indeed exist in the literature. For instance, with the photon isolation algorithm~\cite{Frixione:1998jh}, NLO computations for processes with final tagged photon(s) can be achieved by invoking the mixed-scheme approach~\cite{Pagani:2021iwa}, which is based on the idea that $\alpha$ should be renormalized in the $\alpha(0)$ scheme only for the final-state isolated photons, while other EW interactions should be renormalized in the $\alpha(M_Z^2)$ or $G_\mu$ scheme. This procedure has been automated in the \mgamcshort\ framework. In this approach, concerning the purely QED part of NLO EW corrections, besides effects that are formally beyond NLO and related to the scale evolution of fragmentation functions, a calculation performed in a $\msbar$-like renormalization scheme employing the photon fragmentation function leads to the same result as a calculation performed in the $\alpha(0)$-scheme with isolated-photons~\cite{Frederix:2016ost}. As mentioned earlier, this solution is valid only up to a certain $\mathrm{NLO}_{i_0}$ term, where $i_0\geq 2$ for cases involving both isolated photons and (democratic-)jets or dressed leptons (see section 2.2.3 in ref.~\cite{Pagani:2021iwa}). Let me elaborate a bit more on this. Consider a general process involving $n_\gamma$ isolated photons:
\begin{equation}
pp \to n_\gamma \gamma_{\mathrm{iso}}+X.
\end{equation}
In this case, we should have $\left(\alpha(0)\right)^{n_\gamma}$ in the matrix elements, while for the rest, we should use $\alpha$ either in the $\alpha(M_Z^2)$ or $G_\mu$ scheme. The substitution $\alpha\to\alpha(0)$ occurs only globally, which is important for maintaining gauge invariance and IR cancellation. However, using the $\alpha(0)$ scheme also implies that the standard procedure for generating real-radiation diagrams should be amended. In particular, since the final-state photon in a diagram coincides with a physical observable object, \ie\,the
isolated-photon, final-state QED branching $\gamma_{\mathrm{iso}}\to f\bar{f}$ should be vetoed, where $f$ is a massless charged fermion. Therefore, the definition of the FKS counterterms should also be modified (see section 2.2.2 in ref.~\cite{Pagani:2021iwa}), which amounts to defining
\begin{equation}
C_{\mathrm{EW}}(\gamma_{\mathrm{iso}})=\gamma_{\mathrm{EW}}(\gamma_{\mathrm{iso}})=\gamma^\prime_{\mathrm{EW}}(\gamma_{\mathrm{iso}})=0
\end{equation}
in $\hat{\sigma}^{(C,n)}_{(p,q),{\rm FIN}}(r)$ (\cf\ eq.\eqref{eq:hatsigmaCnpqFIN}).

A second example that has been explored is the use of the photon-to-jet conversion function in the context of NLO EW corrections for processes with jets in the final state, which involves contributions from low-virtuality photons~\cite{Denner:2019zfp}. The conversion function is similar to jet fragmentation functions. However, the long-distance contributions to the conversion function must be extracted from experimental data, which are not yet available. Therefore, an ansatz based on the dispersion relations for the $R$ ratio of the cross sections $\sigma(e^-e^+\to \mathrm{hadrons})/\sigma(e^-e^+\to \mu^-\mu^+)$ is adopted in ref.~\cite{Denner:2019zfp}.

On the other hand, for the purpose of collinear safety, massless charged leptons can either be treated as part of a jet in the democratic jet clustering procedure or as a dressed object through photon recombination. In the latter case, a collinear lepton-photon system is treated as a quasi-particle, also known as a dressed lepton. For final-state electrons, photon recombination is automatically incorporated during their experimental reconstruction from electromagnetic showers detected by calorimeters. Conversely, due to its greater mass compared to the electron and its relatively long lifetime in particle physics, the muon can be observed directly as a bare lepton from its track in the muon chamber. To match theoretical calculations with experiments, the mass of the muon must be retained in perturbative calculations to regularize collinear singularities. However, this introduces potentially large collinear logarithms from final-state radiation. To mitigate significant final-state radiation corrections, the muon is sometimes reconstructed as a dressed muon via photon recombination. The advantage of using dressed leptons instead of  bare leptons is that collinear logarithms cancel out, leaving the cross section largely independent of the lepton mass.

\section{Process generation in \mgamclong}

Along with the publication of the paper~\cite{Frederix:2018nkq}, we released a major upgrade of the \mgamcshort\ code to enable complete NLO calculations (including QCD, EW, and subleading corrections) for arbitrary processes~\footnote{Strictly speaking, this statement only applies to processes with not-too-high particle multiplicity, given the limitations of computing resources. This consideration does not cover loop-induced processes. Features related to fragmentation functions have not been released yet. Therefore, in the context of this dissertation, when we refer to a similar statement, it should be interpreted in a loose sense.} in the SM and is capable of handling mixed-coupling expansions in general. The code is also compatible with other theories, such as the SM effective field theory (SMEFT), provided NLO-compatible UFO models.

The process generation syntax in \mgamcshort\ for a NLO QCD and EW computation may read as follows:
\vskip 0.25truecm
\noindent
~~\prompt\ {\tt ~set~complex\_mass\_scheme~True}

\noindent
~~\prompt\ {\tt ~import~model~myNLOmodel\_w\_qcd\_qed}

\noindent
~~\prompt\ {\tt ~generate~p$_1$ p$_2$ > p$_3$ p$_4$ p$_5$ p$_6$ 
aS=$\nmax$ aEW=$\mmax$ [QCD QED]}

\vskip 0.25truecm
\noindent
with {\tt p$_i$}~\footnote{{\tt p$_i$} can also be a tagged photon $\gamma_{\mathrm{iso}}$ within the $\alpha(0)$ scheme, as discussed in section \ref{sec:taggedphotonlepton}. In this case, one would have ${\tt p}_i={\tt !a!}$, first introduced in ref.~\cite{Pagani:2021iwa}.} being (multi)particles that belong to the particle spectrum of the NLO model\\
 {\tt myNLOmodel\_w\_qcd\_qed}, a placeholder specified by the user. The syntax above implies
calculating the following LO and NLO contributions~\footnote{Instead of using {\tt aS} and {\tt aEW}, one can also specify the Born amplitude-level $\ampnt_{(p,q)}(r)$ coupling order constraint $g_s^p e^q$ ($p\leq \nmax, q\leq \mmax$) via {\tt QCD=}$\nmax$ {\tt QED=}$\mmax$, and/or the LO matrix-element-level $\ampsqnt_{(p,q)}(r)$ coupling orders $g_s^{2p}e^{2q}$ ($2p\leq \nmax, 2q\leq\mmax$) via {\tt QCD\textasciicircum2=}$\nmax$ {\tt QED\textasciicircum2=}$\mmax$, consistent with the original convention in \mgamcshort.}:
\beqn
&&{\rm LO}:\phantom{Naaa}
\alpha_s^p\alpha^q\,,\;\;\;\;\;p\le\nmax\,,\;\;\phantom{+1,}\;\,
q\le\mmax\,,\;\;\phantom{+1,}\;\,\,p+q=k_0\,,
\label{LOsynt}
\\*
&&{\rm NLO}:\phantom{aaa}\,
\alpha_s^p\alpha^q\,,\;\;\;\;\;p\le\nmax+1\,,\;\;\;
q\le\mmax+1\,,\;\;\;\,p+q=k_0+1\,.
\label{NLOsynt}
\eeqn
We point out that the largest power of $\alpha_s$ in eq.\eqref{NLOsynt} is exactly one unit larger than its LO counterpart in eq.\eqref{LOsynt} due to the presence of the keyword {\tt [QCD]} in the process generation command. This instructs the code to consider all possible Feynman diagrams that contribute to the virtual/real matrix elements with two additional QCD vertices compared to those present at LO. Similarly, the keyword {\tt [QED]} determines the largest power of $\alpha$ appearing at the NLO level is \mbox{$\mmax+1$}, as shown in eq.\eqref{NLOsynt}. The keyword {\tt [QED]} is purely conventional. It implies that both electromagnetic and weak effects~\footnote{The separation of electromagnetic and weak effects is not always legitimate. In processes involving $W^\pm$ at LO, this separation can break gauge invariance. Conversely, in other processes, such as those with $t\bar{t}h$~\cite{Frixione:2014qaa} or $t\bar{t}Z$~\cite{Mangano:2015aow} final states, the division poses no issue.}, \ie, the complete $\mathcal{O}(\alpha)$ corrections, are included, as long as they are supported by the model. In this sense, the keywords {\tt QCD} and {\tt QED} are better interpreted as the couplings $g_s$ (or $\alpha_s$) and $e$ (or $\alpha$) rather than the particles in the vertices. Either keyword can be omitted to ensure backward compatibility of \mgamcshort. The values of $\nmax$ and $\mmax$ can be freely set by the user, and the keywords {\tt aS=}$\nmax$ and  {\tt aEW=}$\mmax$ can both be omitted. In such cases, \mgamcshort\ generates the process with the smallest possible power of the QED coupling $\alpha$ at LO, following the hierarchy $\alpha\ll \alpha_s$. Finally, we note that the first command line, ``{\tt set complex\_mass\_scheme True}", instructs the code to use the CM scheme, provided that the model supports it. The default value is {\tt False}.
 
Because IR safety in NLO EW computations is non-trivial, it is worth commenting on how to define Born processes $\allprocn=\{${\tt p}$_1$ {\tt p}$_2$ {\tt > p}$_3$ {\tt p}$_4$ {\tt p}$_5$ {\tt p}$_6 \}$ using the multiparticle definition in \mgamcshort, as shown in the following command lines:
\vskip 0.25truecm
\noindent
~~\prompt\ {\tt ~define p = g d d\~{} u u\~{} s s\~{} c c\~{} b b\~{} a}

\noindent
~~\prompt\ {\tt ~define j = g d d\~{} u u\~{} s s\~{} c c\~{} b b\~{} a}

\vskip 0.25truecm
\noindent
They define the two multiparticles {\tt p} and {\tt j}. While multiparticle names can be freely chosen, in practice the two most commonly used ones--{\tt p} and {\tt j}--are conventionally associated with the incoming hadrons and outgoing jets, respectively. These names help to define the parton components of these objects.
To ensure IR safety in EW corrections, the photon {\tt a} must be included in the definitions of {\tt p} and {\tt j}. For charged massless leptons, this depends on the specific processes and which terms are considered, even when the lepton PDFs are set to zero. In particular, for photon-induced Born processes, at least one charged lepton-induced process must be included to ensure the correct implementation of collinear subtraction counterterms. In conclusion, whenever computing the following NLO corrections:
\begin{equation}
\Sigma_{\mathrm{NLO}_{i-k}}+\ldots+\Sigma_{\mathrm{NLO}_{i}}, \quad 0\leq k\leq i-1, \quad 1 \leq i \leq \Delta(k_0)+2\label{eq:Sconds}
\end{equation}
we recommend defining the multiparticles {\tt p} and {\tt j} as summarized in table~\ref{tab:recs}, where ``jets" in the table refers to democratic jets. In the table, ``\mbox{PDF($qg$)}" denotes PDF sets limited to light (anti)quarks and gluons, while ``\mbox{PDF($qg\gamma$)}" includes photons in addition to light (anti)quark and gluons. When PDF sets feature non-zero charged-lepton distributions, these (anti)leptons must always be included in the definition of {\tt p}.

\begin{table}
\begin{center}
\resizebox{\textwidth}{!}{
\begin{tabular}{llllll}
\toprule
 & \multicolumn{2}{c}{Processes without jets}
 & \multicolumn{2}{c}{Processes with jets}
 & \multirow{2}{*}{Physical objects}\\
 & PDF($q g$) & PDF($q g\gamma$) & PDF($q g$) & PDF($q g\gamma$) & \\
\midrule
\multirow{2}{*}{$i=1$} & \multirow{2}{*}{\texttt{p = q g}} &
\multirow{2}{*}{\texttt{p = q g a}} & \texttt{p = q g} &
\texttt{p = q g a} & $j(qg)$, $\gamma$, $l$, $\nu$,\\
 & & & \texttt{j = q g} & \texttt{j = q g} & massive particles\\
\midrule
\multirow{2}{*}{$i=2$} & \multirow{2}{*}{inconsistent} &
\multirow{2}{*}{\texttt{p = q g a}} & \multirow{2}{*}{inconsistent} &
\texttt{p = q g a} & $j(qg\gamma)$, $l$, $\nu$, \\
 & & & & \texttt{j = q g a} & massive particles\\
\midrule
\multirow{2}{*}{$i\ge 3$} & \multirow{2}{*}{inconsistent} & 
\multirow{2}{*}{\texttt{p = q g a}} & \multirow{2}{*}{inconsistent} &
\texttt{p = q g a l} & $j(qg\gamma l)$, $\nu$,\\
 & & & & \texttt{j = q g a l} & massive particles\\
\bottomrule
\end{tabular}
}
\caption{\label{tab:recs}
Recommendations for defining the multiparticles {\tt p} and {\tt j}
in computations of the NLO corrections given
in eq.\eqref{eq:Sconds} in \mgamcshort\ to ensure IR safety. {\tt q} stands for all the massless quarks and
anti-quarks in the model, {\tt g} is the gluon, {\tt a} is 
the photon, and {\tt l} encompasses all massless charged-leptons 
(\eg, {\tt l = e+ e- mu+ mu- ta+ ta-}). In the rightmost column,
$j$, $\gamma$, $l$, and $\nu$ refer to jets, photons, charged leptons,
and neutrinos, respectively. The symbols $q$ and $g$ denote light (anti)quarks and gluons. The table is same as table 4 in ref.~\cite{Frederix:2018nkq}.
}
\end{center}
\end{table}

In order to validate our implementation in \mgamcshort, extensive cross-checks at both the matrix-element and cross-section levels have been carried out across dozens of processes. These include numerous self-consistency checks as well as cross-checks with other independent calculations. In particular, the virtual matrix elements and (integrated/differential) cross sections have been compared with those from other groups for four lepton production processes during the Les Houches Workshop on ``Physics at TeV Colliders" in 2017 (see section 7 of the first chapter in ref.~\cite{Proceedings:2018jsb}), and for $t\bar{t}h$ during the same workshop in 2015~\cite{Andersen:2016qtm}. Perfect agreement among the different groups was achieved.

\section{Going beyond}

In a somewhat loose sense, the automation of the fixed-order NLO computations with both QCD and EW radiative corrections has been established, at least for inclusive reactions in $pp$ and $e^-e^+$ collisions. It is time to look ahead. Beyond NLO, two prominent issues need to be addressed in the context of EW corrections: the resummation of EW Sudakov logarithms and the matching to parton shower Monte Carlo programs. I will discuss these in this section before closing the chapter.

\subsection{EW Sudakov logarithms\label{sec:EWSudakovLog}}

It has long been known that at sufficiently high energies relative to the internal and external particle masses, perturbative computations suffer from large Sudakov or high-energy logarithms~\cite{Sudakov:1954sw}. This is generally true in gauge theories, including the SM. In the context of EW corrections in the SM, Sudakov logarithms appear in the form of $\frac{\alpha}{s_W^2}\log^2{\frac{Q^2}{M_W^2}}$ (double logarithm) and $\frac{\alpha}{s_W^2}\log{\frac{Q^2}{M_W^2}}$ (single logarithm), where $Q$ is the typical energy scale of the considered process. These logarithms enhance the size of the EW corrections when $Q^2\gg M_W^2$. Unlike in QCD (or QED), where the soft-collinear double logarithms and soft single logarithms cancel out between virtual and real corrections, and where the remaining collinear single logarithms are absorbed into PDFs and/or fragmentation functions, HBR contributions in the EW corrections are usually not considered. This omission is justified, as the masses of the $W^\pm$ and $Z$ bosons provide an IR cutoff, and HBR can be experimentally reconstructed to a large extent. The separation of HBR from the EW corrections renders the EW corrections very significant in the multi-TeV kinematic regime, resulting in corrections on the order of tens of percent. 

At the one-loop level, the double and single EW logarithms stemming from soft and/or collinear limits in the virtual amplitudes can be expressed in a universal form, as derived in refs.~\cite{Denner:2000jv,Denner:2001gw} for the broken phase of the SM. This universal form is valid only in the Sudakov regime, where, using the same notation previously presented, all Mandelstam invariants for any Born-like process $r\in\allprocn$ should satisfy
\begin{equation}
\left|s_{kl}\right|=\left|\left[(-1)^{s(\ident_k)}k_k+(-1)^{s(\ident_l)}k_l\right]^2\right|\gg M_W^2,\quad \forall\ k,l, \quad 1\leq k<l\leq n+2, \label{eq:Sudakovregime}
\end{equation}
where $s(\ident)$ has been defined in eq.\eqref{eq:sIdef}. In the EW leading (power) approximation (EWLA), without mass-suppressed contributions, the one-loop EW Sudakov amplitude can be generically expressed as
\begin{equation}
\ampnl_{(p,q),\mathrm{EWLA}}(r)=\sum_{\tilde{r}\in \tilde{\allproc}_n(r)}{\ampnt_{(p,q-2)}(\tilde{r})\Delta_{\mathrm{EWLA}}(\tilde{r}, r)}.
\end{equation}
Here, $\Delta_{\mathrm{EWLA}}(\tilde{r}, r)$ contains the EW Sudakov logarithms expressed as follows:
\begin{equation}
L(|s_{kl}|,M^2)=\frac{\alpha}{4\pi}\log^2{\frac{|s_{kl}|}{M^2}},\quad l(|s_{kl}|,M^2)=\frac{\alpha}{4\pi}\log{\frac{|s_{kl}|}{M^2}},
\end{equation}
where $M$ can be $M_t,M_h,M_W$, or $M_Z$. The double logarithms $L(|s_{kl}|,M^2)$ arise from loop diagrams where soft–collinear gauge bosons are exchanged between pairs of external legs. In contrast, the single logarithms $l(|s_{kl}|,M^2)$ originate from three sources: subleading soft-collinear, collinear, and parameter renormalization.
The set of new $2\to n$ processes $\tilde{\allproc}_n(r)$ is constructed as follows: it begins with the set $\left\{r\right\}$. For any two external particles $(\ident_k,\ident_l)$ in $r$ (considering their spin/helicity dependence, as the same identity with different helicities might belong to different representations of SU$(2)_L$), apply the SU$(2)_L$ generator to obtain three possible combinations: $(\tilde{\ident}_k,\ident_l)$, $(\ident_k,\tilde{\ident}_l)$, and $(\tilde{\ident}_k,\tilde{\ident}_l)$, where $\tilde{\ident}$ is the SU$(2)_L$ partner of $\ident$. It is understood that if $\ident$ is an SU$(2)_L$ singlet, then $\tilde{\ident}=\ident$. Substituting $\ident_k$ and $\ident_l$ in $r$ with their partners yields three new processes. If these processes are allowed by conservation laws, they are added to the set $\tilde{\allproc}_n(r)$. This procedure iterates until all possible processes are exhausted. In this context, while all external particles in the original process $r$ ($r\in\allprocn$) are physical, the external particles of $\tilde{r}$ with $\tilde{r}\in\tilde{\allproc}_n(r)$ may involve unphysical degrees of freedom, such as Goldstone bosons, because the longitudinal modes of $W^\pm$ and $Z$ bosons are identified as the corresponding Goldstone bosons using the Goldstone-boson equivalence theorem, which holds in high-energy limits.

Since the one-loop structure of EW logarithms is rather simple and involves only Born-like amplitudes, it is appealing from a practical standpoint to use EW Sudakov logarithms as a means to approximate EW corrections in the Sudakov regime, rather than conducting full-fledged but CPU-expensive NLO computations. Consequently, one-loop EW Sudakov logarithmic results have been implemented in several generators: \alpgen~\cite{Chiesa:2013yma},~\footnote{The original reference of \alpgen\ is ref.~\cite{Mangano:2002ea}.} \mcfm~\cite{Campbell:2016dks}, \sherpa~\cite{Bothmann:2020sxm}, \mgamcshort~\cite{Pagani:2021vyk}, and \openloops~\cite{Lindert:2023fcu}. In particular, ref.~\cite{Pagani:2021vyk} revises the original formalism to include the logarithms of two invariants, $L(|s_{kl}|,|s_{k^\prime l^\prime}|)$ and $l(|s_{kl}|,|s_{k^\prime l^\prime}|)$, enabling the approximation to be valid even in cases where $|s_{kl}|\gg |s_{k^\prime l^\prime}| \gg M_W^2$. However, there are limitations to using the one-loop EW logarithms to approximate full NLO EW computations, which are inherently superior when feasible. Firstly, the EW logarithmic formalism does not account for the mass-suppressed logarithms, which may have a different structure. Consequently, the derived EW logarithm approximation cannot be applied to mass-suppressed processes, such as Higgs boson production in vector boson fusion or Higgs-strahlung off a vector boson.~\footnote{The latter case is, however, more involved. As pointed out in ref.~\cite{Lindert:2023fcu}, the amplitude for the transversely polarized $Z$ boson is mass-suppressed, whereas that for the longitudinally polarized $Z$ boson is not (\cf\ figure 19 in ref.~\cite{Lindert:2023fcu}). Nevertheless, the EW Sudakov logarithmic approximation remains valid when the spin of the $Z$ boson is summed, as illustrated in figure 5 of ref.~\cite{Ma:2024ayr}.}  Secondly, the applicability of the EW logarithmic approximation strictly relies on the validity of the condition expressed in eq.\eqref{eq:Sudakovregime}. In processes involving unstable particles and their subsequent decays, this condition cannot be fulfilled, as the invariant masses of the decay products dominate in the resonant regions of the unstable particles. In other words, processes that include unstable particles and their decays cannot be treated under this approximation. Moreover, even in the high energy limit, there are also exceptions where cross sections are not primarily dominated by the Sudakov regime. For example, in Drell-Yan-like processes~\cite{Dittmaier:2001ay,Dittmaier:2009cr}, cross sections receive substantial contributions in the Regge limit, where the Mandelstam variable $t=s_{13}$ remains small  while  $s=s_{12}$ becomes large compared to the EW scale. Overall, a careful assessment of the quality of the high-energy logarithmic approximation always requires comparison with full NLO results.

On the other hand, detailed knowledge of the EW high-energy logarithms should aid in their resummation, allowing us to go beyond NLO calculations. For leading logarithms (LL), these can be resummed simply via exponentiation~\cite{Fadin:1999bq}. For next-to-leading logarithms (NLL), the resummed results have been conjectured to be calculable using renormalization group equations~\cite{Kuhn:1999nn,Kuhn:2001hz}. Based on soft–collinear effective theory (SCET)~\cite{Bauer:2000ew,Bauer:2000yr,Bauer:2001ct,Bauer:2001yt}, a general method for resumming these logarithms in the unbroken phase of the SM has been developed in refs.~\cite{Chiu:2007yn,Chiu:2008vv}. With the approach, it is feasible to obtain resummed EW corrections for all hard scattering processes that are not mass-suppressed and do not involve intermediate resonances at NLL order~\cite{Chiu:2009mg,Chiu:2009ft,Fuhrer:2010eu}. Recently, this method has been implemented in a Monte Carlo integration code based on \recola~\cite{Denner:2024yut}.

\subsection{Matching to parton showers}

In order to extend the scope of NLO computations, it is desirable to combine parton-level perturbative calculations with particle-level parton shower Monte Carlo (PSMC) simulations, which have proven extremely useful for NLO QCD in LHC physics. Analogous to the NLO QCD case,  a consistent combination of these calculations is highly non-trivial and must be carefully devised to avoid double counting and maintain NLO accuracy. However, a general solution for mixed QCD and EW corrections is still absent in the literature. A prominent challenge arises in assigning color flows for the LO$_i$ ($1<i<\Delta(k_0)+1$) terms, as these typically result from amplitude interferences rather than from squaring the same amplitudes. The kinematics and color structure of the hard process from matrix elements provides the necessary initial conditions in a PSMC to generate a shower. Both a QCD shower on top of LO$_i$  and a QED shower on top of LO$_{i-1}$ contribute to the NLO$_i$ corrections when $1<i<\Delta(k_0)+2$. To address this issue, a method for assigning color flows in interferences has been proposed in ref.~\cite{Frixione:2021yim}. Additionally, several approximated methods have been introduced in the literature~\cite{Kallweit:2015dum,Granata:2017iod,Gutschow:2018tuk,Brauer:2020kfv,Bothmann:2021led,Pagani:2023wgc}. These methods often disregard real emission from matrix elements at NLO$_2$ and include the (partial) virtual EW corrections by amending the ingredients in NLO QCD matched to parton shower calculations. For example, ref.~\cite{Pagani:2023wgc} utilizes EW Sudakov logarithms to approximate virtual corrections, while refs.~\cite{Kallweit:2015dum,Granata:2017iod,Gutschow:2018tuk,Brauer:2020kfv,Bothmann:2021led} apply the IR subtracted virtual matrix elements in the dipole subtraction scheme. All of these approaches employ the MC@NLO-type matching scheme. For processes with $\Delta(k_0)=0$, exact matching of NLO QCD and EW corrections to parton shower is possible in a manner analogous to that implemented for NLO QCD.  For instance, within the POWHEG approach~\cite{Nason:2004rx,Frixione:2007vw}, matching has been performed for Drell-Yan-like processes~\cite{Barze:2012tt,Barze:2013fru,Bernaciak:2012hj,Muck:2016pko,CarloniCalame:2016ouw}, diboson production~\cite{Chiesa:2020ttl}, and same-sign $W$-boson scattering~\cite{Chiesa:2019ulk}.
 
Finally, I would like to comment on an issue regarding the matching at $e^-e^+$ colliders. As aforementioned in section \ref{sec:photonleptonPDF}, the phase space generation of momenta for NLO EW corrections in $e^-e^+$ collisions does not rely on event projection. However, this new generation is not compatible with an MC@NLO-type matching for the initial-state QED shower if the QED shower in a PSMC operates in the same way as it does in hadronic collisions. In such cases, the matching between NLO EW computations and parton showers at $e^-e^+$ colliders would present an additional challenge.

\chapter{Relevance of Electroweak Corrections in Collider Physics}
\label{SEC:PHENO}

In this chapter, I will discuss the relevance of EW corrections for understanding the physics at the LHC. My aim is not to be exhaustive but rather selective, as this topic has been studied for decades and a vast amount of results exists in the literature. Reviewing each one individually would be tedious. Instead,  I believe this chapter will be more engaging if I illustrate my points with specific examples.

\section{NLO EW corrections\label{sec:NLOEWres}}

We begin by presenting some illustrative NLO EW results from ref.~\cite{Frederix:2018nkq} to showcase the typical sizes of the EW corrections for processes at the LHC. The goal of this section is to demonstrate the capabilities of full NLO EW automation in \mgamcshort. To facilitate direct comparisons among the processes, the same minimal conditions will be imposed on all of them 

Before discussing the results, let me first specify the setup for our calculations. We work in the $5$-quark flavor scheme, where all fermion masses, except the top quark $t$, are set to zero, and the CKM matrix is taken to be the identity. Additionally, the fine-structure constant $\alpha$  is renormalized in the $G_\mu$ scheme (\cf\ section \ref{sec:renormalization}) with 
\begin{equation}
G_\mu=1.16639\times 10^{-5}~\mathrm{GeV}^{-2}.
\end{equation}
In addition, we use a standard UFO model {\tt loop\_qcd\_qed\_sm\_Gmu} in the \mgamcshort\ framework. This model includes all UV and $R_2$ counterterms for NLO QCD and EW corrections in the SM. It is compatible with both the OS (\cf\ section \ref{sec:renormalization}) and the CM (\cf\ section \ref{sec:CMS}) schemes, as well as their mix, for all massive and unstable particles, namely the $W^\pm$, $Z$, $h$ bosons, and the top quark $t$. To avoid violating the unitarity of the $S$-matrix, all external particles, whether massive or massless, should be treated as stable (\ie, undecayed) particles. Thus, external particles are always renormalized in the OS scheme, and their widths (if non-zero) should be set to zero. If an intermediate particle has a (non-)zero width, it will be renormalized in the (CM) OS scheme. Therefore, we must avoid scenarios where an external particle can be on-shell in an intermediate propagator. In such cases, we must decay the external particle and introduce a (at least NLO-accurate) width for it. In this section, prior to process generation, we execute the following commands:

\vskip 0.25truecm
\noindent
~~\prompt\ {\tt ~set~complex\_mass\_scheme~true}

\noindent
~~\prompt\ {\tt ~import~model~loop\_qcd\_qed\_sm\_Gmu}

\noindent
~~\prompt\ {\tt ~define p = g d d\~{} u u\~{} s s\~{} c c\~{} b b\~{} a}

\noindent
~~\prompt\ {\tt ~define j = g d d\~{} u u\~{} s s\~{} c c\~{} b b\~{} a}

\vskip 0.25truecm
\noindent
We have included the photon {\tt a}, the gluon {\tt g}, and the massless (anti)quarks in the multiparticles {\tt p} and {\tt j} as outlined in section \ref{sec:photonleptonPDF}. Since we are only interested in the NLO$_2$ (\ie\ NLO EW) term in this section, we do not necessarily include the charged leptons in {\tt p} and {\tt j} (see table \ref{tab:recs}).

Let us consider the LHC Run II case in $pp$ collisions at a center-of-mass energy of 13 TeV. We adopt the input parameters listed in table~\ref{tab:EWparaminput}. The pole masses and widths of the $W^\pm$ and $Z$ bosons have been converted from their OS masses and widths using eq.\eqref{eq:polemasswidthfromOS}. As explained above, the widths of external particles must be set to zero. Therefore, if a $W^\pm$ (or $Z$) boson is in the final state, we first use eq.\eqref{eq:polemasswidthfromOS} to obtain its OS mass and then set its width to zero. Because the Higgs width is very small in the SM, we use a zero width in all processes except for the following two:
\begin{equation}
pp\to e^+\nu_e jj, \quad pp\to e^+e^-jj,\label{eq:proc2l2j}
\end{equation} 
These two processes receive contributions from a subset of one-loop diagrams that feature an $s$-channel Higgs boson propagator. At NLO$_2$,  their Born diagrams do not include Higgs propagators. Therefore, even with a zero Higgs width, the virtual matrix elements remain integrable. In this sense, we use a non-zero Higgs width in eq.\eqref{eq:proc2l2j} solely to improve the behavior of the numerical integration. The PDFs used are the central ones from the {\tt LUXqed\_plus\_PDF4LHC15\_nnlo\_100} set~\cite{Manohar:2016nzj,Butterworth:2015oua}, which are accessible via \lhapdf. The $\alpha_s$ value at $91.1876$ GeV is reported in table \ref{tab:EWparaminput}. The renormalization and factorization scales are set as
\begin{equation}
\mu_R=\mu_F=\frac{H_T}{2}=\frac{1}{2}\sum_{i}{\sqrt{p_{T,i}^2+m_i^2}},
\end{equation}
where the sum runs over all final-state partons. I would like to point out that we do not vary scales and PDFs here, but if one wishes to do so, they can be easily obtained without additional computational cost in \mgamcshort.

\begin{table}[b]
\begin{center}
\begin{tabular}{cl|cl}\toprule
Parameter & value & Parameter & value
\\\midrule
$\alpha_{G_\mu}^{-1}$ & \texttt{132.292} & $\alpha_s(91.1876^2~\mathrm{GeV}^2)$ & $0.118$
\\
$\bar{M}_W$ & \texttt{80.358}  & $\bar{\Gamma}_{W}$ & \texttt{2.0843}
\\
$\bar{M}_Z$ & \texttt{91.1535}  & $\bar{\Gamma}_{Z}$ & \texttt{2.49427}
\\
$\bar{M}_t$ & \texttt{173.34}  & $\bar{\Gamma}_{t}$ & \texttt{1.3691}
\\
$\bar{M}_h$ & \texttt{125.0}  & $\bar{\Gamma}_{h}$ & \texttt{$0.00407$} 
\\\bottomrule
\end{tabular}
\caption{\label{tab:EWparaminput}
Parameters used in NLO EW calculations. All masses and widths are expressed in units of $\gev$, which has been suppressed for brevity. In some processes, widths are set to zero (see text for details).
}
\end{center}
\end{table}

In order to ensure that the cross sections are defined in an IR-safe manner, we must introduce fiducial cuts. We adopt minimal selection cuts on the final particles as follows:
\begin{itemize}
\item \textbf{Photon recombination}: As pointed out in section \ref{sec:taggedphotonlepton}, we define the dressed charged fermions through the photon recombination procedure. Specifically, all photons are recombined with the closed charged fermion according to $\Delta R_{f\gamma}$ if $\Delta R_{f\gamma}\leq 0.1$. 
\item \textbf{Cuts on charged dressed leptons}: Charged dressed leptons, denoted as $\ell^\pm$, are required to satisfy $p_T(\ell^\pm)>10$ GeV and $|\eta(\ell^\pm)|<2.5$. For any pair of oppositely signed and same-flavor leptons, we impose the cuts $\Delta R_{\ell^+\ell^-}>0.4$ and $m_{\ell^+\ell^-}>30$ GeV.
\item \textbf{Cuts on jets}: Jets, denoted as $j$, are reconstructed from gluons, (dressed) (anti)quarks, and (uncombined) photons using the anti-$k_T$ clustering algorithm~\cite{Cacciari:2008gp} in \fastjet~\cite{Cacciari:2011ma,Cacciari:2005hq} with the jet radius of $R=0.4$. They must satisfy $p_T(j)>30$ GeV and $|\eta(j)|<4.5$.
\end{itemize}

\begin{figure}[htbp!]
\centering
\includegraphics[width=0.9\textwidth]{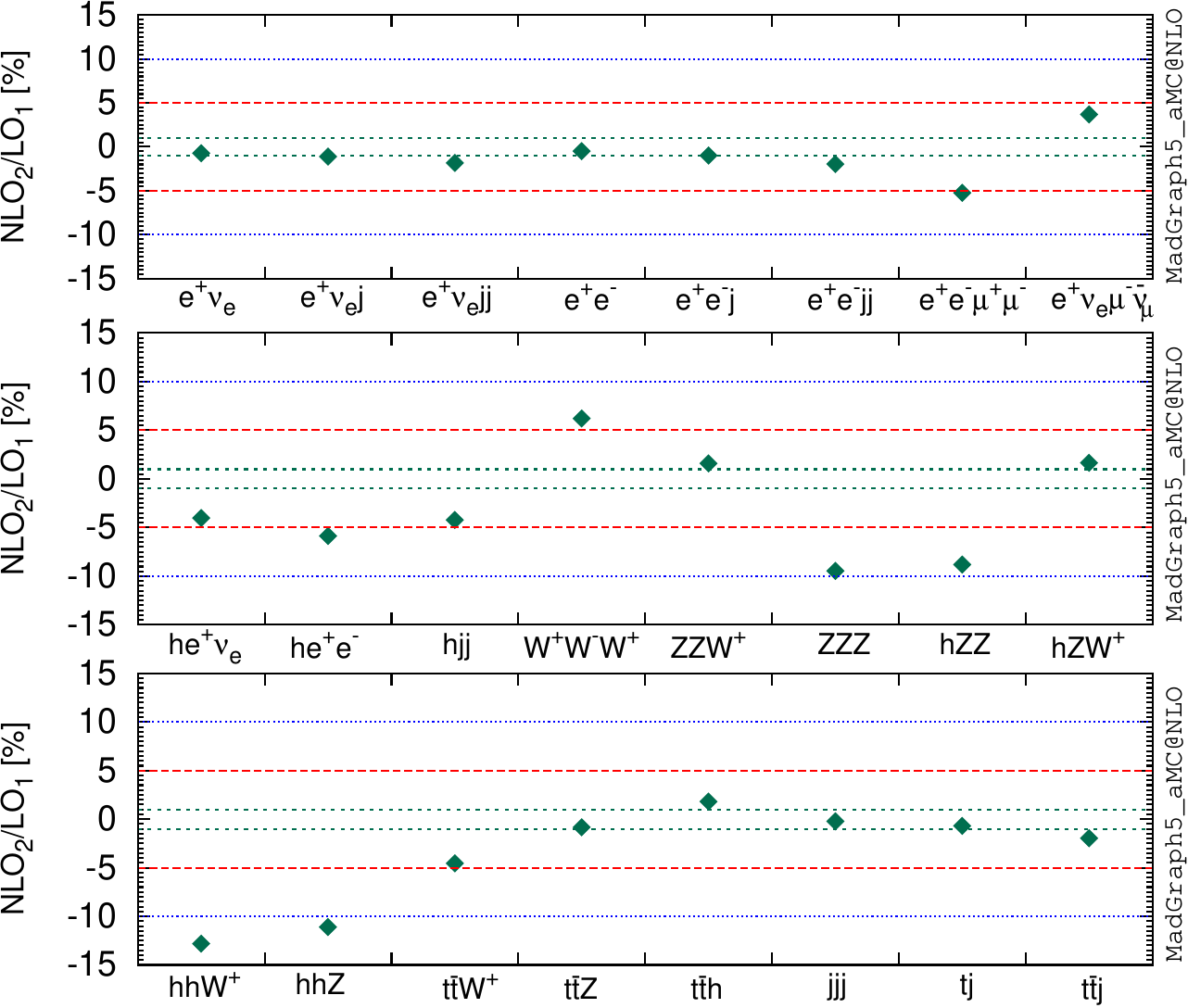}
\caption{A summary of NLO EW corrections relative to LO$_1$ cross sections for $24$ LHC processes. The green, red, and blue dashed horizontal lines represent corrections of $\pm1\%$, $\pm5\%$, and $\pm 10\%$, respectively. The specific values are also provided in the last column of table \ref{tab:NLOEWres}. \label{fig:NLOEWres}}
\end{figure}

\begin{figure}[htbp!]
\centering
\includegraphics[width=0.9\textwidth]{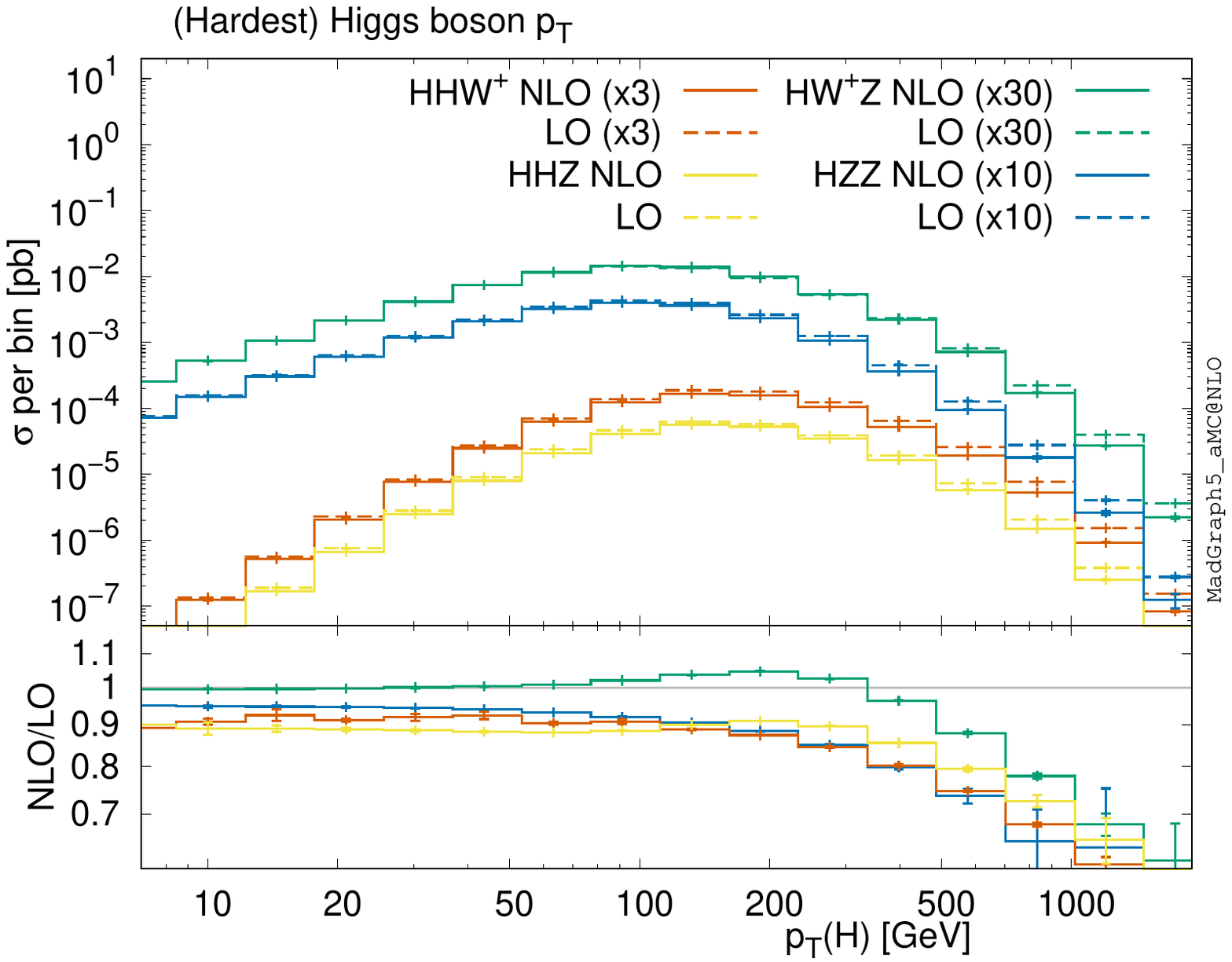}
\caption{Transverse momentum distributions of the hardest Higgs boson in the processes $pp\to hhW^+$ (red), $pp\to hhZ$ (yellow), $pp\to hW^+Z$ (green), and $pp\to hZZ$ (blue) at LO$_1$ (dashed) and NLO$_2$+LO$_1$ (solid).  The lower inset displays the ratios $(\Sigma_{\mathrm{NLO_2}}+\Sigma_{\mathrm{LO_1}})/\Sigma_{\mathrm{LO_1}}$. Some of the histograms in the main frame are rescaled, as indicated in the legend, to enhance visibility. The plot is from ref.~\cite{Frederix:2018nkq}. \label{fig:NLOEWdiff}}
\end{figure}

We report the leading LO (LO$_1$) and second-leading NLO (NLO$_2$, NLO EW) integrated cross sections for $24$ processes in table \ref{tab:NLOEWres} with the above setup. The relative fractions of $\Sigma_{\mathrm{NLO}_2}/\Sigma_{\mathrm{LO}_1}$ can be found in the last column of table \ref{tab:NLOEWres}, which are visualized in figure \ref{fig:NLOEWres}.  The generation syntax for computing the reported cross sections in \mgamcshort\ is provided in the second column of table \ref{tab:NLOEWres}. For inclusive cross sections, NLO EW corrections typically reduce LO$_1$ cross sections by a few percent (usually below $5\%$) for most processes. This reduction reflects the fact that NLO$_2$ corrections are largely negative due to virtual contributions for fairly inclusive observables. The absolute values of the EW corrections tend to increase with the number of final-state particles, and it is observed that corrections are generally larger for bosons than for fermions. Some $W^\pm$ production processes are exceptions, exhibiting positive EW corrections. For example, $pp\to W^+W^-W^+$ and $pp\to hZW^+$ receive $+6.2\%$ and $+1.6\%$ corrections, respectively, driven by photon-quark induced real emissions. A similar observation applies to the process $pp\to e^+\nu_e \mu^-\bar{\nu}_\mu$ ($\Sigma_{\mathrm{NLO}_2}/\Sigma_{\mathrm{LO}_1}=+3.67\%$), which is dominated by $W^+W^-$ resonances. Conversely, the largest negative corrections are seen in Higgs and/or $Z$-related processes, with corrections on the order of $-10\%$, and the largest being $pp\to hhW^+$, which sees corrections nearing $-13\%$. It may be unsurprising that EW corrections are larger in processes with greater momentum transfer and more final-state bosons, given the influence of EW Sudakov logarithms and photon-induced processes. As a result, the size of EW corrections to integrated cross sections may depend heavily on the imposed  kinematic cuts and is typically much larger in differential distributions--especially in high-energy tails, as demonstrated in figure \ref{fig:NLOEWdiff}. I refrain from presenting additional differential distributions here and instead refer interested readers to the discussions in section 6.2 of ref.~\cite{Frederix:2018nkq}. Finally, I point out that only the physical cross sections, rather than the relative fractions $\Sigma_{\mathrm{NLO}_2}/\Sigma_{\mathrm{LO}_1}$, are physical. The values for $\Sigma_{\mathrm{LO}_1}$ and corrections $\Sigma_{\mathrm{NLO}_2}$ can vary depending on the choice of renormalization scheme (\eg, $\alpha(M_Z^2)$ versus $G_\mu$) and/or renormalization and factorization scales. These differences are reduced only when comparing absolute cross sections from LO to NLO, as demonstrated in ref.~\cite{Frixione:2015zaa}. While the reduction in scale uncertainty is well-known in NLO QCD computations, the dependence on the $\alpha$ renormalization scheme is unique to higher-order EW computations.

\begin{landscape}
\begin{table}
\begin{center}
\begin{small}
\begin{tabular}{llr@{$\,\,\pm\,\,$}lr@{$\,\,\pm\,\,$}lr@{$\,\,\pm\,\,$}l}
\toprule
Process & Generation syntax & \multicolumn{4}{c}{Cross section (in pb)} &  \multicolumn{2}{c}{Correction (in \%)}\\
 && \multicolumn{2}{c}{$\Sigma_{\mathrm{LO}_1}$} &  \multicolumn{2}{c}{$\Sigma_{\mathrm{LO}_1}+\Sigma_{\mathrm{NLO}_2}$} & \multicolumn{2}{c}{$\Sigma_{\mathrm{NLO}_2}/\Sigma_{\mathrm{LO}_1}$} \\
\midrule
$pp \to e^+ \nu_e$ & \verb|p p > e+ ve aS=0 aEW=2 [QED]| & $5.2498 $&$ 0.0005\,\cdot 10^{3}$ & $5.2113 $&$ 0.0006\,\cdot 10^{3}$ & $-0.73 $&$ 0.01$\\
$pp \to e^+ \nu_e j$ & \verb|p p > e+ ve j aS=1 aEW=2 [QED]| & $9.1468 $&$ 0.0012\,\cdot 10^{2}$ & $9.0449 $&$ 0.0014\,\cdot 10^{2}$ & $-1.11 $&$ 0.02$\\
$pp \to e^+ \nu_e jj$ & \verb|p p > e+ ve j j aS=2 aEW=2 [QED]| & $3.1562 $&$ 0.0003\,\cdot 10^{2}$ & $3.0985 $&$ 0.0005\,\cdot 10^{2}$ & $-1.83 $&$ 0.02$\\
$pp \to e^+ e^-$ & \verb|p p > e+ e- aS=0 aEW=2 [QED]| & $7.5367 $&$ 0.0008\,\cdot 10^{2}$ & $7.4997 $&$ 0.0010\,\cdot 10^{2}$ & $-0.49 $&$ 0.02$\\
$pp \to e^+ e^-j$ & \verb|p p > e+ e- j aS=1 aEW=2 [QED]| & $1.5059 $&$ 0.0001\,\cdot 10^{2}$ & $1.4909 $&$ 0.0002\,\cdot 10^{2}$ & $-1.00 $&$ 0.02$\\
$pp \to e^+ e^-jj$ & \verb|p p > e+ e- j j aS=2 aEW=2 [QED]| & $5.1424 $&$ 0.0004\,\cdot 10^{1}$ & $5.0410 $&$ 0.0007\,\cdot 10^{1}$ & $-1.97 $&$ 0.02$\\
$pp \to e^+ e^- \mu^+ \mu^-$ & \verb|p p > e+ e- mu+ mu- aS=0 aEW=4 [QED]| & $1.2750 $&$ 0.0000\,\cdot 10^{-2}$ & $1.2083 $&$ 0.0001\,\cdot 10^{-2}$ & $-5.23 $&$ 0.01$\\
$pp \to e^+ \nu_e \mu^- \bar{\nu}_{\mu}$ & \verb|p p > e+ ve mu- vm~ aS=0 aEW=4 [QED]| & $5.1144 $&$ 0.0007\,\cdot 10^{-1}$ & $5.3019 $&$ 0.0009\,\cdot 10^{-1}$ & $+3.67 $&$ 0.02$\\
$pp \to h e^+ \nu_e$ & \verb|p p > h e+ ve aS=0 aEW=3 [QED]| & $6.7643 $&$ 0.0001\,\cdot 10^{-2}$ & $6.4914 $&$ 0.0012\,\cdot 10^{-2}$ & $-4.03 $&$ 0.02$\\
$pp \to h e^+ e^-$ & \verb|p p > h e+ e- aS=0 aEW=3 [QED]| & $1.4554 $&$ 0.0001\,\cdot 10^{-2}$ & $1.3700 $&$ 0.0002\,\cdot 10^{-2}$ & $-5.87 $&$ 0.02$\\
$pp \to h j j$ & \verb|p p > h j j aS=0 aEW=3 [QED]| & $2.8268 $&$ 0.0002\,\cdot 10^{0}$ & $2.7075 $&$ 0.0003\,\cdot 10^{0}$ & $-4.22 $&$ 0.01$\\
$pp \to W^+W^-W^+$ & \verb|p p > w+ w- w+ aS=0 aEW=3 [QED]| & $8.2874 $&$ 0.0004\,\cdot 10^{-2}$ & $8.8017 $&$ 0.0012\,\cdot 10^{-2}$ & $+6.21 $&$ 0.02$\\
$pp \to Z Z W^+$ & \verb|p p > z z w+ aS=0 aEW=3 [QED]| & $1.9874 $&$ 0.0001\,\cdot 10^{-2}$ & $2.0189 $&$ 0.0003\,\cdot 10^{-2}$ & $+1.58 $&$ 0.02$\\
$pp \to ZZZ$ & \verb|p p > z z z aS=0 aEW=3 [QED]| & $1.0761 $&$ 0.0001\,\cdot 10^{-2}$ & $0.9741 $&$ 0.0001\,\cdot 10^{-2}$ & $-9.47 $&$ 0.02$\\
$pp \to hZZ$ & \verb|p p > h z z aS=0 aEW=3 [QED]| & $2.1005 $&$ 0.0003\,\cdot 10^{-3}$ & $1.9155 $&$ 0.0003\,\cdot 10^{-3}$ & $-8.81 $&$ 0.02$\\
$pp \to hZW^+$ & \verb|p p > h z w+ aS=0 aEW=3 [QED]| & $2.4408 $&$ 0.0000\,\cdot 10^{-3}$ & $2.4809 $&$ 0.0005\,\cdot 10^{-3}$ & $+1.64 $&$ 0.02$\\
$pp \to hhW^+$ & \verb|p p > h h w+ aS=0 aEW=3 [QED]| & $2.7827 $&$ 0.0001\,\cdot 10^{-4}$ & $2.4259 $&$ 0.0027\,\cdot 10^{-4}$ & $-12.82 $&$ 0.10$\\
$pp \to hhZ$ & \verb|p p > h h z aS=0 aEW=3 [QED]| & $2.6914 $&$ 0.0003\,\cdot 10^{-4}$ & $2.3926 $&$ 0.0003\,\cdot 10^{-4}$ & $-11.10 $&$ 0.02$\\
$pp \to t \bar{t} W^+$ & \verb|p p > t t~ w+ aS=2 aEW=1 [QED]| & $2.4119 $&$ 0.0003\,\cdot 10^{-1}$ & $2.3025 $&$ 0.0003\,\cdot 10^{-1}$ & $-4.54 $&$ 0.02$\\
$pp \to t \bar{t} Z$ & \verb|p p > t t~ z aS=2 aEW=1 [QED]| & $5.0456 $&$ 0.0006\,\cdot 10^{-1}$ & $5.0033 $&$ 0.0007\,\cdot 10^{-1}$ & $-0.84 $&$ 0.02$\\
$pp \to t \bar{t} h$ & \verb|p p > t t~ h aS=2 aEW=1 [QED]| & $3.4480 $&$ 0.0004\,\cdot 10^{-1}$ & $3.5102 $&$ 0.0005\,\cdot 10^{-1}$ & $+1.81 $&$ 0.02$\\
$pp \to jjj$ & \verb|p p > j j j aS=3 aEW=0 [QED]| & $7.9639 $&$ 0.0010\,\cdot 10^{6}$ & $7.9472 $&$ 0.0011\,\cdot 10^{6}$ & $-0.21 $&$ 0.02$\\
$pp \to t j$ & \verb|p p > t j aS=0 aEW=2 [QED]| & $1.0613 $&$ 0.0001\,\cdot 10^{2}$ & $1.0539 $&$ 0.0001\,\cdot 10^{2}$ & $-0.70 $&$ 0.02$\\
$pp \to t \bar{t} j$ & \verb|p p > t t j aS=3 aEW=0 [QED]| & $3.0277 $&$ 0.0003\,\cdot 10^{2}$ & $2.9683 $&$ 0.0004\,\cdot 10^{2}$ & $-1.96 $&$ 0.02$\\
\bottomrule
\end{tabular}
\caption{\label{tab:NLOEWres}\protect
The integrated cross sections at LO$_1$ (third column) and LO$_1$+NLO$_2$ (fourth column) for various processes. The second column provides the generation syntax in \mgamcshort. The fifth column presents the NLO EW corrections relative to LO$_1$. All uncertainties are due to Monte Carlo numerical integration errors. The table is from ref.~\cite{Frederix:2018nkq}.}
\end{small}
\end{center}
\end{table}
\end{landscape}

\begin{figure}[htbp!]
\centering
\includegraphics[width=0.9\textwidth]{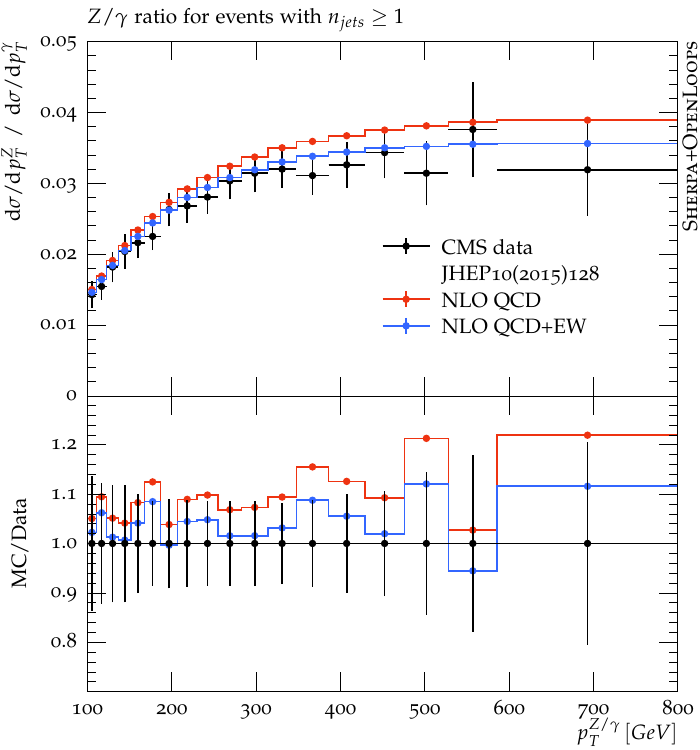}
\caption{Transverse momentum distribution of the cross section ratio for the processes $pp\to Z+\mathrm{jets}$ and $pp\to \gamma+\mathrm{jets}$ at 8 TeV. The CMS data is sourced from ref.~\cite{CMS:2015onn}, and the plot is taken from ref.~\cite{Blumenschein:2018gtm}. \label{fig:Zgammaratio}}
\end{figure}

Finally, one may wonder whether current LHC data already demonstrate the necessity of systematically including NLO EW corrections. The answer is yes. An example is provided in ref.~\cite{Blumenschein:2018gtm}, which compares the NLO QCD+EW calculation, performed by the \sherpa+\openloops\ group, for the cross sections ratio $\sigma(pp\to Z+\mathrm{jets})/\sigma(pp\to \gamma+\mathrm{jets})$ with CMS measurement~\cite{CMS:2015onn} in $pp$ collisions at $8$ TeV. The comparison, shown in figure \ref{fig:Zgammaratio}, is done for the ratio as a function of the transverse momentum of the $Z$ or $\gamma$. The results indicate an improved agreement between data and theory when EW corrections are included. Thus, the inclusion of EW corrections at the LHC is no longer just an academic exercise.

\section{Subleading NLO corrections}

In this section, I will discuss the subleading NLO corrections, \ie, NLO$_i$ with $i>2$. Based on the naive power counting of the couplings $\alpha_s$ and $\alpha$, these corrections are expected to be subdominant compared to NLO$_1$ (NLO QCD) and NLO$_2$ (NLO EW) corrections (\cf\ eq.\eqref{eq:hierachy0}). Here, I will demonstrate when subleading NLO terms should be considered. Specifically, I will highlight a few interesting cases where the naive hierarchy in eq.\eqref{eq:hierachy0} breaks down.

\subsection{Subleading NLO corrections are small}

My first example is Higgs boson production in association with a top quark pair, $t\bar{t}$, at the LHC:
\begin{equation}
pp\to t\bar{t}h.
\end{equation}
This process is crucial for directly probing the top-Higgs Yukawa coupling $y_t$ and has been observed by both ATLAS~\cite{ATLAS:2018mme} and CMS~\cite{CMS:2018uxb}. On the precision calculation side, the NLO QCD corrections~\cite{Beenakker:2001rj,Beenakker:2002nc,Dawson:2002tg,Dawson:2003zu} have been known for more than two decades. NLO QCD matching to PSMC~\cite{Frederix:2011zi,Garzelli:2011vp,Hartanto:2015uka} has also been available for quite some time. NLO weak~\cite{Frixione:2014qaa} and EW~\cite{Zhang:2014gcy,Frixione:2015zaa} corrections were calculated early in the development of NLO EW automation.  Further refinements to the cross section include off-shell effects~\cite{Denner:2015yca,Denner:2016wet,Stremmer:2021bnk,Denner:2020orv,Bevilacqua:2022twl}, soft-gluon threshold logarithm resummation~\cite{Kulesza:2015vda,Broggio:2015lya,Broggio:2016lfj,Kulesza:2017ukk,vanBeekveld:2020cat}, Coulomb resummation~\cite{Ju:2019lwp}, and approximated NNLO QCD corrections~\cite{Catani:2022mfv}. 

Using the same setup as in section \ref{sec:NLOEWres}, the complete NLO result for this process was first obtained in ref.~\cite{Frederix:2018nkq} with the following \mgamcshort\ generation command:
\vskip 0.25truecm

\noindent
~~\prompt\ {\tt ~generate p p > t t\~{} h aS=2 aEW=3 [QCD QED]}
\vskip 0.25truecm
\noindent
Using the notations introduced in section \ref{eq:EWgeneral}, we have $k_0=3$, $c_s(k_0)=0$, $c(k_0)=1$, and $\Delta(k_0)=2$, meaning the process includes three LO terms
\begin{equation}
\Sigma_{\mathrm{LO}_1}~(\mathcal{O}(\alpha_s^2\alpha)),\quad \Sigma_{\mathrm{LO}_2}~(\mathcal{O}(\alpha_s\alpha^2)),\quad \Sigma_{\mathrm{LO}_3}~(\mathcal{O}(\alpha^3)),
\end{equation}
and four NLO terms
\begin{equation}
\Sigma_{\mathrm{NLO}_1}~(\mathcal{O}(\alpha_s^3\alpha)), \quad \Sigma_{\mathrm{NLO}_2}~(\mathcal{O}(\alpha_s^2\alpha^2)),\quad \Sigma_{\mathrm{NLO}_3}~(\mathcal{O}(\alpha_s\alpha^3)),\quad \Sigma_{\mathrm{NLO}_4}~(\mathcal{O}(\alpha^4)).
\end{equation}
The inclusive cross sections at each order are displayed in the second column of table \ref{tab:ttBNLO} for $pp$ collisions at 13 TeV. As expected from the simple power counting of the couplings, all contributions except LO$_1$ and NLO$_1$ are small. The NLO terms follow the hierarchy in equation \eqref{eq:hierachy0}. However, the LO terms deviate from this hierarchy because the LO$_2$ term is too small. This results from the lack of interference between the quark-antiquark-initiated QCD and QED diagrams due to color. The only partonic channels contributing at LO$_2$ are $\gamma g \to t\bar{t}h$ and $b\bar{b}\to t\bar{t}h$, where the photon PDF is suppressed by $\mathcal{O}(\alpha)$, as explained in section \ref{sec:photonleptonPDF}, and the bottom quark PDF is suppressed by its heavy mass. On the other hand, at NLO, color no longer plays a similar role for the quark-antiquark channels. This observation remains true at the differential level (see, \eg, figure 13 in ref.~\cite{Frederix:2018nkq} or the results for dijet hadroproduction in ref.~\cite{Frederix:2016ost}).

\begin{table}
\begin{center}
%\begin{small}
\begin{tabular}{lr@{$\,\,\pm\,\,$}lr@{$\,\,\pm\,\,$}lr@{$\,\,\pm\,\,$}l}
\toprule
 & \multicolumn{2}{c}{$pp \to t \bar{t} h$} & \multicolumn{2}{c}{$pp \to t \bar{t} Z$} & \multicolumn{2}{c}{$pp \to t \bar{t} W^+$} \\
\midrule
$\Sigma_{\mathrm{LO}_1}$ ($10^{-1}$ pb) & \multicolumn{2}{c}{$3.4483 \pm 0.0003$}  & \multicolumn{2}{c}{$5.0463 \pm 0.0003$} & \multicolumn{2}{c}{$2.4116 \pm 0.0001$} \\
$\Sigma_{\mathrm{LO}_2}/\Sigma_{\mathrm{LO}_1}$ ($\%$) & $+0.406 $&$ 0.001$ & $-0.691 $&$ 0.001$ & $+0.000 $&$ 0.000$  \\
$\Sigma_{\mathrm{LO}_3}/\Sigma_{\mathrm{LO}_1}$ ($\%$) & $+0.702 $&$ 0.001$ & $+2.259 $&$ 0.001$ & $+0.962 $&$ 0.000$ \\
$\Sigma_{\mathrm{NLO}_1}/\Sigma_{\mathrm{LO}_1}$ ($\%$) & $+28.847 $&$ 0.020$ & $+44.809 $&$ 0.028$ & $+49.504 $&$ 0.015$\\
$\Sigma_{\mathrm{NLO}_2}/\Sigma_{\mathrm{LO}_1}$ ($\%$) & $+1.794 $&$ 0.005$ & $-0.846 $&$ 0.004$ & $-4.541 $&$ 0.003$\\
$\Sigma_{\mathrm{NLO}_3}/\Sigma_{\mathrm{LO}_1}$ ($\%$) & $+0.483 $&$ 0.008$ & $+0.845 $&$ 0.003$ & $+12.242 $&$ 0.014$\\
$\Sigma_{\mathrm{NLO}_4}/\Sigma_{\mathrm{LO}_1}$ ($\%$) & $+0.044 $&$ 0.000$ & $-0.082 $&$ 0.000$ & $+0.017 $&$ 0.003$ \\
\bottomrule
\end{tabular}
\caption{\label{tab:ttBNLO} Cross sections for the $t\bar{t}+B$ processes, where $B=h,Z,W^+$, are calculated using the setup described in section~\ref{sec:NLOEWres}. The quoted uncertainties are purely statistical, arising from the Monte Carlo integration over the phase space.
}
%\end{small}
\end{center}
\end{table}

\subsection{Subleading NLO corrections become comparable}

An example of subleading NLO terms comparable to the NLO EW term I present here is the $Z$ boson production in association with a top quark pair:
\begin{equation}
pp\to t\bar{t}Z.
\end{equation}
This process often serves as a background for the $t\bar{t}h$ process at the LHC. In \mgamcshort, the complete NLO calculation for this process can be generated using the following command: 
\vskip 0.25truecm

\noindent
~~\prompt\ {\tt ~generate p p > t t\~{} z aS=2 aEW=3 [QCD QED]}
\vskip 0.25truecm
\noindent
Like the $pp\to t\bar{t}h$ process, there are 3 LO and 4 NLO contributions, with their inclusive cross sections shown in the third column of table \ref{tab:ttBNLO}. The complete NLO result was first reported in ref.~\cite{Frederix:2018nkq}. As in the $t\bar{t}h$ case, the fact that only $\gamma g$- ($\gamma g \to t\bar{t}Z$) and $b\bar{b}$-initiated ($b\bar{b}\to t\bar{t}Z$) partonic channels contribute to LO$_2$ makes $\Sigma_{\mathrm{LO}_2}$ particularly small. However, unlike in the $t\bar{t}h$ case, we observe that $\Sigma_{\mathrm{NLO}_3}$ is of the same order as $\Sigma_{\mathrm{NLO}_2}$ but with an opposite sign. This results from accidental cancellations at NLO$_2$ in the current setup, although such cancellations depend on the renormalization scheme. For example, using the $\alpha(M_Z^2)$ scheme, $\Sigma_{\mathrm{NLO}_2}/\Sigma_{\mathrm{LO}_1}$ decreases to around $-4\%$, as shown in ref.~\cite{Frixione:2015zaa}. 

There are certainly many similar examples in the literature where subleading NLO terms cancel with NLO EW corrections. For instance, in $pp\to t\bar{t}\gamma_{\mathrm{iso}}$~\cite{Pagani:2021iwa}, the sum of LO$_2$, LO$_3$, and NLO$_3$ largely cancels the impact of the NLO$_2$ term (see able 2 in ref.~\cite{Pagani:2021iwa}). Another example worth mentioning is triple jet hadroproduction~\cite{Reyer:2019obz}, where  $\Sigma_{\mathrm{NLO}_2}$ was found to largely cancel with  $\Sigma_{\mathrm{LO}_2}$, but only when the leading jet has a transverse momentum greater than 2 TeV (see table 2 in ref.~\cite{Reyer:2019obz}). I must emphasize again that such cancellations are fragile and may vary depending on the specific setup of the calculations. Therefore, it is always advisable to check all subleading LO and NLO contributions whenever possible.

\subsection{Subleading NLO corrections surpass NLO EW corrections}

An interesting example that showcases how subleading NLO terms can be larger than NLO EW corrections is the process $pp\to t\bar{t}W^+$. The complete NLO can be generated as:
\vskip 0.25truecm

\begin{figure}[htbp!]
\centering
\includegraphics[width=0.9\textwidth]{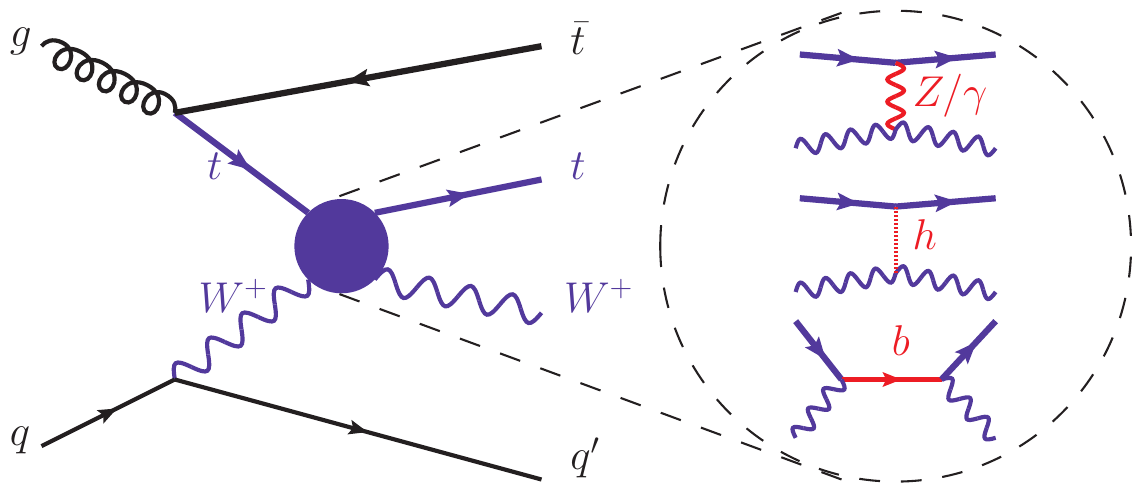}
\caption{The $tW^+\to tW^+$ scattering subgraphs contributing to $pp\to t\bar{t}W^+$ at NLO$_3$. The inset shows the possible subdiagrams for $tW^+\to tW^+$. \label{fig:tWscattering}}
\end{figure}

\noindent
~~\prompt\ {\tt ~generate p p > t t\~{} w+ aS=2 aEW=3 [QCD QED]}
\vskip 0.25truecm
\noindent
in \mgamcshort. The charge-conjugate process $pp\to t\bar{t}W^-$ can be calculated similarly. In this case, there are 3 LO and 4 NLO terms to be computed. NLO EW and complete NLO corrections for this process were first reported in ref.~\cite{Frixione:2015zaa} and ref.~\cite{Frederix:2017wme}, respectively. The inclusive cross sections for $pp\to t\bar{t}W^+$ are listed in the last column of table \ref{tab:ttBNLO}, using the same setup defined in section \ref{sec:NLOEWres}. For this process, there are no $\gamma g$ or $b\bar{b}$ initiated subprocesses, so $\Sigma_{\mathrm{LO}_2}$ is exactly zero. Additionally, $\Sigma_{\mathrm{NLO}_3}$ is significantly larger than $\Sigma_{\mathrm{NLO}_2}$ in absolute terms. The former contributes approximately $+12\%$ relative to LO$_1$, whereas the latter contributes only $-4.5\%$ of $\Sigma_{\mathrm{LO}_1}$. $\Sigma_{\mathrm{NLO}_2}$ is of typical size for an NLO EW correction, while $\Sigma_{\mathrm{NLO}_3}$ is anomalously large. This can be understood by the opening of new channels at NLO$_3$. Specifically, $tW^+\to tW^+$ subdiagrams in the real emission contribution give rise to the process $pp\to t\bar{t}W^+j$ at NLO$_3$ (depicted in figure \ref{fig:tWscattering}), which are enhanced as highlighted in ref.~\cite{Dror:2015nkp}. These new channels are physical, and thus the enhancement persists regardless of changes to the calculational setup.

Large quantum corrections due to new channels opening at higher order are not unique to NLO EW calculations. In fact, this phenomenon occurs in many known examples of higher-order QCD corrections. A classical example is $pp\to V_1V_2$ with $V_1V_2=\gamma\gamma,\gamma Z, ZZ, W^+W^-$. The $gg\to V_1V_2$ partonic channels only contribute at NNLO QCD and beyond, but not at LO or NLO, often leading to sizable NNLO QCD corrections. In some cases, new channels at higher order can result in giant $K$ factors, which we will discuss further in section \ref{sec:GiantK}.

\subsection{Subleading NLO corrections as the primary quantum corrections}

Finally, there are instances when we need to reverse the hierarchy described in  eq.\eqref{eq:hierachy0}:
\begin{equation}
\Sigma_{{\mathrm{N}^p\mathrm{LO}}_i}\lesssim \Sigma_{{\mathrm{N}^p\mathrm{LO}}_{i+1}},\quad \forall\ i, p.\label{eq:hierachy2}
\end{equation}
A known example of such kind is the vector-boson scattering (VBS) processes~\cite{Biedermann:2016yds,Biedermann:2017bss,Denner:2019tmn,Denner:2020zit,Denner:2022pwc,Denner:2024tlu}, particularly when dedicated VBS event selections are applied. Let us consider the same-sign $W^+W^+$ VBS process as a concrete example. Its Born process is defined as:
\begin{equation}
pp\to e^+\nu_e \mu^+\nu_\mu jj.
\end{equation}
In this case, we have $k_0=6$, $c_s(k_0)=0$, $c(k_0)=4$, and $\Delta(k_0)=2$. Consequently, there are three LO terms:
\begin{equation}
\Sigma_{\mathrm{LO}_1}~(\mathcal{O}(\alpha_s^2\alpha^4)),\quad \Sigma_{\mathrm{LO}_2}~(\mathcal{O}(\alpha_s\alpha^5)),\quad \Sigma_{\mathrm{LO}_3}~(\mathcal{O}(\alpha^6)),
\end{equation}
and four NLO terms:
\begin{equation}
\Sigma_{\mathrm{NLO}_1}~(\mathcal{O}(\alpha_s^3\alpha^4)), \quad \Sigma_{\mathrm{NLO}_2}~(\mathcal{O}(\alpha_s^2\alpha^5)),\quad \Sigma_{\mathrm{NLO}_3}~(\mathcal{O}(\alpha_s\alpha^6)),\quad \Sigma_{\mathrm{NLO}_4}~(\mathcal{O}(\alpha^7)).
\end{equation}
This process can also be generated within \mgamcshort\ as follows:
\vskip 0.25truecm
%\noindent
%~~\prompt\ {\tt ~define p = p e+ e- mu+ mu- ta+ ta-}
%\vskip 0.25truecm

%\noindent
%~~\prompt\ {\tt ~define j = p}
%\vskip 0.25truecm

\begin{figure}[htbp!]
\centering
\subfloat[]{\includegraphics[width=0.32\textwidth]{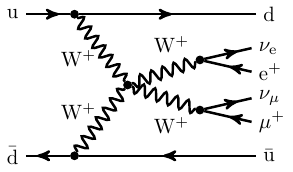}\label{fig:VBSWWdiaga}}
\subfloat[]{\includegraphics[width=0.32\textwidth]{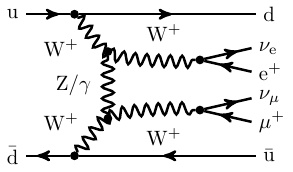}\label{fig:VBSWWdiagb}}
\subfloat[]{\includegraphics[width=0.32\textwidth]{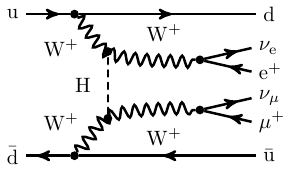}\label{fig:VBSWWdiagc}}\\
\subfloat[]{\includegraphics[width=0.32\textwidth]{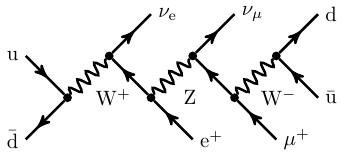}\label{fig:VBSWWdiagd}}
\subfloat[]{\includegraphics[width=0.32\textwidth]{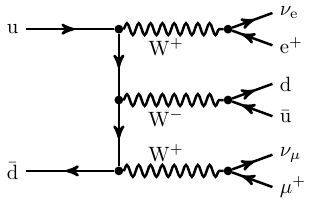}\label{fig:VBSWWdiage}}
\subfloat[]{\includegraphics[width=0.32\textwidth]{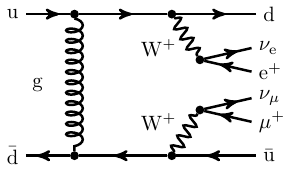}\label{fig:VBSWWdiagf}}
\caption{Representative Born-level Feynman diagrams for same-sign $W^+W^+$ production at the LHC. The diagrams are taken from ref.~\cite{Biedermann:2017bss}. \label{fig:VBSWWdiags}}
\end{figure}

\noindent
~~\prompt\ {\tt ~generate p p > e+ ve mu+ vm j j aS=2 aEW=6 [QCD QED]}
\vskip 0.25truecm
\noindent
NLO QCD corrections have been calculated in ref.~\cite{Ballestrero:2018anz}. However, the complete NLO computation is quite CPU intensive. Therefore, I will utilize the results from ref.~\cite{Biedermann:2017bss}, which were calculated by the \recola\ group. Sample Born-level Feynman diagrams for the partonic subprocess $u\bar{d}\to e^+\nu_e\mu^+\nu_\mu \bar{u}d$ are displayed in figure \ref{fig:VBSWWdiags}. The first five diagrams contribute to $\ampnt_{(0,6)}(r)\propto e^6$, while the last diagram contributes to $\ampnt_{(2,4)}(r)\propto g_s^2 e^4$. The first three diagrams in the top row represent the genuine VBS process of interest. Specifically, their longitudinal modes probe the unitarity of the scattering amplitude, which is protected by the Higgs mechanism in the SM, or alternatively, the unitarity violation effects introduced by any BSM theories. The diagrams in figures \ref{fig:VBSWWdiagd} and \ref{fig:VBSWWdiage} lead to a three-vector-boson final state, which we wish to suppress. Additionally, contributions from QCD-induced diagrams, such as that shown in figure \ref{fig:VBSWWdiagf}, should also be vetoed. The latter two types of diagrams are referred to as the EW-induced and QCD-induced irreducible backgrounds, respectively. The signal we aim to probe originates from the VBS-like topologies.

In order to enhance signal purity, typical VBS event selections must be applied. Following section 3.1 in ref.~\cite{Biedermann:2017bss}, these selections are:
\begin{itemize}
\item \textbf{Photon recombination}: Dressed charged fermions are defined through a photon recombination procedure. All photons are recombined with the nearest charged fermion if  $\Delta R_{f\gamma}\leq 0.1$ and $|\eta(\gamma,f)|<5$. 
\item \textbf{Cuts on charged dressed leptons}: Charged dressed leptons, denoted as $\ell^+=e^+,\mu^+$, must satisfy $p_T(\ell^+)>20$ GeV, $|y(\ell^+)|<2.5$, and $\Delta R_{\ell^+\ell^+}>0.3$.
\item \textbf{Cuts on neutrinos}: The missing transverse energy is required to fulfill $E_{T,\mathrm{miss}}=|\vec{p}_{T,\mathrm{miss}}|>40$ GeV, where $\vec{p}_{T,\mathrm{miss}}$ is the vector sum of the transverse momenta of the two neutrinos.
\item \textbf{Cuts on jets}: Jets $j$ are reconstructed from gluons and (dressed) (anti)quarks using the anti-$k_T$ clustering algorithm with a jet radius of $R=0.4$. A QCD parton system after recombination is classified as a jet if it meets the criteria $p_T(j)>30$ GeV, $|y(j)|<4.5$, and $\Delta R_{j\ell^+}>0.3$.
\item \textbf{VBS cuts}: The two leading jets must satisfy $m_{jj}>500$ GeV and $|\Delta y_{jj}|>2.5$.
\end{itemize}
In particular, the last VBS cuts enhance the signal purity by suppressing both the EW-induced and QCD-induced backgrounds (see, \eg, ref.~\cite{Ballestrero:2018anz} for a detailed study). For the remainder of the calculational setup, I refer to section 3.1 in ref.~\cite{Biedermann:2017bss}. 

\begin{table}[htpb!]
\begin{center} \small
\begin{tabular}{|c|c|c|c|c|c|c|}
\hline
$\Sigma_{\mathrm{LO}_3}$ & $\Sigma_{\mathrm{LO}_1}/\Sigma_{\mathrm{LO}_3}$ & $\Sigma_{\mathrm{LO}_2}/\Sigma_{\mathrm{LO}_3}$ & $\Sigma_{\mathrm{NLO}_1}/\Sigma_{\mathrm{LO}_3}$ & $\Sigma_{\mathrm{NLO}_2}/\Sigma_{\mathrm{LO}_3}$ & $\Sigma_{\mathrm{NLO}_3}/\Sigma_{\mathrm{LO}_3}$ & $\Sigma_{\mathrm{NLO}_4}/\Sigma_{\mathrm{LO}_3}$ \\\hline
 $1.4178$ fb & $+12.15~\%$ & $+3.40~\%$ & $-0.44~\%$ & $-0.02~\%$ & $-4.01~\%$ & $-15.30~\%$ \\
\hline
\end{tabular}
\caption{Cross sections for the $pp\to e^+\nu_e \mu^+\nu_\mu jj$ process in $pp$ collisions at 13 TeV. The values are taken from ref.~\cite{Biedermann:2017bss}. \label{tab:WWNLOxs}}
\end{center}
\end{table}

Due to the event selection cuts, $\Sigma_{\mathrm{LO}_3}$ becomes the dominant LO contribution, accounting for $87\%$ of the total LO cross section as shown in table \ref{tab:WWNLOxs}.  In contrast, $\Sigma_{\mathrm{LO}_1}$ contributes approximately $10\%$ of the LO cross section, while $\Sigma_{\mathrm{LO}_2}$ is suppressed due to the color structure of the interferences. These interferences arise only when diagrams with different quark flows between the initial and final states are multiplied together. For example, in figure \ref{fig:VBSWWdiags}, the contraction of the QCD-induced diagram (see figure \ref{fig:VBSWWdiagf}) with the VBS diagrams (top row) vanishes due to the color structure, whereas the corresponding contraction with the EW background diagrams (see figures \ref{fig:VBSWWdiagd} and \ref{fig:VBSWWdiage}) yields a non-zero interference contribution at order $\mathcal{O}(\alpha_s\alpha^5)$. Consequently, rather than normalizing to $\Sigma_{\mathrm{LO}_1}$, we have taken the ratio over $\Sigma_{\mathrm{LO}_3}$ in table \ref{tab:WWNLOxs}.

At NLO, the $\mathcal{O}(\alpha)$ correction to $\Sigma_{\mathrm{LO}_3}$, namely $\Sigma_{\mathrm{NLO}_4}$, results in the dominant quantum correction, amounting to $-15.30\%$. This can be understood primarily by considering the EW Sudakov logarithms~\cite{Biedermann:2016yds}. The second-largest NLO correction, contributing $-4.01\%$ with respect to $\Sigma_{\mathrm{LO}_3}$, comes from NLO$_3$. The remaining NLO corrections are negligible. Photon-induced processes have not been included in the NLO terms; if they were, $\Sigma_{\mathrm{NLO}_4}/\Sigma_{\mathrm{LO}_3}$ would become $-13.55\%$ using the \luxqed\ photon PDF, while the other NLO terms would remain unchanged. Overall, the NLO corrections approximately follow the reverse hierarchy given in eq.\eqref{eq:hierachy2}.

\section{Giant $K$ factors\label{sec:GiantK}}

\begin{figure}[!t]
    \centering
    \subfloat[]
    {
        \includegraphics[width=4cm, height=3.5cm]{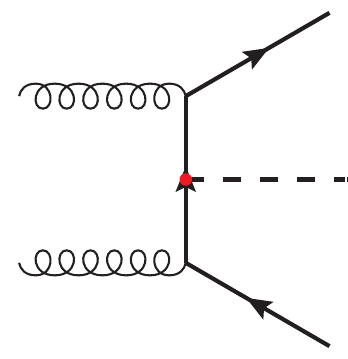}
        \label{fig:ggHbb}
    }
    \subfloat[]
    {
        \includegraphics[width=4cm, height=3.5cm]{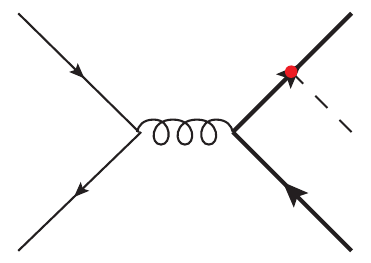}
        \label{fig:qqHbb}
    }
    \subfloat[]
    {
        \includegraphics[width=4cm, height=3.5cm]{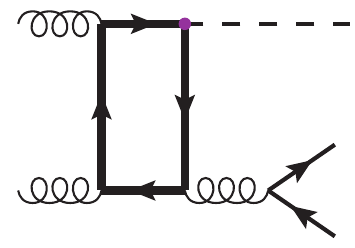}
        \label{fig:Yt}
    }\\
    \subfloat[]
    {
        \includegraphics[width=4cm, height=3.5cm]{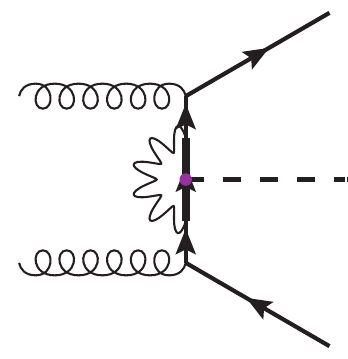}
        \label{fig:gg2hbb_yt}
    }
    \subfloat[]
    {
        \includegraphics[width=4cm, height=3.5cm]{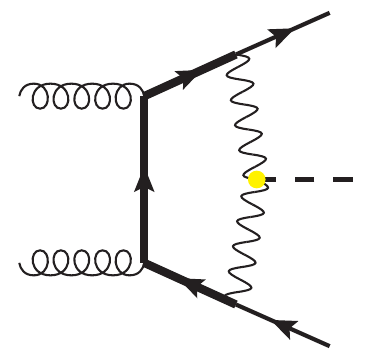}
        \label{fig:ggHbb_W}
    }
    \subfloat[]
    {
        \includegraphics[width=4cm, height=3.5cm]{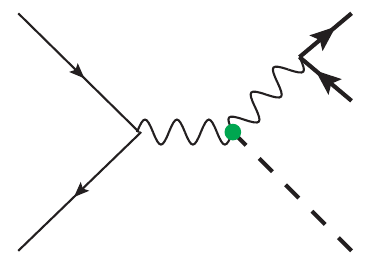}
        \label{fig:VH}
    }\\

    \subfloat[]
    {
        \includegraphics[width=4cm, height=3.5cm]{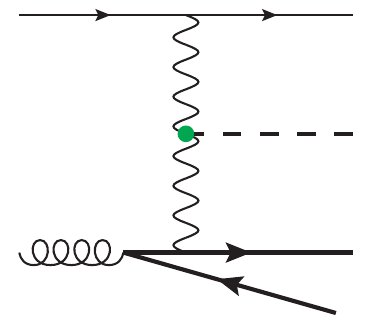}
        \label{fig:VBF}
    }
    \caption{Sample Feynman diagrams appearing in the complete NLO calculation for $pp\to b\bar{b}h$ production. The thick, medium-thick, and thin solid lines represent top, bottom, and light (anti)quarks, respectively. The dashed lines denote the Higgs boson, the curly lines indicate gluons, and the wiggly lines represent the weak bosons ($W^\pm$ and $Z$). The red, violet, green, and yellow bullets correspond to $b\bar{b}h$, $t\bar{t}h$, $hZZ$, and $hW^+W^-$ interactions, respectively. The figure is adapted from ref.~\cite{Pagani:2020rsg}.}
    \label{fig:sample_bbh_subfigures}
\end{figure}

My final example is the Higgs boson production in association with a bottom quark pair, $pp\to b\bar{b}h$, which illustrates how NLO EW corrections can sometimes yield giant $K$ factors when new channels open up. The $b\bar{b}h$ production process is well known for directly probing the bottom-Higgs Yukawa coupling $y_b$, complementing the extraction of $y_b$ based on the decay $h\to b\bar{b}$. However, it is also affected by significant irreducible backgrounds from other Higgs production channels, which I will elucidate in the following. Representative Feynman diagrams can be found in figure \ref{fig:sample_bbh_subfigures}.

The first complete NLO calculation was carried out in ref.~\cite{Pagani:2020rsg} using \mgamcshort\ within the $4$-quark flavor number (4FS) scheme. The Born process is given by
\begin{equation}
pp\to b\bar{b} h.
\end{equation}
To generate the complete NLO calculation in  \mgamcshort, one can execute the following commands:
\vskip 0.25truecm
\noindent
~~\prompt\ {\tt ~set~complex\_mass\_scheme~true}

\noindent
~~\prompt\ {\tt ~import~model~loop\_qcd\_qed\_sm\_Gmu\_4FS-with\_b\_mass}

\noindent
~~\prompt\ {\tt ~define p = g d d\~{} u u\~{} s s\~{} c c\~{} a}

\noindent
~~\prompt\ {\tt ~generate p p > b b\~{} h aS=2 aEW=3 [QCD QED]}
\vskip 0.25truecm
\noindent
The NLO UFO model {\tt loop\_qcd\_qed\_sm\_Gmu\_4FS-with\_b\_mass}~\footnote{Strictly speaking, it represents the UFO model {\tt loop\_qcd\_qed\_sm\_Gmu\_4FS} with the {\tt restrict\_with\_b\_mass.dat} restriction card applied, where the latter serves as a restriction card within the model.} is used for the complete NLO calculations in the 4FS scheme and the $G_\mu$ scheme. Note that since we keep the mass of the bottom quark in the 4FS scheme,  the bottom (anti)quark is not included in the PDF definition. In this process, we have $k_0=3$, $c_s(k_0)=0$, $c(k_0)=1$, and $\Delta(k_0)=2$, as defined in section \ref{eq:EWgeneral}. There are three LO terms:
\begin{equation}
\Sigma_{\mathrm{LO}_1}~(\mathcal{O}(\alpha_s^2\alpha)),\quad \Sigma_{\mathrm{LO}_2}~(\mathcal{O}(\alpha_s\alpha^2)),\quad \Sigma_{\mathrm{LO}_3}~(\mathcal{O}(\alpha^3)),
\end{equation}
and four NLO terms:
\begin{equation}
\Sigma_{\mathrm{NLO}_1}~(\mathcal{O}(\alpha_s^3\alpha)), \quad \Sigma_{\mathrm{NLO}_2}~(\mathcal{O}(\alpha_s^2\alpha^2)),\quad \Sigma_{\mathrm{NLO}_3}~(\mathcal{O}(\alpha_s\alpha^3)),\quad \Sigma_{\mathrm{NLO}_4}~(\mathcal{O}(\alpha^4)).
\end{equation}

By considering complete NLO predictions, new topologies emerge alongside the genuine $b\bar{b}h$ contribution, which is proportional to $y_b^2$ at LO$_1$ (\eg, the Born diagrams shown in figures \ref{fig:ggHbb} and \ref{fig:qqHbb}). These contributions are summarized in  table \ref{table:Hbb_orders}. Let me explain them one by one:
\begin{itemize}
\item The LO$_1$ originates solely from the genuine $b\bar{b}h$ production via gluon-gluon fusion (see diagram \ref{fig:ggHbb}) and quark-antiquark annihilation (see diagram \ref{fig:qqHbb}). This term is proportional to $y_b^2$. At the LHC, the gluon-gluon fusion channel is dominant due to the size of the partonic luminosity. Photon-induced partonic channels are present in LO$_2$ and LO$_3$.
\item At LO$_3$, a new topology arises, namely the $Zh$ associated production, where the $Z$ boson subsequently decays into a $b\bar{b}$ pair (illustrated in figure \ref{fig:VH}). Since the $Z$ boson is typically on-shell, the LO$_3$ contribution is not expected to be suppressed relative to LO$_1$ by a factor of $\alpha^2/\alpha_s^2$.  At the differential level, events from LO$_1$ and LO$_3$ populate very different phase space regions. The $Zh$ topology, which features the $hZZ$ coupling rather than  $y_b$, also appears at NLO$_3$ and NLO$_4$. The two orders can be viewed as NLO QCD and EW corrections to LO$_3$ respectively. 
\item NLO$_1$ (NLO QCD) receives a contribution from an additional topology. The $gg\to b\bar{b}h$ Born diagrams (like that in figure \ref{fig:ggHbb}) interfere with gluon-fusion Higgs production at one-loop with an additional emission of a bottom-quark pair (see diagram \ref{fig:Yt}), denoted as ``$gg\mathrm{F}+b\bar{b}$". This interference leads to a term proportional to $y_by_t$, where the top-Higgs Yukawa coupling $y_t$ is significantly larger than $y_b$~\cite{Wiesemann:2014ioa}. In fact, the square of $gg\mathrm{F}+b\bar{b}$ contributes to a NNLO QCD term  (\ie, NNLO$_1$, following the same convention as in section \ref{eq:EWgeneral}) featuring $y_t^2$. The $y_ty_b$ term is non-negligible compared to the term proportional to $y_b^2$ originating from genuine $b\bar{b}h$ production, and the $y_t^2$ term at NNLO QCD is much larger than the $y_b^2$ contribution. Both the $y_by_t$ and $y_t^2$ terms have been calculated at NLO QCD~\cite{Deutschmann:2018avk}. If at least one $b$-jet is required, the $y_ty_b$ term is around $-20\%$ of the $y_b^2$ term, while the $y_t^2$ term is even four times larger than the $y_b^2$ contribution. The interference contribution from $gg\mathrm{F}+b\bar{b}$ is also present at NLO$_2$, where the bottom-quark pair arises from a photon or $Z$-boson propagator instead of the gluon propagator shown in figure \ref{fig:Yt}. Other $y_by_t$ terms at NLO$_2$ can be induced by the interference between one-loop diagrams (such as the one in figure \ref{fig:gg2hbb_yt}) and the $gg\to b\bar{b}h$ Born diagrams. Similarly, one-loop diagrams (like figure \ref{fig:ggHbb_W}) induced by the $hW^+W^-$ vertex can interfere with $gg\to b\bar{b}h$ at NLO$_2$.~\footnote{An interesting aspect of the $gg\to b\bar{b}h$ process involves the so-called anomalous thresholds, or leading Landau singularities, as discussed in refs.~\cite{Boudjema:2008zn,Boudjema:2007uh} in the context of one-loop diagrams, such as figure \ref{fig:ggHbb_W}, but with an $s$-channel gluon attached to the box loop. These anomalous thresholds occur when $M_h>2M_W$, a condition not satisfied in the SM.  The Landau singularities can be mitigated by incorporating the widths of the internal loop particles. Identifying these anomalous thresholds in the $S$-matrix is essential for unambiguously establishing genuine resonances and understanding the analyticity of the $S$-matrix~\cite{Hannesdottir:2022bmo}. Their impact on other LHC processes has been explored in ref.~\cite{Passarino:2018wix}.} Additionally, one-loop diagrams for di-Higgs production, where a Higgs boson subsequently decays into $b\bar{b}$, can also interfere with the Born $gg\to b\bar{b}h$ diagrams, contributing to NLO$_2$.
\item The vector-boson fusion (VBS) topology, illustrated in diagram \ref{fig:VBF}, emerges at NLO$_3$ in the real emission contribution, where an initial gluon splits into a bottom quark pair. These $t$-channel diagrams can potentially lead to a very large NLO$_3$ contribution with distinct differential distributions compared to other contributions. If the initial gluon is replaced with a photon, a similar VBF topology contributes at NLO$_4$.
\end{itemize}
In general, EW corrections can induce sensitivity to any other SM EW interactions, particularly in the case of the Higgs boson, affecting interactions different from $y_b$ that have much larger coupling constants.  Thus, all perturbative orders are in principle non-negligible and exhibit different shapes at the differential level.

\begin{table}[!t]
\begin{center}
\begin{tabular}{? l| l ? l| l?}
\hlinewd{1pt}
Order & Topologies & Order & Topologies \\
\hlinewd{1pt}
LO$_1$ $(\mathcal{O}(\alpha_s^2\alpha))$ & $gg,q\bar q \to b\bar{b} h$ & NLO$_1$  $(\mathcal{O}(\alpha_s^3\alpha))$ & $b\bar{b}h$, $\cancel{gg{\rm F}+b \bar{b}}$ \\\hline
LO$_2$ $(\mathcal{O}(\alpha_s\alpha^2))$ & $\gamma g\to b\bar{b} h$   & NLO$_2$  $(\mathcal{O}(\alpha_s^2\alpha^2))$ & $b\bar{b}h$, $gg\mathrm{F}+b \bar{b}$ \\\hline
\multirow{2}{*}{LO$_3$ $(\mathcal{O}(\alpha^3))\phantom{\alpha^2}$} & $q\bar q \to Zh (Z \to b \bar{b} )$ & NLO$_3$  $(\mathcal{O}(\alpha_s\alpha^3))$ & $Zh$, VBF  \\\cline{3-4}
& $ q\bar q, \gamma \gamma\to b\bar{b} h$ & NLO$_4$  $(\mathcal{O}(\alpha^4))\phantom{\alpha^2}$ & $Zh$, VBF \\
\hlinewd{1pt}
\end{tabular}
\end{center}
\caption{ Topologies of the $pp\to b\bar{b}h$ process at LO (left), with specified initial states, and at NLO (right). The term proportional to $y_b y_t$ at NLO$_1$, arising from the interference between the $b\bar{b}h$ and $gg\mathrm{F}+b\bar{b}$ topologies, has been excluded from the calculation.} 
\label{table:Hbb_orders} 
\end{table}

\begin{figure}[t!]
\centering
\subfloat[]{\includegraphics[width=.4\textwidth,draft=false]{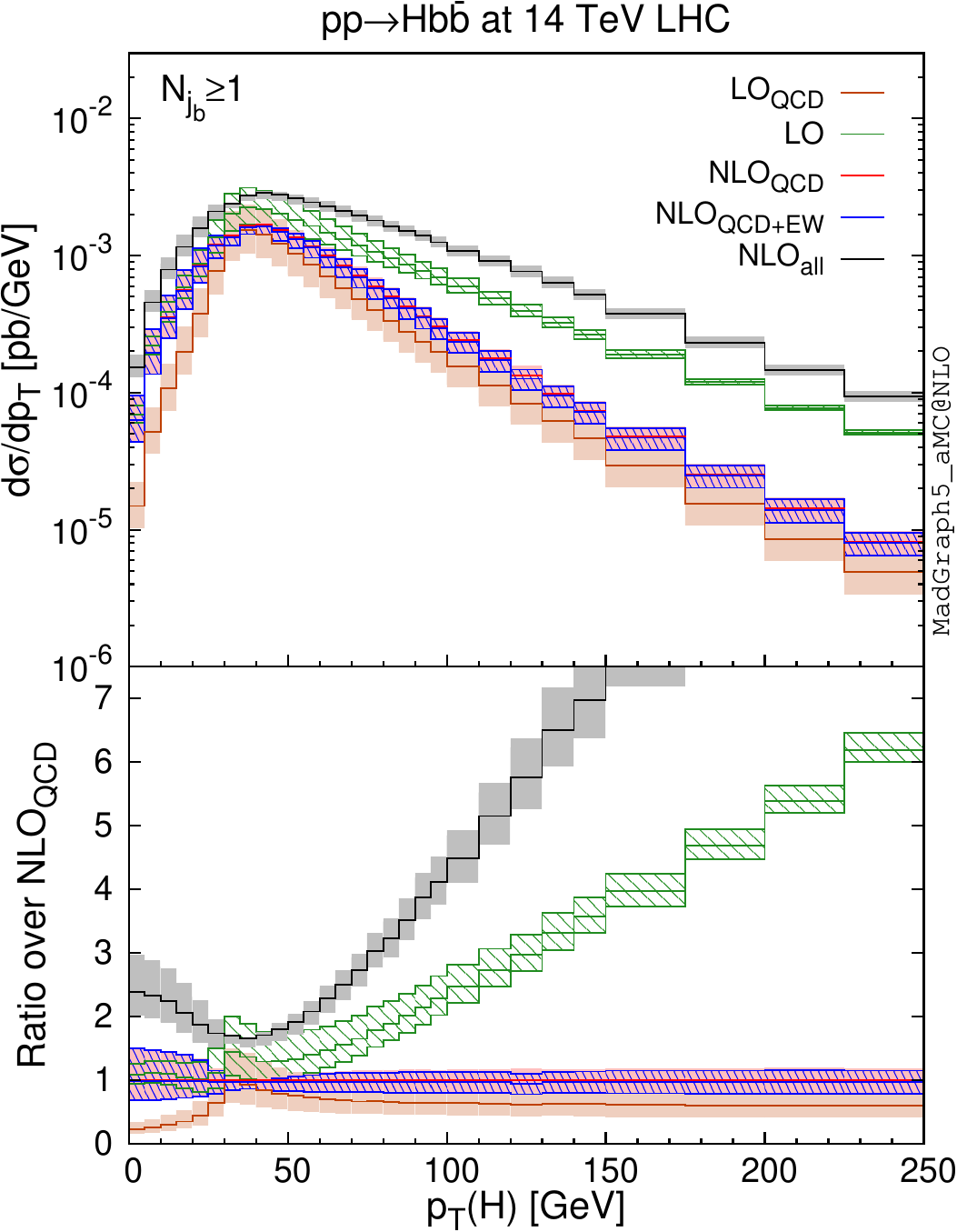}\label{fig:Hbb_ptha}}
\subfloat[]{\includegraphics[width=.4\textwidth,draft=false]{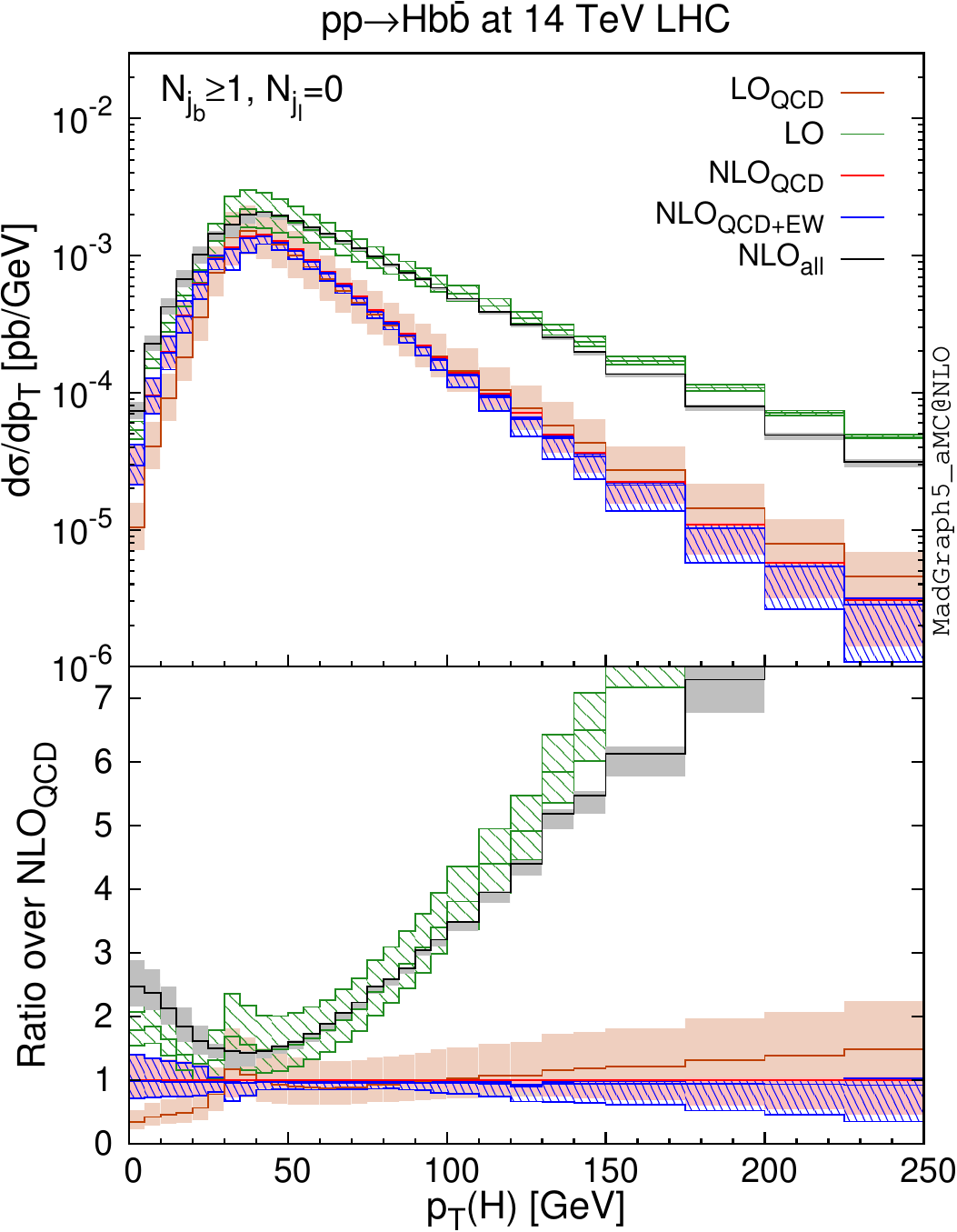}\label{fig:Hbb_pthb}}\\
%\vspace*{-5mm}
\subfloat[]{\includegraphics[width=.4\textwidth,draft=false]{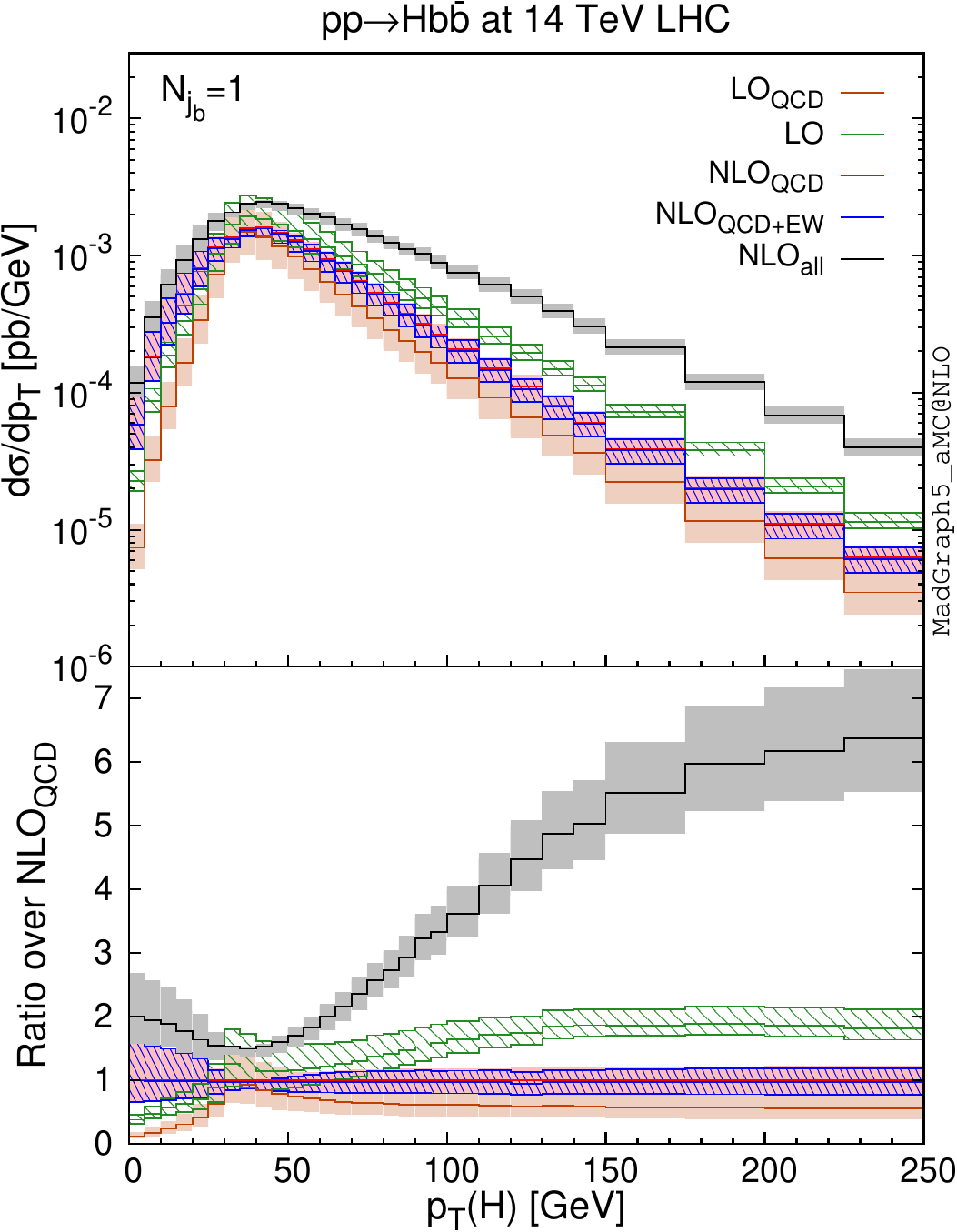}\label{fig:Hbb_pthc}}
\subfloat[]{\includegraphics[width=.4\textwidth,draft=false]{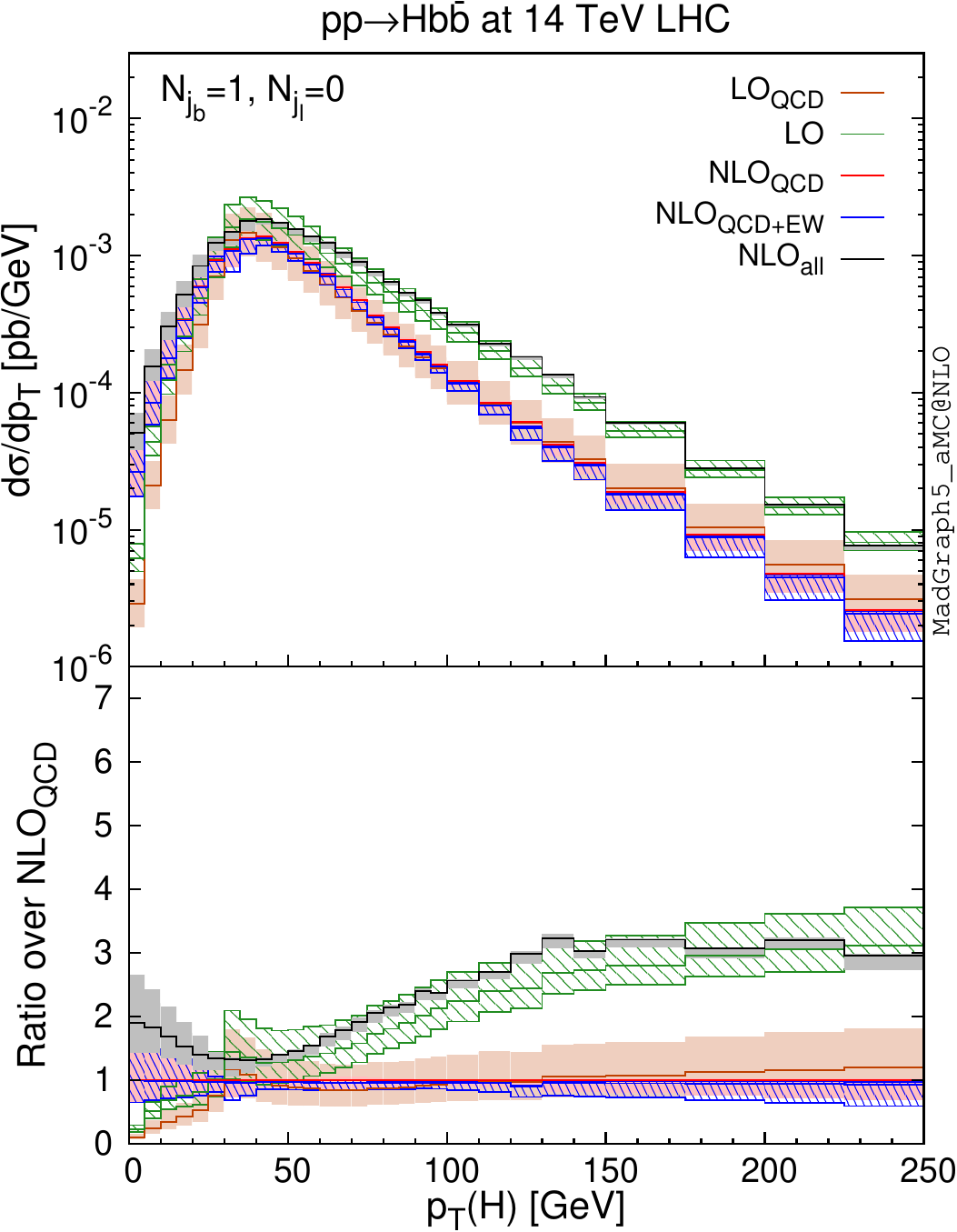}\label{fig:Hbb_pthd}}\\
%\vspace*{-5mm}
\caption{Transverse momentum distributions of the Higgs boson in the process $pp\to b \bar{b} h$ at the 14 TeV LHC. The top row corresponds to the case where $N_{j_b}\ge1$, while the bottom row is for $N_{j_b}=1$. The right plots include a light-jet veto. The figure is from ref.~\cite{Pagani:2020rsg}.\label{fig:Hbb_pth}}
\end{figure}

In order to be clear and specific, let me present some physical results for the process in $pp$ collisions at the LHC with a center-of-mass energy of 14 TeV. The results are taken from ref.~\cite{Pagani:2020rsg}. Using the new notations introduced in section 2.3 and the setup specified in section 3.1 of the paper, I present the transverse momentum distributions of the Higgs boson in figure \ref{fig:Hbb_pth}. In all cases, the NLO$_{\mathrm{QCD}}$ term (red curve), which is the sum of $\Sigma_{\mathrm{LO}_1}$ and $\Sigma_{\mathrm{NLO}_1}$, is close to the NLO$_{\mathrm{QCD}+\mathrm{EW}}$ term (blue curve), which includes $\Sigma_{\mathrm{LO}_1}$, $\Sigma_{\mathrm{NLO}_1}$, and $\Sigma_{\mathrm{NLO}_2}$. This indicates that the NLO$_2$ contribution does not yield a giant $K$ factor. Conversely, the LO (green curve) and NLO$_{\mathrm{all}}$ (black curve) contributions are highly sensitive to the imposed cuts. Note that, according to section 2.3 in ref.~\cite{Pagani:2020rsg}, LO corresponds to the sum of all three LO blobs, while NLO$_{\mathrm{all}}$ is the sum of the three LO contributions and four NLO contributions. Therefore, the enhanced contributions when comparing LO with LO$_{\mathrm{QCD}}$ (from LO$_1$), NLO$_{\mathrm{QCD}}$, or NLO$_{\mathrm{QCD}+\mathrm{EW}}$ arise from the $Zh$ topology.  Meanwhile, the VBF topology is only included in NLO$_{\mathrm{all}}$. For the minimal cuts requiring at least one $b$ jet in the events, figure \ref{fig:Hbb_ptha} shows that both the $Zh$ and VBF topologies produce giant $K$ factors, enhancing the NLO$_{\mathrm{QCD}+\mathrm{EW}}$ cross section by an order of magnitude at $p_T(h)\sim 250$ GeV. The contributions from the $Zh$ topology can be mitigated by vetoing the second $b$ jet, as evidenced by comparing the green curves in figures \ref{fig:Hbb_pthc} and \ref{fig:Hbb_ptha}. This can be understood since the two $b$ quarks from the decay of the $Z$ boson are primarily produced centrally. In contrast, for events from the VBF topology, the two bottom quarks originate from an initial gluon splitting, with a significant chance that one $b$ (anti)quark has a rapidity too large to be detected. To remove the $Zh$ contribution, one can also reconstruct two $b$ jets and impose the invariant mass cuts on them. For the VBF topology, a light jet veto can effectively suppress its contribution. This distinction is clearer when comparing the black curves in the right plots with those in the left plots of figure \ref{fig:Hbb_pth}. 

In conclusion, even when not aimed at improving the precision of theoretical results, NLO EW radiative corrections should not be overlooked, as they can significantly affect predictions by introducing new channels that change the order of magnitude. Identifying these corrections can sometimes be challenging. In the case of $b\bar{b}h$, while the $Zh$ topology is straightforward to recognize, the VBF contribution is less obvious, as it represents a higher-order radiative correction effect. The occurrence of giant $K$ factors is not a novel phenomenon exclusive to EW corrections; several examples in the context of QCD have been documented in the literature, demonstrating substantial higher-order corrections. Notable cases include $Z$ boson production in association with a jet and quarkonium hadroproduction. Dedicated methods have been proposed to address these significant QCD radiative corrections, as discussed in ref.~\cite{Rubin:2010xp} for $Z$+jet and ref.~\cite{Shao:2018adj} for quarkonium.

%\begin{table}[!t]
%\begin{center}
%\begin{tabular}{l|  p{5cm}}
%\toprule
%Order & Topologies \\
%\midrule
%LO$_1$ $(\mathcal{O}(\alpha_s^2\alpha))$ & $gg,q\bar q \to b\bar{b} h$ \\
%LO$_2$ $(\mathcal{O}(\alpha_s\alpha^2))$ & $\gamma g\to b\bar{b} h$   \\
%\multirow{2}{*}{LO$_3$ $(\mathcal{O}(\alpha^3))\phantom{\alpha^2}$} & $q\bar q \to Zh (Z \to b \bar{b} )$,  \\
%& $ q\bar q, \gamma \gamma\to b\bar{b} h$\\ 
%\midrule
%Order & Topologies \\
%\midrule
%NLO$_1$  $(\mathcal{O}(\alpha_s^3\alpha))$ & $b\bar{b}h$, $\cancel{gg{\rm F}+b \bar{b}}$\\
%NLO$_2$  $(\mathcal{O}(\alpha_s^2\alpha^2))$ & $b\bar{b}h$, $gg\mathrm{F}+b \bar{b}$ \\ 
%NLO$_3$  $(\mathcal{O}(\alpha_s\alpha^3))$ & $Zh$, VBF \\
%NLO$_4$  $(\mathcal{O}(\alpha^4))\phantom{\alpha^2}$ & $Zh$, VBF \\
%\bottomrule
%\end{tabular}
%\end{center}
%\caption{ Topologies entering at LO, with initial states that are explicitly specified, and at NLO. As discussed in section~\ref{sec:msbar}, the terms proportional to $y_b y_t$ at NLO$_1$, emerging from the interference of $b\bar{b}h$ and $gg\mathrm{F}+b\bar{b}$ topologies, are not taken into account in our calculation.} 
%\label{table:Hbb_orders} 
%\end{table}

%$HW$ and $Hb\bar{b}$

%\subsection{$HW$}

%\subsection{$Hb\bar{b}$}

\chapter{Outlook}
\label{SEC:outlook}

The discovery of a Higgs boson with a mass of approximately 125 GeV at the CERN LHC has fundamentally shifted the perspective of particle physics. Alongside the null results from direct and indirect searches for BSM phenomena at colliders and other experiments, all tests of strong and EW interactions described by the SM have passed without indicating the need for any BSM effects. This suggests that, while many observed particle phenomena remain poorly understood, they can be explained within the framework of the SM. Consequently, any potential BSM effects that may be accessible to current and future experiments are likely to be challenging to detect, as new particles are either heavy or interact weakly with SM particles. In this context, precision has become paramount in both experimental analyses and theoretical predictions. Improving the precision of theoretical calculations by incorporating more perturbative corrections, including EW radiative corrections, is essential for the success of current and future collider programs. Unlike traditional methods of exploring physics on a case-by-case basis, automation through advanced computational techniques allows researchers to systematically analyze experimental data and explore new ideas. This capability enables the generation of comprehensive insights and results much more efficiently across a wide range of scenarios, particularly in the LHC era, where thousands of physicists conduct numerous analyses every day.

The dissertation reports a step forward in this direction, specifically the automation of EW correction computations within the \mgamcshort\ framework. It introduces established techniques and highlights the key features that underpin NLO calculations performed by \mgamcshort\ in the context of quantum corrections arising from both QCD and EW interactions in the SM. To illustrate the phenomenological relevance of EW corrections at the LHC, I first present the NLO EW corrections to the integrated cross sections for a variety of processes at the 13 TeV LHC. While the typical size of these corrections ranges from a few percent to more than 10\%, making them comparable to NNLO QCD corrections, they can be enhanced through various mechanisms. Several known examples demonstrating these enhancements are provided in this dissertation. From a phenomenological perspective, one of the most intriguing findings of this study is that the numerical effects of various subleading terms are often challenging to predict using simple power counting arguments based on the hierarchy of the couplings.

The primary application of this new development in the near future will be the systematic computation of QCD+EW corrections for processes of interest, as is gradually being observed in the current LHC data analyses. However, there are a couple of limitations in the current implementation that need to be addressed in the future. The most pressing issue is to enable particle level predictions by matching the complete NLO calculations to PSMC simulations, including both QCD and QED, as well as weak-boson parton showers. This would make the complete NLO results directly applicable to low-level experimental data analysis, such as in correcting detector efficiencies. Secondly, given the interesting challenges posed by tagged photons and leptons, exploring the fragmentation function approach in the context of full QCD and EW corrections in \mgamcshort\ would be valuable. This exploration could start with theoretically motivated functions associated with photons and leptons to be extracted from experimental data. Moreover, applications in BSM theories, either in UV-complete models or effective field theories, require careful consideration of renormalization and the appropriate choice of input parameter schemes. Finally, the automation of EW Sudakov logarithm resummation up to NLL accuracy would be essential for the physics at the future colliders, such as a multi-TeV muon collider or a 100 TeV $pp$ collider. 

In order to fully explore the physics potential of the LHC, we must look ahead. Firstly, the LHC has proven to be a highly versatile machine. In addition to its role in precision tests of the SM and searches for BSM physics via inclusive hard reactions, it hosts extensive heavy-ion programs aimed at studying new phenomena, such as soft physics, small-$x$ physics, and collective effects. A notable example is photon-photon physics, which can lead to novel BSM and SM investigations in ultraperipheral collisions at the LHC. Another area of my research focuses on the physics of quarkonium, where quarkonium serves as an intriguing probe for studying largely unexplored aspects of the strong interaction. Extending the capabilities of the \mgamcshort\ framework to encompass these cases will form a part of my research in the coming years. 

%Finally, I would like to point out that I am also interested in developing resummation and NNLO techniques. Whether their calculations can be automated like in the NLO case is still unclear now. The timing and the direction of this kind of research will depend on the experimental developments, and cannot be fore- seen in much detail. 

\newpage
\addcontentsline{toc}{chapter}{Bibliography}
\bibliographystyle{utphys}
\bibliography{HDR}

\end{document}